# EVALUATION OF INTEGRALS RELATED TO THE MAGNETIC FIELD INTEGRAL EQUATION OVER BILINEAR QUADRILATERALS

by

John S. Asvestas

10 July 2016

SUMMARY

We present a new method for computing the impedance matrix elements in the method of moments for geometries described by bilinear quadrilaterals (BQ) and for higher-order basis functions (HOBF). Our method is restricted to the Magnetic Field Integral Equation (MFIE) and focuses on the self-elements of the impedance matrix and elements for which the observation point (OP) is near the integration BQ. The method is based on the simple idea of analytical integration along one of the BQ's parameters and numerical integration along the remaining one. For the singular (or nearly so) part of the integral, we show through analysis and examples that our method can provide precision up to fifteen significant digits (SD).

# TABLE OF CONTENTS

# LIST OF TABLES

# LIST OF FIGURES

# LIST OF ACRONYMS

BQ: Bilinear Quadrilateral.
DEQ: Double Exponential Quadrature (a.k.a. TSQ).
DOA: Digits Of Agreement
DP: Double Precision.
EFIE: Electric Field Integral Equation.
GKQ: Gauss-Kronrod Quadrature.
MFIE: Magnetic Field Integral Equation.
NOR: Number Of Recursions.
OP: Observation Point.
SD: Significant Digit(s).
SP: Single Precision.
TSQ: Tanh-Sinh Quadrature (a.k.a. DEQ).

# 1. INTRODUCTION

In this report we explore ways to compute the singular integral of the Magnetic Field Integral Equation (MFIE) over a Bilinear Quadrilateral (BQ). Despite the importance of MFIE in boundary-integral-equation formulations of electromagnetic problems, either on its own for closed bodies ([1], pp. 354-355) or in combination with the Electric Field Integral Equation (EFIE) in avoiding resonances ([2], p. 500), we have not been able to find any mention in the open literature on how to evaluate this integral over a BQ [3, 4]. We follow here the approach we used in evaluating the corresponding singular integral for EFIE [5]. As we did in [5], we extract the singular part of the integral in question, we integrate it analytically along one of the variables, and we use numerical experiments to ascertain that the numerical algorithms we use provide precision up to 15 significant digits (SD) along the other variable.

In Section 2, we state the problem and we remove the non-smooth part of the integral that appears in MFIE. We then proceed to define geometric entities in BQ coordinates, and introduce higher-order basis functions to represent the linear surface-current density. We introduce these currents to the singular part of the integral and catalog the various forms the integrals that result and that we must consider. In Sections 3 through 8, we discuss how to integrate integrals that contain the inverse of the distance function to the third power. In Section 9, we consider the integrals that contain the inverse of the distance function to the first power. In Section 10, we consider the case in which the Observation Point (OP) projects to the boundary of the BQ. In Section 11, we offer some concluding remarks and suggestions for additional investigations. We also include material from [5] in Appendices A and B. In Appendix A, we provide a mathematical description of a BQ while, in Appendix B, we present three BQs that may be used for testing purposes.

As we mentioned above, the approach we take in evaluating the two dimensional integral is to integrate analytically along one of the dimensions and numerically along the other. All numerical calculations as well as graphs were made using Mathematica® 7 [6]. We employed two numerical methods in Mathematica®: the Gauss-Kronrod [7] and the double exponential [7], [8]. They are two very distinct numerical integration methods and employing both provides a degree of confidence when aiming at Double Precision (DP) (15 SD). High-precision is required when simulating metallic enclosures in which energy can creep in as well as in applications where tracking phase accurately is of importance.

## 2. STATEMENT OF THE PROBLEM AND INTEGRALS OF INTEREST

For harmonic time dependence, $+i\check{S}t$, the MFIE for a perfectly conducting closed surface is ([1], pp. 354-5)

$$\frac{1}{2}\mathbf{J}^t(\mathbf{r}) + \int_S \hat{n}' \times \left[\mathbf{J}^t(\mathbf{r}) \times \nabla g(\mathbf{r}, \mathbf{r}')\right] dS = \mathbf{J}^i(\mathbf{r}) \tag{2.1}$$

where $g$ is the free-space Green's function defined by

$$g(\mathbf{r}, \mathbf{r}') = -\frac{e^{-ik|\mathbf{r}-\mathbf{r}'|}}{4f|\mathbf{r}-\mathbf{r}'|} \quad , \quad k = \check{S}\sqrt{v_0 \tilde{~}_0} \,. \tag{2.2}$$

Here the primed point is the Observation Point (OP) and lies on the closed surface $S$. The contribution of the singularity has been taken into account in deriving MFIE; thus, the integral is a direct value integral and is absolutely integrable provided the linear current density is continuous over $S$ ([9], pp. 194-195). We assume that the surface $S$ is made up of a finite number of BQs and that the linear current density is continuous over the closure of each BQ. In this report, we examine ways to compute the integral in (2.1) when the OP is on or near the integration BQ.

The first item we investigate is the number of terms we must extract from the Green's function before we have a remainder that is numerically stable. We let

$$R = |\mathbf{r} - \mathbf{r}'| \tag{2.3}$$

and calculate

$$\nabla\left[\frac{e^{-ikR}}{R}\right] = \nabla\left[\frac{\cos(kR) - i\sin(kR)}{R}\right] = \nabla\left[\frac{1}{R} - \frac{k^2}{2}R + \frac{k^4}{4!}R^3 - \ldots\right] - ik\nabla\left[1 - \frac{k^2}{3!}R^2 + \ldots\right]$$

$$= \nabla\left(\frac{1}{R}\right) - \frac{k^2}{2}\frac{\mathbf{r}-\mathbf{r}'}{R} + \frac{3k^4}{4!}R^2\frac{\mathbf{r}-\mathbf{r}'}{R} - \ldots + i\frac{k^3}{3!}2R\frac{\mathbf{r}-\mathbf{r}'}{R} - \ldots$$

$$= \nabla\left(\frac{1}{R}\right) - \frac{k^2}{2}\frac{\mathbf{r}-\mathbf{r}'}{R} + \frac{3k^4}{4!}R(\mathbf{r}-\mathbf{r}') - \ldots + i\frac{k^3}{3!}2(\mathbf{r}-\mathbf{r}') - \ldots \tag{2.4}$$

As $\mathbf{r}$ tends to $\mathbf{r}'$, the first term above becomes infinite while the direction of the unit vector in the second is undefined. The third term and the last one tend to zero. We reach then the conclusion that, in order to stabilize the Green's function, we must extract the first two terms from the cosine term.

We let

$$\mathbf{f}(\mathbf{r},\mathbf{r}') = k\nabla \left[ \frac{\cos(kR) - 1 + \frac{(kR)^2}{2} - i\sin(kR)}{kR} \right]. \tag{2.5}$$

For small values of $R$, we write

$$\mathbf{f}(\mathbf{r},\mathbf{r}') = k\nabla \left[ \frac{\cos(kR) - 1 + \frac{(kR)^2}{2} - i\sin(kR)}{kR} \right] = k\nabla \left[ \frac{\cos(kR) - 1 + \frac{(kR)^2}{2}}{kR} \right] - ik\nabla \left[ \frac{\sin(kR)}{kR} \right]$$

$$= k\nabla \left[ \frac{1 - \frac{(kR)^2}{2} + \frac{(kR)^4}{4!} - \frac{(kR)^6}{6!} + \ldots - 1 + \frac{(kR)^2}{2}}{kR} \right] - ik\nabla \left[ 1 - \frac{(kR)^2}{3!} + \frac{(kR)^4}{5!} - \ldots \right]$$

$$= k\nabla \left[ \frac{(kR)^3}{4!} - \frac{(kR)^5}{6!} + \ldots \right] - ik\nabla \left[ 1 - \frac{(kR)^2}{3!} + \frac{(kR)^4}{5!} - \ldots \right]$$

$$= k^2 \left[ \frac{3(kR)^2}{4!} - \frac{5(kR)^4}{6!} + \ldots \right] \nabla R + ik^2 \left[ \frac{2(kR)}{3!} - \frac{4(kR)^3}{5!} + \ldots \right] \nabla R. \tag{2.6}$$

The term in the first bracket is $O(R^2)$ while that of the second is $O(R)$, as $R$ tends to zero. Since the gradient of $R$ is $\nabla R = (\mathbf{r} - \mathbf{r}')/R$, then we can write for (2.5) that

$$\mathbf{f}(\mathbf{r},\mathbf{r}') = \frac{k^2}{2}(\mathbf{r} - \mathbf{r}') O(R) + i\frac{k^2}{2}(\mathbf{r} - \mathbf{r}') O(1), \quad R \to 0 \tag{2.7}$$

and, hence, $\mathbf{f}$ tends to zero with $R$. We note that the analysis in (2.6) clearly displays the need for extracting not only the first but also the second term in the Maclaurin expansion of the cosine function. If we omit this term, then the first bracket in (2.6) becomes $O(1)$ and, thus, the entire first term in (2.6) is undefined when $R$ is equal to zero.

We can then write

$$\nabla \left[ \frac{e^{-ikR}}{R} \right] = k\nabla \left[ \frac{\cos(kR) - 1 + \frac{(kR)^2}{2} - i\sin(kR)}{kR} \right] + \nabla \left[ \frac{1}{R} - \frac{k^2}{2} R \right] \tag{2.8}$$

and for the integral

$$\int_S \hat{n}' \times \left[ \mathbf{J}^t(\mathbf{r}) \times \nabla g(\mathbf{r}, \mathbf{r}') \right] dS = -\frac{k}{4f} \int_S \hat{n}' \times \left\{ \mathbf{J}^t(\mathbf{r}) \times \nabla \left[ \frac{\cos(kR) - 1 + \frac{(kR)^2}{2} - i\sin(kR)}{kR} \right] \right\} dS$$

$$-\frac{k}{4f} \int_S \hat{n}' \times \left\{ \mathbf{J}^t(\mathbf{r}) \times \nabla \left[ \frac{1}{R} - \frac{k^2}{2} R \right] \right\} dS . \tag{2.9}$$

The first integral on the right can be evaluated using traditional methods. When $R$ is smaller than a pre-specified value, say $R_0$, we may wish to replace the trigonometric functions by their Maclaurin series truncated to the number of terms so that, for $R$ less or equal to $R_0$, the truncated sums represent exactly the trigonometric functions to the required number of SD. We indicate how this is done at the end of this section.

From Theorem 43, p. 194, in [9], we see that, as the OP approaches the BQ, the second integral on the right of (2.9) is equal to its direct value plus or minus a contribution from the singularity. In forming MFIE, we have already taken care of the singularity. In (2.9) then, we are considering the direct value of the integral, *i.e.*, with the normal moved inside the integral. According to Müller ([9], p. 195), this integral is absolutely integrable. In the BQ in which the OP lies, the integral over this BQ is treated as a principal value integral. This means that from the BQ we exclude a neighborhood of the OP and use a limiting process to extend the remaining surface to the OP.

In Appendix A, we present the mathematical description and properties of a BQ. If $(p', q')$ is a point in the domain of definition of the BQ, then, by (A.23), we can write the position vector to any point of the BQ in the form

$$\mathbf{r}(p,q) = \mathbf{r}_0 + \left(\mathbf{r}_p + \mathbf{r}_{pq} q'\right)(p - p') + \left(\mathbf{r}_q + \mathbf{r}_{pq} p'\right)(q - q') + \mathbf{r}_{pq}(p - p')(q - q') \tag{2.10}$$

where, by (A.22),

$$\mathbf{r}_0 = \mathbf{r}(p', q') = \mathbf{r}_{00} + \mathbf{r}_p p' + \mathbf{r}_q q' + \mathbf{r}_{pq} p' q', \quad |p'| \leq 1, \quad |q'| \leq 1. \tag{2.11}$$

If we introduce the covariant or unitary ([10], p. 39) basis vectors

$$\mathbf{p}(q) = \frac{\partial \mathbf{r}(p,q)}{\partial p} = \mathbf{r}_p + \mathbf{r}_{pq} q, \quad \mathbf{q}(p) = \frac{\partial \mathbf{r}(p,q)}{\partial q} = \mathbf{r}_q + \mathbf{r}_{pq} p \tag{2.12}$$

we can write for (2.10)

$$\mathbf{r}(p,q) = \mathbf{r}_0 + \mathbf{p}'(p - p') + \mathbf{q}'(q - q') + \mathbf{r}_{pq}(p - p')(q - q') \tag{2.13}$$

where

$$\mathbf{p}' = \mathbf{p}(q'), \quad \mathbf{q}' = \mathbf{q}(p'). \tag{2.14}$$

If we also make the transformation

$$u = p - p', \quad v = q - q'; \quad -(1+p') \leq u \leq 1-p', \quad -(1+q') \leq v \leq 1-q' \qquad (2.15)$$

then we can write for (2.13) that

$$\mathbf{r}(u,v) = \mathbf{r}_0 + \mathbf{p}'u + \mathbf{q}'v + \mathbf{r}_{pq}uv; \quad -(1+p') \leq u \leq 1-p', \quad -(1+q') \leq v \leq 1-q' \qquad (2.16)$$

where we have intentionally not changed the letter **r** for the position vector. The covariant vectors are not necessarily orthogonal and we assume that they are not co-linear. At the point $(p', q')$, the first is tangent to the curve $q = q'$, while the second to the curve $p = p'$.

The position vector to the BQ is given by (2.9). We define the unit normal at the point $(p, q)$ by

$$\hat{n}(p,q) = \frac{\mathbf{p}(q) \times \mathbf{q}(p)}{|\mathbf{p}(q) \times \mathbf{q}(p)|}. \qquad (2.17)$$

We consider an OP that projects to a point $(p', q')$ of the BQ. For its position vector, we can write

$$\mathbf{r}' = \mathbf{r}(p',q') + \hat{n}'h = \mathbf{r}_{00} + \mathbf{r}_p p' + \mathbf{r}_q q' + \mathbf{r}_{pq} p'q' + \hat{n}(p',q')h \qquad (2.18)$$

where $h$ is the projection distance. For the distance vector between the IP and the OP, we use (2.10) to write

$$\mathbf{R} = \mathbf{r} - \mathbf{r}' = \mathbf{r}_p(p-p') + \mathbf{r}_q(q-q') + \mathbf{r}_{pq}(pq - p'q') - \hat{n}(p',q')h$$
$$= \mathbf{p}'(p-p') + \mathbf{q}'(q-q') + \mathbf{r}_{pq}(p-p')(q-q') - \hat{n}(p',q')h \qquad (2.19)$$

or, with (2.15)

$$\mathbf{R}(u,v,h) = \mathbf{p}'u + \mathbf{q}'v + \mathbf{r}_{pq}uv - \hat{n}'h, \quad \hat{n}' = \hat{n}(p',q') \qquad (2.20)$$

and

$$|\mathbf{R}(u,v,h)|^2 = |\mathbf{p}'|^2 u^2 + |\mathbf{q}'|^2 v^2 + |\mathbf{r}_{pq}|^2 (uv)^2 + 2\left[(\mathbf{p}' \cdot \mathbf{q}') + (\mathbf{p}'u + \mathbf{q}'v - \hat{n}'h) \cdot \mathbf{r}_{pq}\right]uv + h^2 \qquad (2.21)$$

From this last expression, we see that the square of the distance is a second degree polynomial in $u$ and in $v$. We will take advantage of this in subsequent sections.

We express the linear current density in the form

$$\mathbf{J}^l(\mathbf{r}) = \frac{\mathbf{j}(\mathbf{r})}{J(p,q)}, \quad J(p,q) = \left|\frac{\partial \mathbf{r}(p,q)}{\partial p} \times \frac{\partial \mathbf{r}(p,q)}{\partial q}\right|. \qquad (2.22)$$

We return to (2.9) and write

$$\hat{n}' \times \left[ \mathbf{J}^t(\mathbf{r}) \times \nabla \left( \frac{1}{R} - \frac{k^2}{2} R \right) \right] = \frac{\hat{n}'}{J(p,q)} \times \left[ \mathbf{j}(\mathbf{r}) \times \nabla \left( \frac{1}{R} - \frac{k^2}{2} R \right) \right]$$

$$= \frac{\hat{n}'}{J(p,q)} \times \left\{ [\mathbf{j}(\mathbf{r}) - \mathbf{j}'] \times \nabla \left( \frac{1}{R} - \frac{k^2}{2} R \right) \right\} + \frac{\hat{n}'}{J(p,q)} \times \left\{ \mathbf{j}' \times \nabla \left( \frac{1}{R} - \frac{k^2}{2} R \right) \right\} \quad (2.23)$$

where $R$ is the magnitude of $\mathbf{R}$, and

$$\mathbf{j}' = \mathbf{j}(\mathbf{r})\big|_{(p',q',0)}. \quad (2.24)$$

For the first term on the right, we have

$$\frac{\hat{n}'}{J(p,q)} \times \left\{ [\mathbf{j}(\mathbf{r}) - \mathbf{j}'] \times \nabla \left( \frac{1}{R} - \frac{k^2}{2} R \right) \right\}$$

$$= \frac{\mathbf{j}(\mathbf{r}) - \mathbf{j}'}{J(p,q)} \hat{n}' \cdot \nabla \left( \frac{1}{R} - \frac{k^2}{2} R \right) - \frac{\hat{n}' \cdot [\mathbf{j}(\mathbf{r}) - \mathbf{j}']}{J(p,q)} \nabla \left( \frac{1}{R} - \frac{k^2}{2} R \right)$$

$$= \frac{-\hat{n}' \times \{\hat{n}' \times [\mathbf{j}(\mathbf{r}) - \mathbf{j}']\} + \hat{n}' \{\hat{n}' \cdot [\mathbf{j}(\mathbf{r}) - \mathbf{j}']\}}{J(p,q)} \hat{n}' \cdot \nabla \left( \frac{1}{R} - \frac{k^2}{2} R \right)$$

$$- \frac{\hat{n}' \cdot [\mathbf{j}(\mathbf{r}) - \mathbf{j}']}{J(p,q)} \left[ -\hat{n}' \times \left\{ \hat{n}' \times \nabla \left( \frac{1}{R} - \frac{k^2}{2} R \right) \right\} + \hat{n}' \left\{ \hat{n}' \cdot \nabla \left( \frac{1}{R} - \frac{k^2}{2} R \right) \right\} \right]$$

$$= -\frac{\hat{n}' \times \{\hat{n}' \times [\mathbf{j}(\mathbf{r}) - \mathbf{j}']\}}{J(p,q)} \left[ \hat{n}' \cdot \nabla \left( \frac{1}{R} - \frac{k^2}{2} R \right) \right] + \frac{\hat{n}' \cdot [\mathbf{j}(\mathbf{r}) - \mathbf{j}']}{J(p,q)} \left\{ \hat{n}' \times \left[ \hat{n}' \times \nabla \left( \frac{1}{R} - \frac{k^2}{2} R \right) \right] \right\} \quad (2.25)$$

The general form of the two components of the non-normalized linear current density is

$$\mathbf{j}_p(\mathbf{r}) = \mathbf{p}(q) \sum_{l=0}^{L_p} \sum_{m=0}^{M_p} P_{l,m} p^l q^m, \quad \mathbf{j}_q(\mathbf{r}) = \mathbf{q}(p) \sum_{l=0}^{L_q} \sum_{m=0}^{M_q} Q_{l,m} p^l q^m \quad (2.26)$$

where $P_{l,m}$ and $Q_{l,m}$ are constants. From (2.12), (2.14) and (2.15)

$$\mathbf{p}(q) = \mathbf{r}_p + \mathbf{r}_{pq}(q - q') + \mathbf{r}_{pq} q' = \mathbf{r}_p + \mathbf{r}_{pq} q' + \mathbf{r}_{pq} v = \mathbf{p}' + \mathbf{r}_{pq} v \quad (2.27)$$

and

$$\mathbf{q}(p) = \mathbf{r}_q + \mathbf{r}_{pq}(p - p') + \mathbf{r}_{pq} p' = \mathbf{r}_q + \mathbf{r}_{pq} p' + \mathbf{r}_{pq} u = \mathbf{q}' + \mathbf{r}_{pq} u. \quad (2.28)$$

Employing (2.15) and the last two equations in (2.21), we can write

$$\mathbf{j}_p(\mathbf{r}) = (\mathbf{p}' + \mathbf{r}_{pq}v) \sum_{l=0}^{L_p} \sum_{m=0}^{M_p} \overline{P}_{l,m}(p',q') u^l v^m, \quad \mathbf{j}_q(\mathbf{r}) = (\mathbf{q}' + \mathbf{r}_{pq}u) \sum_{l=0}^{L_q} \sum_{m=0}^{M_q} \overline{Q}_{l,m}(p',q') u^l v^m. \quad (2.29)$$

We note that

$$\mathbf{j}_p' = \mathbf{j}_p(\mathbf{0}) = \mathbf{p}' \overline{P}_{0,0}(p',q'), \quad \mathbf{j}_q' = \mathbf{j}_q(\mathbf{0}) = \mathbf{q}' \overline{Q}_{0,0}(p',q'). \quad (2.30)$$

For the *p*-component of the linear current density and from the last two equations, we have

$$\mathbf{j}_p(\mathbf{r}) - \mathbf{j}_p' = (\mathbf{p}' + \mathbf{r}_{pq}v) \sum_{l=0}^{L_p} \sum_{m=0}^{M_p} \overline{P}_{l,m}(p',q') u^l v^m - \mathbf{p}' \overline{P}_{0,0}(p',q')$$

$$= \mathbf{p}' \left\{ \sum_{l=1}^{L_p} \overline{P}_{l,0}(p',q') u^l + \sum_{m=1}^{M_p} \overline{P}_{0,m}(p',q') v^m + \sum_{l=1}^{L_p} \sum_{m=1}^{M_p} \overline{P}_{l,m}(p',q') u^l v^m \right\} + \mathbf{r}_{pq} v \sum_{l=0}^{L_p} \sum_{m=0}^{M_p} \overline{P}_{l,m}(p',q') u^l v^m$$

$$= \mathbf{p}' \left\{ u \sum_{l=0}^{L_p-1} \overline{P}_{l+1,0}(p',q') u^l + v \sum_{m=0}^{M_p-1} \overline{P}_{0,m+1}(p',q') v^m + uv \sum_{l=0}^{L_p-1} \sum_{m=0}^{M_p-1} \overline{P}_{l+1,m+1}(p',q') u^l v^m \right\}$$

$$+ \mathbf{r}_{pq} v \sum_{l=0}^{L_p} \sum_{m=0}^{M_p} \overline{P}_{l,m}(p',q') u^l v^m. \quad (2.31)$$

For the last term in (2.25), we use (2.17) to write

$$\hat{n}' \cdot \left[ \mathbf{j}_p(\mathbf{r}) - \mathbf{j}_p' \right] = v \frac{(\mathbf{p}' \times \mathbf{q}') \cdot \mathbf{r}_{pq} \sum_{l=0}^{L_p} \sum_{m=0}^{M_p} \overline{P}_{l,m}(p',q') u^l v^m}{|\mathbf{p}' \times \mathbf{q}'|} = v \frac{(\mathbf{r}_p \times \mathbf{r}_q) \cdot \mathbf{r}_{pq} \sum_{l=0}^{L_p} \sum_{m=0}^{M_p} \overline{P}_{l,m}(p',q') u^l v^m}{|\mathbf{p}' \times \mathbf{q}'|} \quad (2.32)$$

Before we substitute the last two expressions in (2.25), we note that

$$\nabla \left( \frac{1}{R} - \frac{k^2}{2} R \right) = -\left( \frac{1}{R^3} + \frac{k^2}{2R} \right)(\mathbf{r} - \mathbf{r}'). \quad (2.33)$$

From (2.20), we have that

$$\hat{n}' \cdot (\mathbf{r} - \mathbf{r}') = \hat{n}' \cdot \mathbf{r}_{pq} uv - h = \frac{(\mathbf{r}_p \times \mathbf{r}_q) \cdot \mathbf{r}_{pq}}{|\mathbf{p}' \times \mathbf{q}'|} uv - h. \quad (2.34)$$

Then in the first term on the right in (2.25), we encounter terms of the form

$$\left(\frac{1}{R^3}+\frac{k^2}{2R}\right)\frac{\left(\hat{n}'\cdot\mathbf{r}_{pq}u^2v-hu\right)}{J(p,q)}, \quad \left(\frac{1}{R^3}+\frac{k^2}{2R}\right)\frac{\left(\hat{n}'\cdot\mathbf{r}_{pq}uv^2-hv\right)}{J(p,q)}, \quad \left(\frac{1}{R^3}+\frac{k^2}{2R}\right)\frac{\left(\hat{n}'\cdot\mathbf{r}_{pq}u^2v^2-huv\right)}{J(p,q)}$$

(2.35)

for the **p'** vector in (2.26), while

$$\left(\frac{1}{R^3}+\frac{k^2}{2R}\right)\frac{\left(\hat{n}'\cdot\mathbf{r}_{pq}uv^2-hv\right)}{J(p,q)}$$

(2.36)

for the $\mathbf{r}_{pq}$ vector. We also have higher-order terms in $u$ and $v$.

For the second term in (2.25), we use (2.20) and (2.33) to write

$$\hat{n}'\times\left[\hat{n}'\times\nabla\left(\frac{1}{R}-\frac{k^2}{2}R\right)\right]=-\left(\frac{1}{R^3}+\frac{k^2}{2R}\right)\hat{n}'\times\left[\hat{n}'\times(\mathbf{r}-\mathbf{r}')\right]$$

$$=-\left(\frac{1}{R^3}+\frac{k^2}{2R}\right)\hat{n}'\times\left\{\hat{n}'\times\left[\mathbf{p}'u+\mathbf{q}'v+\mathbf{r}_{pq}uv-\hat{n}'h\right]\right\}$$

$$=\left(\frac{1}{R^3}+\frac{k^2}{2R}\right)\left[\mathbf{p}'u+\mathbf{q}'v-\hat{n}'\times\left(\hat{n}'\times\mathbf{r}_{pq}\right)uv\right].$$

(2.37)

For the last term in (2.25), and taking (2.32) into consideration, we encounter terms of the form

$$\left(\frac{1}{R^3}+\frac{k^2}{2R}\right)\frac{\left[\mathbf{p}'uv+\mathbf{q}'v^2-\hat{n}'\times\left(\hat{n}'\times\mathbf{r}_{pq}\right)uv^2\right]}{J(p,q)}$$

(2.38)

and higher-order terms in $u$ and $v$.

We turn now to the $q$-component of the surface current density. In place of (2.31), we write

$$\mathbf{j}_q(\mathbf{r})-\mathbf{j}_q'=\left(\mathbf{q}'+\mathbf{r}_{pq}u\right)\sum_{l=0}^{L_q}\sum_{m=0}^{M_q}\overline{Q}_{l,m}(p',q')u^lv^m-\mathbf{q}'\overline{Q}_{0,0}(p',q')$$

$$=\mathbf{q}'\left\{\sum_{l=1}^{L_q}\overline{Q}_{l,0}(p',q')u^l+\sum_{m=1}^{M_q}\overline{Q}_{l,m}(p',q')v^m+\sum_{l=1}^{L_q}\sum_{m=1}^{M_q}\overline{Q}_{l,m}(p',q')u^lv^m\right\}+\mathbf{r}_{pq}u\sum_{l=0}^{L_q}\sum_{m=0}^{M_q}\overline{Q}_{l,m}(p',q')u^lv^m$$

$$=\mathbf{q}'\left\{u\sum_{l=0}^{L_q-1}\overline{Q}_{l,0}(p',q')u^l+v\sum_{m=0}^{M_q-1}\overline{Q}_{l,m}(p',q')v^m+uv\sum_{l=0}^{L_q-1}\sum_{m=0}^{M_q-1}\overline{Q}_{l,m}(p',q')u^lv^m\right\}$$

$$+\mathbf{r}_{pq}u\sum_{l=0}^{L_q}\sum_{m=0}^{M_q}\overline{Q}_{l,m}(p',q')u^lv^m.$$

(2.39)

In place of (2.32), we have

$$\hat{n}'\cdot\left[\mathbf{j}_q(\mathbf{r})-\mathbf{j}_q'\right]=u\frac{(\mathbf{p}'\times\mathbf{q}')\cdot\mathbf{r}_{pq}\sum_{l=0}^{L_q}\sum_{m=0}^{M_q}\bar{Q}_{l,m}(p',q')u^lv^m}{|\mathbf{p}'\times\mathbf{q}'|}=u\frac{(\mathbf{r}_p\times\mathbf{r}_q)\cdot\mathbf{r}_{pq}\sum_{l=0}^{L_q}\sum_{m=0}^{M_q}\bar{Q}_{l,m}(p',q')u^lv^m}{|\mathbf{p}'\times\mathbf{q}'|}$$

(2.40)

Using these results and those in (2.33) and (2.34), we get terms exactly as in (2.35) for the **q'** vector, and terms

$$\left(\frac{1}{R^3}+\frac{k^2}{2R}\right)\frac{(\hat{n}'\cdot\mathbf{r}_{pq}u^2v-hu)}{J(p,q)} \tag{2.41}$$

for the $\mathbf{r}_{pq}$ vector. We also have higher-order terms in $u$ and $v$. For the last term in (2.25), we have from (2.37) and (2.40) terms of the form

$$\left(\frac{1}{R^3}+\frac{k^2}{2R}\right)\frac{\left[\mathbf{p}'u^2+\mathbf{q}'uv-\hat{n}'\times(\hat{n}'\times\mathbf{r}_{pq})u^2v\right]}{J(p,q)}. \tag{2.42}$$

This expression and (2.35) and (2.36) include all the integrands in the first term on the right of (2.23). For the second term, we employ (2.33) and (2.34) to write

$$\frac{\hat{n}'}{J(p,q)}\times\left\{\mathbf{j}'\times\nabla\left(\frac{1}{R}-\frac{k^2}{2}R\right)\right\}=\mathbf{j}'\frac{\hat{n}'\cdot\nabla\left(\frac{1}{R}-\frac{k^2}{2}R\right)}{J(p,q)}$$

$$=-\mathbf{j}'\left(\frac{1}{R^3}+\frac{k^2}{2R}\right)\frac{\hat{n}'\cdot(\mathbf{r}-\mathbf{r}')}{J(p,q)}=-\frac{\mathbf{j}'}{J(p,q)}\left(\frac{1}{R^3}+\frac{k^2}{2R}\right)\left[\frac{(\mathbf{r}_p\times\mathbf{r}_q)\cdot\mathbf{r}_{pq}}{|\mathbf{p}'\times\mathbf{q}'|}uv-h\right]. \tag{2.43}$$

We summarize the integrals we have to deal with. Integrals that involve $h$ as a factor in the numerator are

$$\left(\frac{1}{R^3}+\frac{k^2}{2R}\right)\frac{h\{1,u,v,uv\}}{J(p,q)} \tag{2.44}$$

while the rest are

$$\left(\frac{1}{R^3}+\frac{k^2}{2R}\right)\frac{\{uv,uv^2,u^2,u^2v^2\}}{J(p,q)} \quad\text{and}\quad \left(\frac{1}{R^3}+\frac{k^2}{2R}\right)\frac{\{vu,vu^2,v^2,v^2u^2\}}{J(p,q)}. \tag{2.45}$$

We will examine ways to compute these integrals in the sections that follow.

We re-iterate that, when the OP is on the integration BQ (when $h=0$), then the integral must be treated as a principal value integral; thus, if in general the integral is

$$J(p',q',h) = \int_{-(1+q')}^{1-q'} dv \int_{-(1+p')}^{1-p'} du f(u,v,h) \tag{2.46}$$

then

$$J(p',q',0) = \lim_{v \to 0} \int_{-(1+q')}^{-v} dv \left\{ \lim_{u \to 0} \int_{-(1+p')}^{-u} du f(u,v,0) + \lim_{u \to 0} \int_{u}^{1-p'} du f(u,v,0) \right\}$$

$$+ \lim_{v \to 0} \int_{-(1+q')}^{-v} dv \left\{ \lim_{u \to 0} \int_{-(1+p')}^{-u} du f(u,v,0) + \lim_{u \to 0} \int_{u}^{1-p'} du f(u,v,0) \right\}. \tag{2.47}$$

We conclude with a few words regarding the computation of the first integral on the right of (2.9). When $kR$ gets smaller than a certain value, we proceed as follows. We let

$$\mathbf{f}(R) = \nabla \left[ \frac{\cos(kR) - 1 + \frac{(kR)^2}{2} - i\sin(kR)}{kR} \right]. \tag{2.48}$$

Performing the gradient operation, we get

$$\mathbf{f}(R) = \frac{R\left[-k\sin(kR) + k^2 R\right] - \left[\cos(kR) - 1 + \frac{(kR)^2}{2}\right]}{kR^2} \nabla R - i \frac{(kR)\cos(kR) - \sin(kR)}{kR^2} \nabla R$$

$$= \frac{-(kR)\sin(kR) + (kR)^2 - \cos(kR) + 1 - \frac{(kR)^2}{2}}{kR^3} (\mathbf{r} - \mathbf{r}') - i \frac{(kR)\cos(kR) - \sin(kR)}{kR^3} (\mathbf{r} - \mathbf{r}')$$

$$\tag{2.49}$$

For small values of $kR$, we can replace the trigonometric functions by their Maclaurin series and keep as many terms as needed for a required number of SDs. Suppose we need $N$ terms for the sine and $M$ for the cosine. We can then write

$$\mathbf{f}(R) = \frac{(\mathbf{r} - \mathbf{r}')}{kR^3} \left\{ (kR)^2 \left[ 1 - \sum_{n=0}^{N} (-1)^n \frac{(kR)^{2n}}{(2n+1)!} \right] - \sum_{m=2}^{M} (-1)^m \frac{(kR)^{2m}}{(2m)!} \right\}$$

$$- i \frac{(\mathbf{r} - \mathbf{r}')}{kR^3} \left[ (kR) \sum_{m=0}^{M} (-1)^m \frac{(kR)^{2m}}{(2m)!} - (kR) \sum_{n=0}^{N} (-1)^n \frac{(kR)^{2n}}{(2n+1)!} \right]$$

$$= \frac{(\mathbf{r}-\mathbf{r}')}{kR^3}\left\{-(kR)^2\sum_{n=1}^{N}(-1)^n\frac{(kR)^{2n}}{(2n+1)!}-(kR)^4\sum_{m=0}^{M-2}(-1)^m\frac{(kR)^{2m}}{(2m+4)!}\right\}$$

$$-i\frac{(\mathbf{r}-\mathbf{r}')}{R^2}\left[\sum_{m=1}^{M}(-1)^m\frac{(kR)^{2m}}{(2m)!}-\sum_{n=1}^{N}(-1)^n\frac{(kR)^{2n}}{(2n+1)!}\right]$$

$$=-\frac{k(\mathbf{r}-\mathbf{r}')}{R}\left\{-(kR)^2\sum_{n=0}^{N-1}(-1)^n\frac{(kR)^{2n}}{(2n+3)!}+(kR)^2\sum_{m=0}^{M-2}(-1)^m\frac{(kR)^{2m}}{(2m+4)!}\right\}$$

$$+i\frac{(\mathbf{r}-\mathbf{r}')}{R^2}\left[(kR)^2\sum_{m=0}^{M-1}(-1)^m\frac{(kR)^{2m}}{(2m)!}-(kR)^2\sum_{n=0}^{N-1}(-1)^n\frac{(kR)^{2n}}{(2n+1)!}\right]$$

$$=k^3R(\mathbf{r}-\mathbf{r}')\left[\sum_{n=0}^{N-1}(-1)^n\frac{(kR)^{2n}}{(2n+3)!}-\sum_{m=0}^{M-2}(-1)^m\frac{(kR)^{2m}}{(2m+4)!}\right]$$

$$+ik^2(\mathbf{r}-\mathbf{r}')\left[\sum_{m=0}^{M-1}(-1)^m\frac{(kR)^{2m}}{(2m)!}-\sum_{n=0}^{N-1}(-1)^n\frac{(kR)^{2n}}{(2n+1)!}\right]. \tag{2.50}$$

Once $N$ and $M$ have specific values, we can combine the two sums into one.

# 3. EVALUATION OF INTEGRALS IN $R^{-3}$. PART I

In this section, we consider the first integral in (2.44), namely

$$I_1(p',q',h) = h \int_{-1-q'}^{1-q'} dv \int_{-1-p'}^{1-p'} \frac{du}{R^3(u,v,h)} \tag{3.1}$$

with $R$ the magnitude of **R** in (2.20). Expressing the integral as in (2.47), we see that its value is zero when $h$ is zero

$$I_1(p',q',0) = 0. \tag{3.2}$$

In fact, the contribution of the singularity is accounted for in the derivation of MFIE. When $h$ is different from zero, we proceed as follows. From (2.22), we can write for the distance function

$$R = \sqrt{C^2 u^2 + Bu + A} \tag{3.3}$$

where

$$A(v,h) = |\mathbf{q}'|^2 v^2 + h^2, \quad B(v,h) = 2\left[\mathbf{q}' \cdot (\mathbf{p}' + \mathbf{r}_{pq}v) - \hat{n}' \cdot \mathbf{r}_{pq}h\right]v, \quad C(v) = |\mathbf{p}' + \mathbf{r}_{pq}v|. \tag{3.4}$$

We use Formula 2.264.5, p. 83, in [11] and write

$$\begin{aligned} I_1 &= h \int_{-1-q'}^{1-q'} dv \int_{-1-p'}^{1-p'} \frac{du}{R^3(u,v,h)} = h \int_{-1-q'}^{1-q'} dv \left\{ \frac{2\left[2C^2(v)u + B(v,h)\right]}{\Delta(v,h) R(u,v,h)} \bigg|_{-1-p'}^{1-p'} \right\} \\ &= 2h \int_{-1-q'}^{1-q'} \frac{dv}{\Delta(v,h)} \left\{ \frac{\left[2C^2(v)(1-p') + B(v,h)\right]}{R(1-p',v,h)} - \frac{\left[-2C^2(v)(1+p') + B(v,h)\right]}{R(-(1+p'),v,h)} \right\} \end{aligned} \tag{3.5}$$

with

$$\Delta(v,h) = 4C^2(v) A(v,h) - B^2(v,h). \tag{3.6}$$

We can evaluate the remaining integral analytically using Procedure III, p. 80, in [11][1]. This process, however, is very complex and we are not quite sure how stable the outcome is. As a first step, we must determine the roots of $\Delta$, a fourth degree polynomial. By using Laguerre's iteration method ([12], p. 244), we can show that, for $h$ different from zero, the four roots comprise two pairs of complex roots; thus, $\Delta$ does not become zero in the interval of integration. The same is true for the distance function. The roots of $\Delta$ must be computed numerically. Once we have done so, we can do a partial fraction expansion in terms of each pair of roots of $\Delta$ and end up with two integrals for (3.5), each of which can be evaluated using the aforementioned procedure. We followed this path and, due to a good number of "if..., then..." statements, we decided that this is not a good way to proceed; instead we opted for a straight-forward numerical evaluation of (3.5).

---

[1] All formulas we quote from [11] have been verified by us.

We computed (3.5) in Mathematica® [6] where we used two different numerical integration methods. The first is the Gauss-Kronrod Quadrature (GKQ) [7], while the second is the Double Exponential Quadrature (DEQ) [7], [8], also known as the Sinh-Tanh Quadrature (STQ). For test BQs, we used the second and third ones from Appendix B. The OP is (0.25, 0.25, $h$). From Table 3.1, we see that both quadratures perform well, except for the last value of $h$, where the GKQ appears to be off. The DEQ appears to be very consistent in the number of recursions required as a function of $h$, while the GKQ does not. The same comment holds for the timing. Suffice it to say that we do not expect values of $h$ smaller than, conservatively speaking, $10^{-5}$; thus, for these values of $h$, both quadratures appear to perform well.

Similar comments hold for the third BQ of Appendix B. For both BQs and for the smallest value of $h$, Mathematica® [6] returned 15 SD of accuracy using GKQ. We doubt, however, that the result we obtained is accurate at all since it is out of line with the results preceding it and with the result obtained using DEQ. In fact, the GKQ result is half the DEQ result. We have not been able to explain this. *For this reason, and although DEQ carries a higher time cost, we prefer it over GKQ for this integral.* We also wish to point out that the third BQ of Appendix B constitutes an extreme case and should be avoided in practice.

Table 3.1. Outcome, precision, Number Of Recursions (NOR) and timing of quadratures for (3.5) using the second BQ with $p' = q' = 0.25$.

| SECOND BQ, EQ. (3.5) | | | | | | | | |
|---|---|---|---|---|---|---|---|---|
| $h$ | GKQ | GKQ SD | NOR | GKQ TIME* | DEQ | DEQ SD | NOR | DEQ TIME* |
| 1 | **6.51423052393166E-01** | 15 | 5 | 0.150 | **6.51423052393166E-01** | 15 | 3 | 0.000 |
| 1.E-01 | **9.92017001842781E-01** | 15 | 7 | 0.016 | **9.92017001842780E-01** | 15 | 4 | 0.015 |
| 1.E-03 | **1.03513763174326E+00** | 15 | 7 | 0.016 | **1.03513763174326E+00** | 15 | 5 | 0.032 |
| 1.E-05 | **1.03557217173131E+00** | 15 | 16 | 0.078 | **1.03557217173131E+00** | 15 | 6 | 0.031 |
| 1.E-07 | **1.03557651743578E+00** | 15 | 14 | 0.047 | **1.03557651743578E+00** | 15 | 6 | 0.047 |
| 1.E-09 | **1.03557656089286E+00** | 15 | 8 | 0.109 | **1.03557656089286E+00** | 15 | 7 | 0.078 |
| 1.E-11 | **1.03557656132743E+00** | 15 | 15 | 0.078 | **1.03557656132743E+00** | 15 | 7 | 0.078 |
| 1.E-13 | **1.03557656133177E+00** | 15 | 15 | 0.032 | **1.03557656133177E+00** | 15 | 7 | 0.078 |
| 1.E-15 | **1.03557656133182E+00** | 15 | 19 | 0.062 | **1.03557656133182E+00** | 15 | 7 | 0.078 |
| 1.E-17 | **1.03557656133182E+00** | 15 | 15 | 0.031 | **1.03557656133182E+00** | 15 | 7 | 0.109 |
| 1.E-19 | **5.17788280665909E-01** | ?? | 16 | 0.016 | **1.03557656133182E+00** | 15 | 8 | 0.140 |

*CPU time (in seconds) spent in Mathematica® kernel.

Table 3.2. Outcome, precision, NOR and timing of quadratures for (3.5) using the third BQ with $p' = q' = 0.25$.

| | THIRD BQ, EQ. (3.5) | | | | | | | |
|---|---|---|---|---|---|---|---|---|
| h | GKQ | GKQ SD | NOR | GKQ TIME* | DEQ | DEQ SD | NOR | DEQ TIME* |
| 1 | 5.02587711372941E-01 | 15 | 5 | 0.031 | 5.02587711372941E-01 | 15 | 4 | 0.031 |
| 1.E-01 | 5.44750508705969E-01 | 15 | 7 | 0.031 | 5.44750508705969E-01 | 15 | 4 | 0.032 |
| 1.E-03 | 5.79842620210127E-01 | 15 | 7 | 0.031 | 5.79842620210127E-01 | 15 | 5 | 0.047 |
| 1.E-05 | 5.80256889107114E-01 | 15 | 12 | 0.031 | 5.80256889107114E-01 | 15 | 6 | 0.047 |
| 1.E-07 | 5.80261038974142E-01 | 15 | 6 | 0.063 | 5.80261038974142E-01 | 15 | 6 | 0.047 |
| 1.E-09 | 5.80261080473531E-01 | 15 | 7 | 0.093 | 5.80261080473531E-01 | 15 | 7 | 0.094 |
| 1.E-11 | 5.80261080888525E-01 | 15 | 19 | 0.078 | 5.80261080888525E-01 | 15 | 7 | 0.062 |
| 1.E-13 | 5.80261080892676E-01 | 15 | 15 | 0.031 | 5.80261080892675E-01 | 15 | 7 | 0.094 |
| 1.E-15 | 5.80261080892717E-01 | 15 | 19 | 0.078 | 5.80261080892717E-01 | 15 | 7 | 0.093 |
| 1.E-17 | 5.80261080892718E-01 | 15 | 16 | 0.046 | 5.80261080892717E-01 | 15 | 8 | 0.140 |
| 1.E-19 | 2.90130540446359E-01 | ?? | 4 | 0.031 | 5.80261080892717E-01 | 15 | 8 | 0.156 |

*CPU time (in seconds) spent in Mathematica® kernel.

We conclude this section by examining the integrand of (3.5) for the second and third BQ and for the OP used in the tables above. In Fig. 3.1.1, we display the graph of $2h / \Delta(v, h)$, for the second BQ and for $h = 0.1$, over the interval of integration: [-1.25, 0.75]. It clearly exhibits an impulse-like behavior. This becomes more apparent in Figs. 3.1.2 and 3.1.3, where $h$ is equal to $10^{-5}$ and $10^{-19}$, respectively, and where we have reduced the ordinate interval so as to make the shape of the curve visible. Meanwhile, the part of the integrand in braces exhibits a very smooth behavior and is almost independent of $h$, as we see in Fig. 3.2. There, we have plotted the expression in braces for the three values of $h$ above and we see that the influence of $h$ is minimal. Thus, the entire integrand is controlled by the $2h / \Delta(v, h)$ term. We present graphs of the entire integrand in Figs. 3.3.1 – 3.3.3 and we see that it is not an easy integrand to evaluate numerically. Similar comments hold for the third BQ. We present graphs of the entire integrand for the third BQ in Figs. 3.4.1 – 3.4.3.

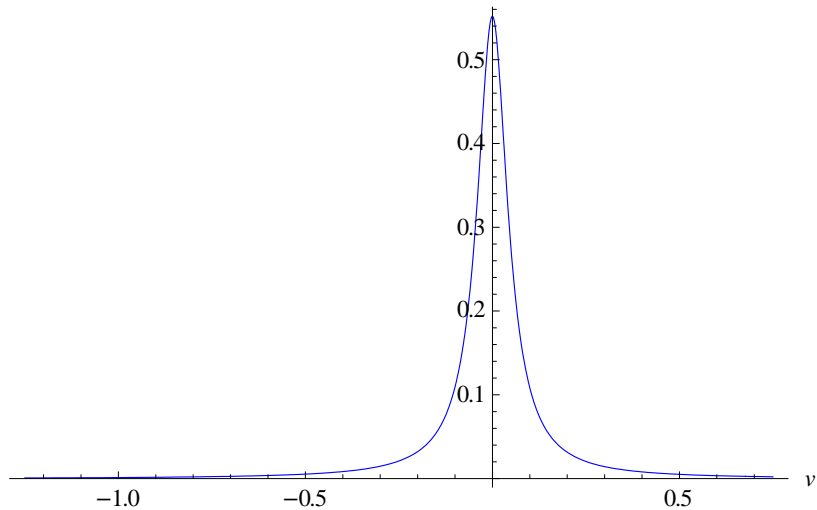

Figure 3.1.1. Graph of $2h / \Delta(v, h)$ for the second BQ and for the OP of Table 3.1. $h = 0.1$.

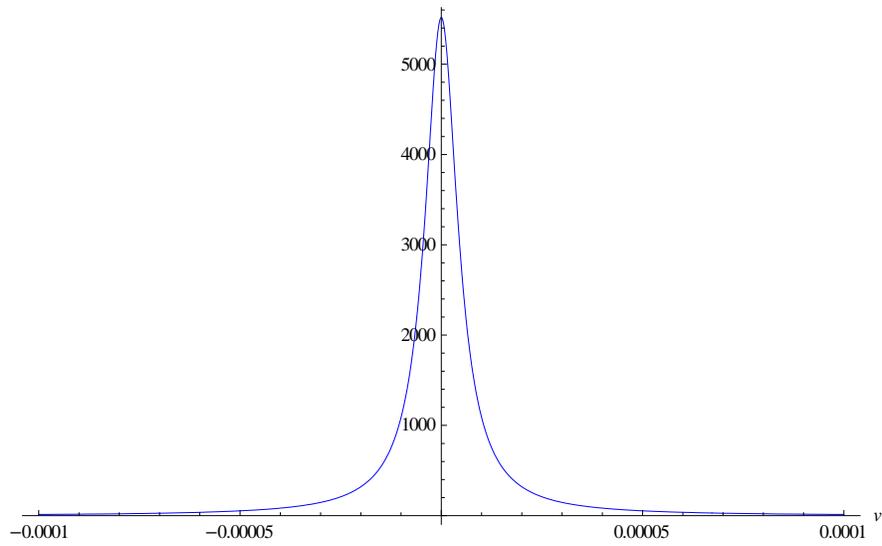

Figure 3.1.2. Graph of $2h / \Delta(v, h)$ for the second BQ and for the OP of Table 3.1. $h = 10^{-5}$.

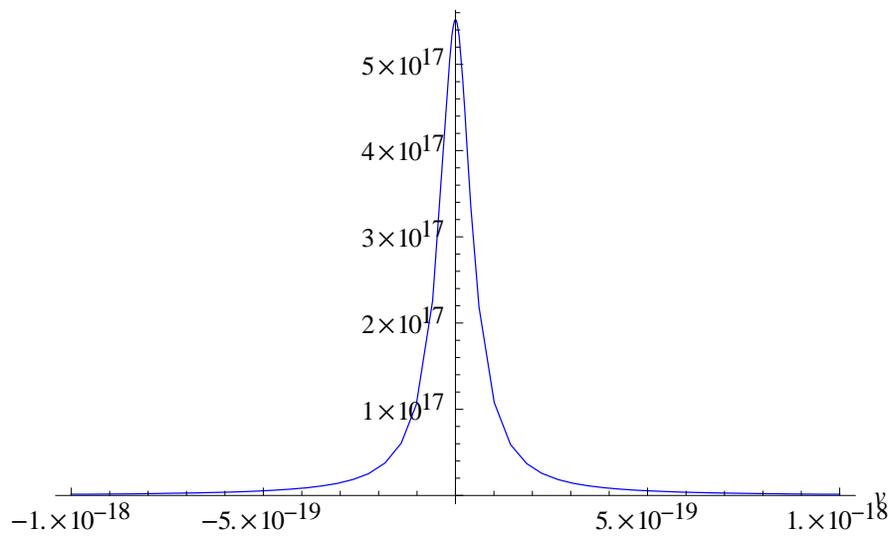

Figure 3.1.3. Graph of $2h / \Delta(v, h)$ for the second BQ and for the OP of Table 3.1. $h = 10^{-19}$.

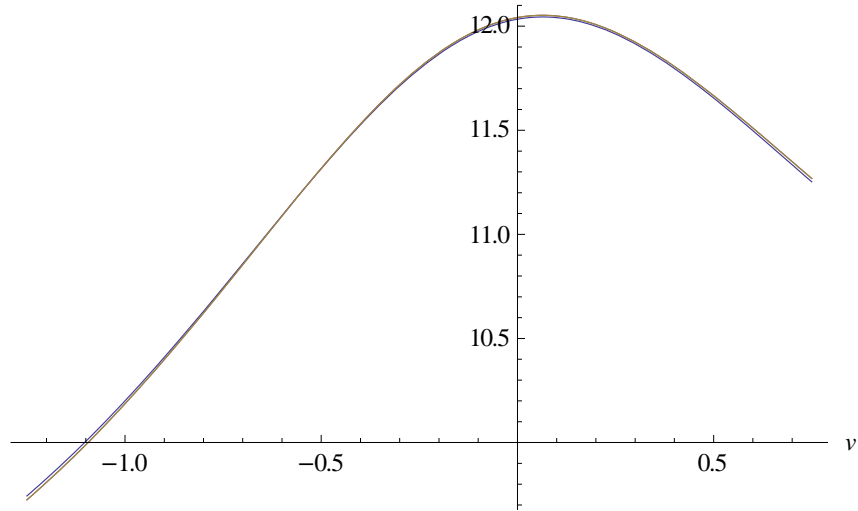

Figure 3.2. Graph of the part of the integrand that appears in braces in (3.5) for the second BQ and for the OP of Table 3.1. There are three curves corresponding to the values $h = 0.1$, $10^{-5}$ and $10^{-19}$.

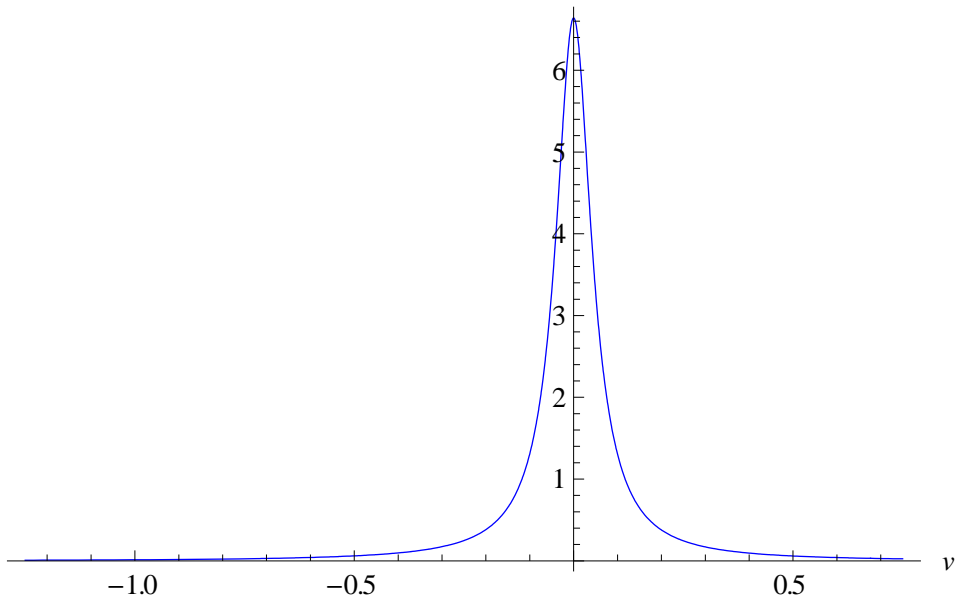

Figure 3.3.1. Graph of the entire integrand in (3.5) for the second BQ and for the OP of Table 3.1. $h = 0.1$.

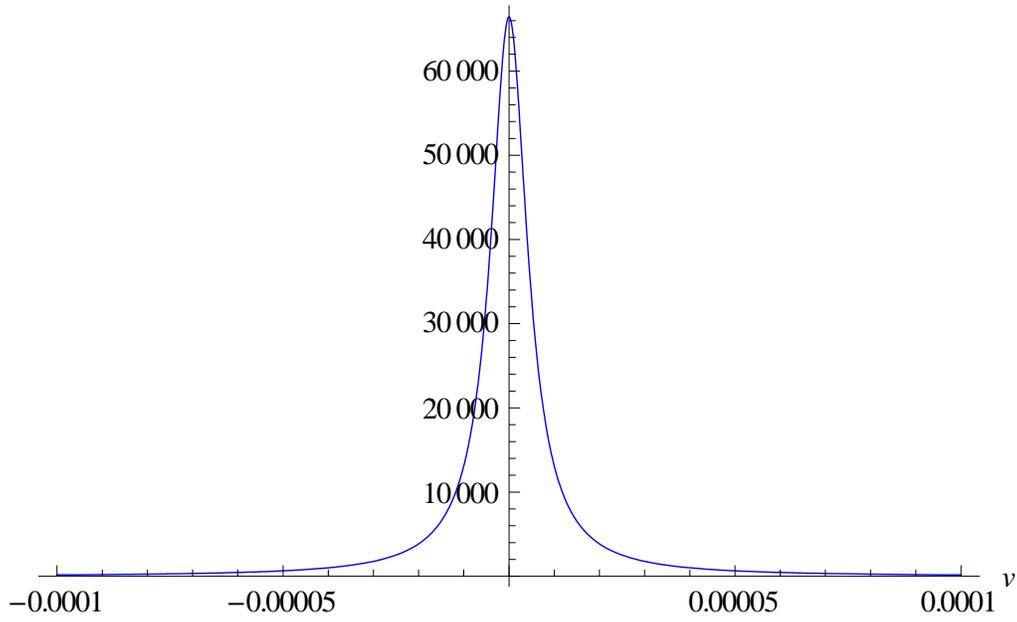

Figure 3.3.2. Graph of the entire integrand in (3.5) for the second BQ and for the OP of Table 3.1. $h = 10^{-5}$.

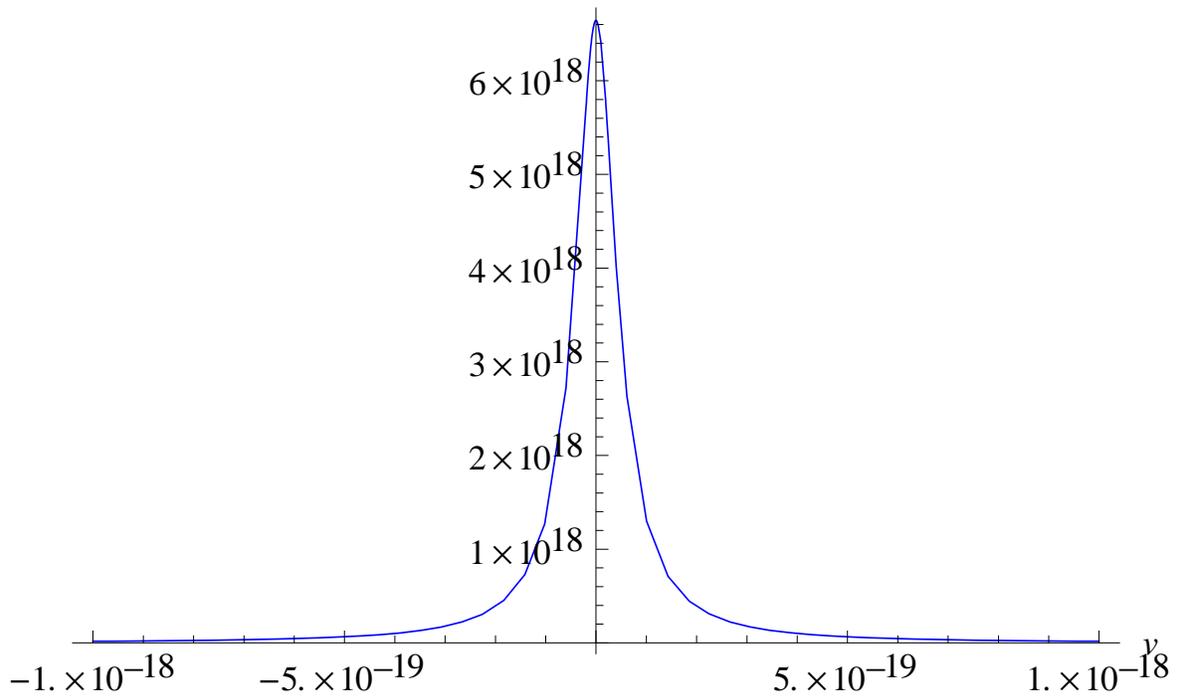

Figure 3.3.3. Graph of the entire integrand in (3.5) for the second BQ and for the OP of Table 3.1. $h = 10^{-19}$.

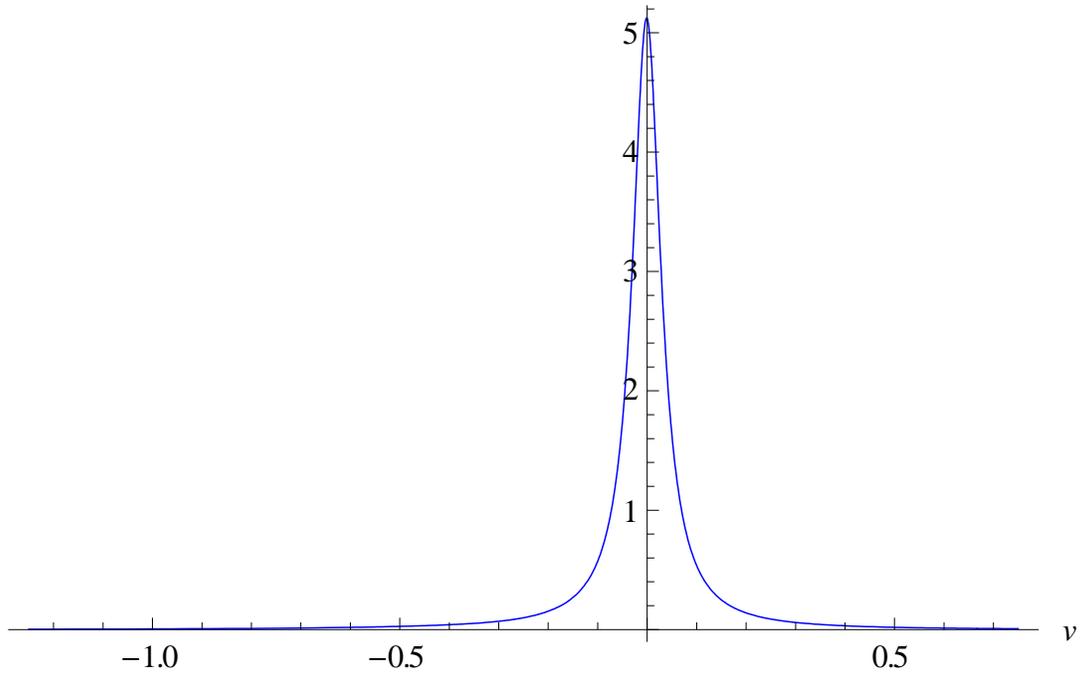

Figure 3.4.1. Graph of the entire integrand in (3.5) for the third BQ and for the OP of Table 3.2. $h = 0.1$.

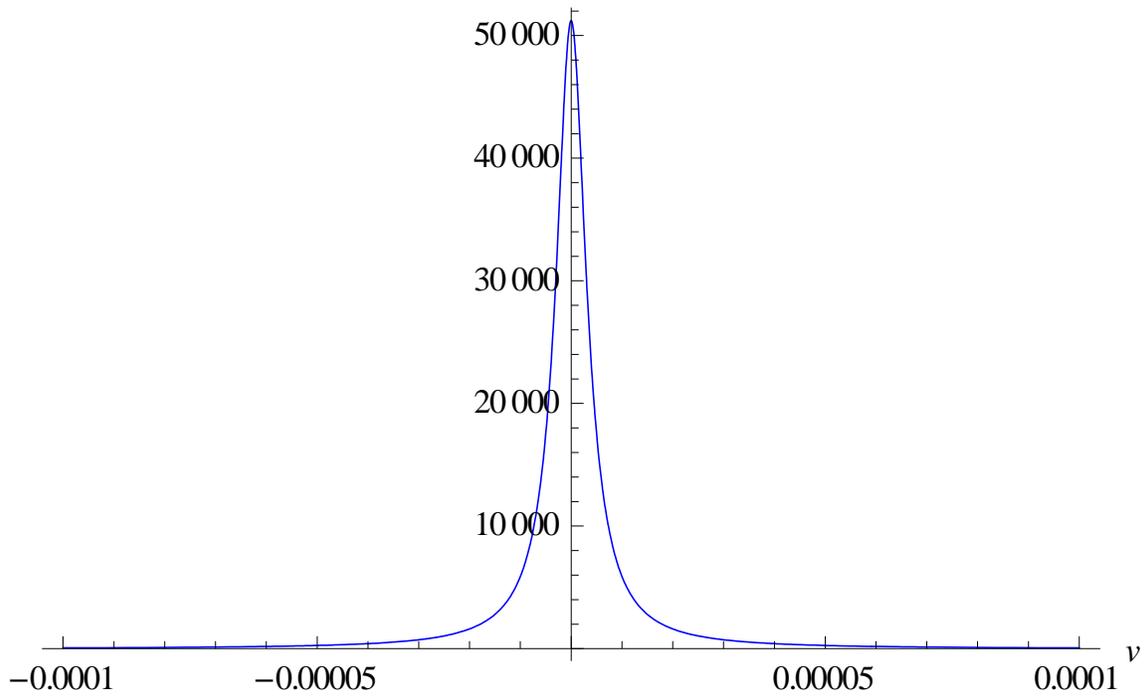

Figure 3.4.2. Graph of the entire integrand in (3.5) for the third BQ and for the OP of Table 3.2. $h = 10^{-5}$.

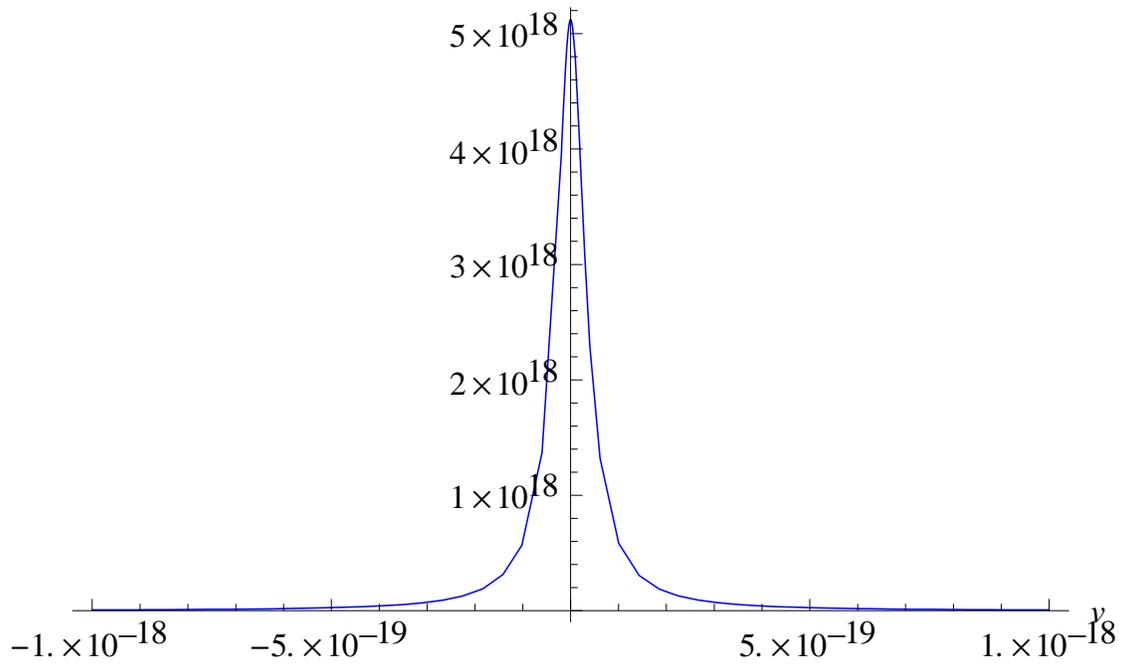

Figure 3.4.3. Graph of the entire integrand in (3.5) for the third BQ and for the OP of Table 3.2. $h = 10^{-19}$.

The conclusion of this section is that we are dealing with an impulse-like integrand that can be efficiently and accurately evaluated using DEQ.

# 4. EVALUATION OF INTEGRALS IN $R^{-3}$. PART II

The second integral we encounter in (2.44) is

$$I_1(p',q',h) = h \int_S \frac{u}{J(p,q)R^3} dS = h \int_{-1-q'}^{1-q'} dv \int_{-1-p'}^{1-p'} \frac{u\,du}{R^3(u,v,h)}.$$ (4.1)

Using (2.47), we find that the (4.1) is zero when $h$ is zero. For other values of $h$, we examine the graph of the integrand so as to get a sense of how delicate this integration is. In Fig. 4.1, we present the graph of $u / R^3$ for $h = 0.1$. We have used the second BQ and the OP is $p' = q' = 0.25$. Graphs for the third BQ and for the same OP are very similar to those for the second OP. We see that the integrand is constant everywhere except in a neighborhood of the origin where, along the $u$-axis, it exhibits a doublet-like behavior, while, along the $v$-axis, an impulse-like behavior at two distinct values of $u$. The two values of $u$ appear to be symmetric about the origin. The behavior of the graph near the origin is better exhibited in Fig. 4.2. For this graph, $h = 10^{-5}$ and the variables are restricted to the square $-10^{-5} < u,v < 10^{-5}$. We note that the peaks are of the order of $10^9$ and they get bigger as $h$ gets smaller.

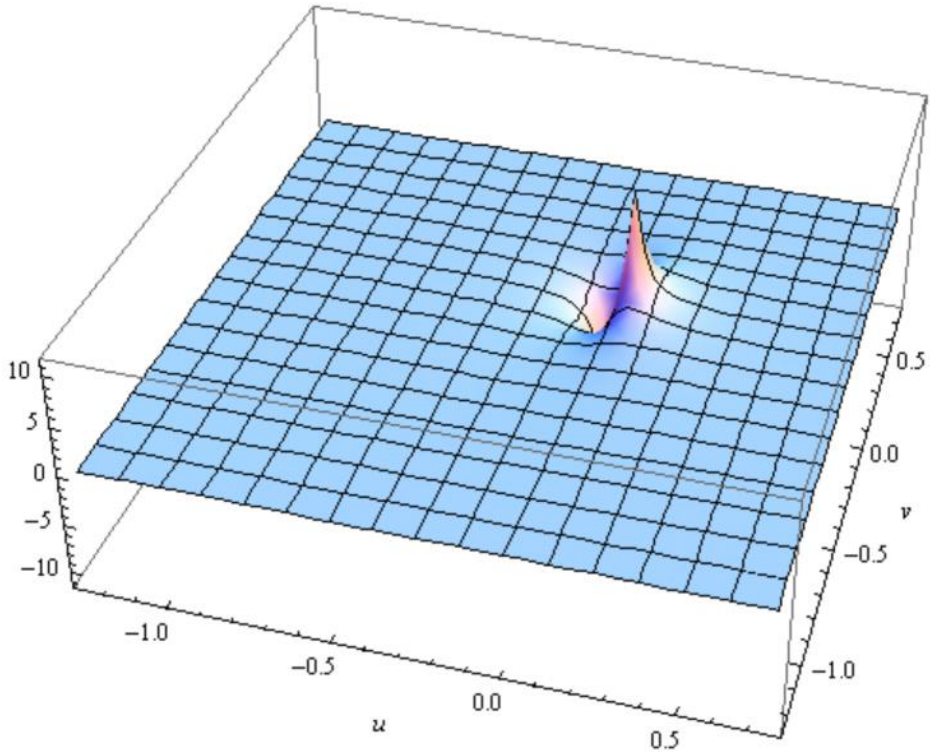

Figure 4.1. Graph of $u / R^3$ for $h = 0.1$. Second BQ. $p' = q' = 0.25$.

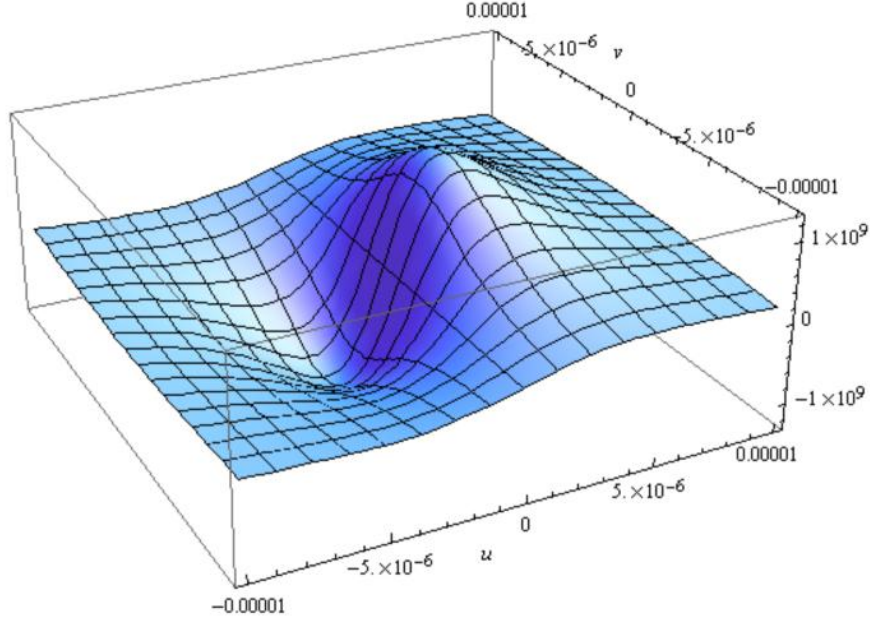

Figure 4.2. Graph of $u/R^3$ for $h = 0.1$ ($|u|, |v| < 10^{-5}$). Second BQ. $p' = q' = 0.25$.

From these two graphs we see that, if we first integrate (4.1) with respect to $u$, we will be integrating a doublet like function while, if we first integrate with respect to $v$, we will be integrating an impulse-like function. Either integration can be performed analytically. We begin with the first case and use (3.3) and (3.4) to write

$$I_1(p',q',h) = h \int_{-1-q'}^{1-q'} dv \int_{-1-p'}^{1-p'} \frac{u\,du}{\left(\sqrt{C^2 u^2 + Bu + A}\right)^3}. \tag{4.2}$$

From Formula 2.264.6, p. 83, in [11], we have that

$$\int \frac{u\,du}{R^3(u,v,h)} = -2\frac{2A(v,h) + B(v,h)u}{\mathsf{U}(v,h)R(u,v,h)}, \quad \mathsf{U}(v,h) = 4A(v,h)C^2(v) - B^2(v,h). \tag{4.3}$$

Applying it to (4.2), we get

$$I_1(p',q',h) = -2h \int_{-1-q'}^{1-q'} dv \left.\frac{2A(v,h) + B(v,h)u}{\mathsf{U}(v,h)R(u,v,h)}\right|_{-1-p'}^{1-p'}$$

$$= 2h \int_{-1-q'}^{1-q'} \frac{dv}{\mathsf{U}(v,h)} \left\{ \frac{2A(v,h) - B(v,h)(1+p')}{R(-(1+p'),v,h)} - \frac{2A(v,h) + B(v,h)(1-p')}{R(1-p',v,h)} \right\}. \tag{4.4}$$

As with the integral of the previous section, we can evaluate this one analytically but, for the same reasons, we avoid this approach.

We next reverse the order of integration in (4.1). From (2.21), we can write

$$R = \sqrt{F^2 v^2 + Ev + D} \tag{4.5}$$

where

$$D(u,h) = |\mathbf{p}'|^2 u^2 + h^2, \quad E(u,h) = 2\left[\mathbf{p}' \cdot (\mathbf{q}' + \mathbf{r}_{pq} u) - h\hat{n}' \cdot \mathbf{r}_{pq}\right]u, \quad F(u) = |\mathbf{q}' + \mathbf{r}_{pq} u|. \tag{4.6}$$

From [11], p. 83, Formula 2.265.5, we have that

$$I_1(p', q', h) = h \int_{-1-p'}^{1-p'} u\,du \int_{-1-q'}^{1-q'} \frac{dv}{\left(\sqrt{F^2 v^2 + Ev + D}\right)^3} = 2h \int_{-1-p'}^{1-p'} \frac{u\,du}{\mathsf{U}(u,h)} \frac{2vF^2(u) + E(u,h)}{\sqrt{F^2 v^2 + Ev + D}}\bigg|_{-1-q'}^{1-q'} \tag{4.7}$$

where

$$\mathsf{U}(u,h) = 4F^2(u)D(u,h) - E^2(u,h). \tag{4.8}$$

Proceeding with (4.7), we have

$$I_1(p', q', h) = h \int_{-1-p'}^{1-p'} \frac{u\,du}{\mathsf{U}(u,h)} \left\{ \frac{2(1-q')F^2(u) + E(u,h)}{\sqrt{(1-q')^2 F^2(u) + (1-q')E(u,h) + D(u,h)}} \right.$$

$$\left. - \frac{-2(1+q')F^2(u) + E(u,h)}{\sqrt{(1+q')^2 F^2(u) - (1+q')E(u,h) + D(u,h)}} \right\}. \tag{4.9}$$

We tested both (4.4) and (4.9) using Mathematica® [6]. The test point in these calculations is $p' = q' = 0.25$. We employ only the second BQ and we ask for 15 SD. We display the results for (4.4) in Table 4.1 and those for (4.9) in Table 4.2. From Table 4.1 we see that both quadratures converge to 15 SD, with GKQ being a lot faster than DEQ. Moreover, the Digits Of Agreement (DOA) between the two quadratures does not drop below 14. The story in Table 4.2 is quite different. We see that, for certain values of $h$, the quadratures were unable to produce 15 SD of accuracy. DEQ is performing better than GKQ but at a significant time expense; thus, it appears that it is better to use (4.4) rather than (4.9) in evaluating (4.1). We take another look at the same data by arranging it according to the quadrature used; thus, in Table 4.3 we display the results of GKQ for both (4.4) and (4.9), while, in Table 4.4 we present the corresponding results for DEQ. We see that the results provided by GKQ are in better agreement than those provided by DEQ.

Although the above is a single example rather than a thorough statistical analysis using a lot of BQs and OPs, we believe that the safe and quick way to proceed is using (4.4) and GKQ. Certainly, for values of $h$ no smaller than $10^{-5}$, DEQ is an excellent alternative.

Table 4.1. Entire expression in (4.4) evaluated by two quadratures at the OP (0.25, 0.25, *h*).

| | SECOND BQ, INTEGRAL IN (4.4) | | | | | | |
|---|---|---|---|---|---|---|---|
| *h* | GKQ | GKQ SD | GKQ TIME* | DEQ | DEQ SD | DEQ TIME* | DOA |
| 1 | **-3.26236684159375E-02** | 15 | 0.016 | **-3.26236684159374E-02** | 15 | 0.015 | 14 |
| 1.E-01 | **-3.68526659187055E-03** | 15 | 0.015 | **-3.68526659187055E-03** | 15 | 0.047 | 15 |
| 1.E-03 | **-3.69122165668271E-05** | 15 | 0.031 | **-3.69122165668271E-05** | 15 | 0.031 | 15 |
| 1.E-05 | **-3.69124782955421E-07** | 15 | 0.046 | **-3.69124782955422E-07** | 15 | 0.078 | 14 |
| 1.E-07 | **-3.69124808810716E-09** | 15 | 0.047 | **-3.69124808810716E-09** | 15 | 0.062 | 15 |
| 1.E-09 | **-3.69124809069238E-11** | 15 | 0.062 | **-3.69124809069237E-11** | 15 | 0.125 | 14 |
| 1.E-11 | **-3.69124809071823E-13** | 15 | 0.063 | **-3.69124809071822E-13** | 15 | 0.125 | 14 |
| 1.E-13 | **-3.69124809071848E-15** | 15 | 0.062 | **-3.69124809071848E-15** | 15 | 0.125 | 15 |
| 1.E-15 | **-3.69124809071849E-17** | 15 | 0.063 | **-3.69124809071849E-17** | 15 | 0.125 | 15 |
| 1.E-17 | **-3.69124809071849E-19** | 15 | 0.046 | **-3.69124809071848E-19** | 15 | 0.218 | 14 |
| 1.E-19 | **-3.69124809071849E-21** | 15 | 0.063 | **-3.69124809071848E-21** | 15 | 0.531 | 14 |

*CPU time (in seconds) spent in Mathematica® kernel

Table 4.2. Entire expression in (4.9) evaluated by two quadratures at the OP (0.25, 0.25, *h*).

| | SECOND BQ, INTEGRAL IN (4.9) | | | | | | |
|---|---|---|---|---|---|---|---|
| *h* | GKQ | GKQ SD | GKQ TIME* | DEQ | DEQ SD | DEQ TIME* | DOA |
| 1 | **-3.26236684159375E-02** | 15 | 0.031 | **-3.26236684159374E-02** | 15 | 0.031 | 14 |
| 1.E-01 | **-3.68526659187055E-03** | 14 | 0.031 | **-3.68526659187055E-03** | 15 | 0.047 | 15 |
| 1.E-03 | **-3.69122165668270E-05** | 14 | 0.063 | **-3.69122165668268E-05** | 14 | 0.063 | 13 |
| 1.E-05 | **-3.69124782955422E-07** | 14 | 0.062 | **-3.69124782955418E-07** | 14 | 0.094 | 13 |
| 1.E-07 | **-3.69124808810717E-09** | 13 | 0.047 | **-3.69124808810718E-09** | 14 | 0.156 | 14 |
| 1.E-09 | **-3.69124809069230E-11** | 13 | 0.062 | **-3.69124809069236E-11** | 13 | 0.141 | 14 |
| 1.E-11 | **-3.69124809071825E-13** | 13 | 0.063 | **-3.69124809071783E-13** | 15 | 0.593 | 12 |
| 1.E-13 | **-3.69124809071844E-15** | 13 | 0.062 | **-3.69124809071864E-15** | 15 | 0.296 | 13 |
| 1.E-15 | **-3.69124809071848E-17** | 13 | 0.046 | **-3.69124809071849E-17** | 15 | 0.530 | 14 |
| 1.E-17 | **-3.69124809071865E-19** | 13 | 0.063 | **-3.69124809071857E-19** | 13 | 0.297 | 13 |
| 1.E-19 | **-3.69124809071847E-21** | 13 | 0.062 | **-3.69124809071843E-21** | 13 | 0.234 | 14 |

*CPU time (in seconds) spent in Mathematica® kernel

Table 4.3. Comparison of (4.4) and (4.9) for GKQ. Second BQ. OP at (0.25, 0.25, $h$).

| | SECOND BQ, GKQ | | | | | | |
|---|---|---|---|---|---|---|---|
| $h$ | FROM (4.4) | SD | TIME* | FROM (4.9) | SD | TIME* | DOA |
| 1 | **-3.26236684159375E-02** | 15 | 0.016 | **-3.26236684159375E-02** | 15 | 0.031 | 15 |
| 1.E-01 | **-3.68526659187055E-03** | 15 | 0.015 | **-3.68526659187055E-03** | 14 | 0.031 | 15 |
| 1.E-03 | **-3.69122165668271E-05** | 15 | 0.031 | **-3.69122165668270E-05** | 14 | 0.063 | 14 |
| 1.E-05 | **-3.69124782955421E-07** | 15 | 0.046 | **-3.69124782955422E-07** | 14 | 0.062 | 14 |
| 1.E-07 | **-3.69124808810716E-09** | 15 | 0.047 | **-3.69124808810717E-09** | 13 | 0.047 | 14 |
| 1.E-09 | **-3.69124809069238E-11** | 15 | 0.062 | **-3.69124809069230E-11** | 13 | 0.062 | 14 |
| 1.E-11 | **-3.69124809071823E-13** | 15 | 0.063 | **-3.69124809071825E-13** | 13 | 0.063 | 14 |
| 1.E-13 | **-3.69124809071848E-15** | 15 | 0.062 | **-3.69124809071844E-15** | 13 | 0.062 | 14 |
| 1.E-15 | **-3.69124809071849E-17** | 15 | 0.063 | **-3.69124809071848E-17** | 13 | 0.046 | 14 |
| 1.E-17 | **-3.69124809071849E-19** | 15 | 0.046 | **-3.69124809071865E-19** | 13 | 0.063 | 13 |
| 1.E-19 | **-3.69124809071849E-21** | 15 | 0.063 | **-3.69124809071847E-21** | 13 | 0.062 | 14 |

*CPU time (in seconds) spent in Mathematica® kernel

Table 4.4. Comparison of (4.4) and (4.9) for DEQ. Second BQ. OP at (0.25, 0.25, $h$).

| | SECOND BQ, DEQ | | | | | | |
|---|---|---|---|---|---|---|---|
| $h$ | FROM (4.4) | SD | TIME* | FROM (4.9) | SD | TIME* | DOA |
| 1 | **-3.26236684159374E-02** | 15 | 0.015 | **-3.26236684159374E-02** | 15 | 0.031 | 15 |
| 1.E-01 | **-3.68526659187055E-03** | 15 | 0.047 | **-3.68526659187055E-03** | 15 | 0.047 | 15 |
| 1.E-03 | **-3.69122165668271E-05** | 15 | 0.031 | **-3.69122165668268E-05** | 14 | 0.063 | 13 |
| 1.E-05 | **-3.69124782955422E-07** | 15 | 0.078 | **-3.69124782955418E-07** | 14 | 0.094 | 13 |
| 1.E-07 | **-3.69124808810716E-09** | 15 | 0.062 | **-3.69124808810718E-09** | 14 | 0.156 | 14 |
| 1.E-09 | **-3.69124809069237E-11** | 15 | 0.125 | **-3.69124809069236E-11** | 13 | 0.141 | 14 |
| 1.E-11 | **-3.69124809071822E-13** | 15 | 0.125 | **-3.69124809071783E-13** | 15 | 0.593 | 12 |
| 1.E-13 | **-3.69124809071848E-15** | 15 | 0.125 | **-3.69124809071864E-15** | 15 | 0.296 | 13 |
| 1.E-15 | **-3.69124809071849E-17** | 15 | 0.125 | **-3.69124809071849E-17** | 15 | 0.530 | 15 |
| 1.E-17 | **-3.69124809071848E-19** | 15 | 0.218 | **-3.69124809071857E-19** | 13 | 0.297 | 13 |
| 1.E-19 | **-3.69124809071848E-21** | 15 | 0.531 | **-3.69124809071843E-21** | 13 | 0.234 | 14 |

*CPU time (in seconds) spent in Mathematica® kernel

We conclude this section by examining the integrands of (4.4) and (4.9). We consider (4.4) first. The graph of the term $2h / \Delta(v, h)$ is the same as in Section 3, resembling that of an impulse function. The graph of the remaining integrand is best visualized using the third BQ. We display it in Fig. 4.3 for three values of $h$ and, as in Section 3, we observe that it does not change much with $h$. What is significant, however, is the fact that, over a short interval to the left of the origin, it takes on positive values. This results in the total integrand exhibiting a doublet-like behavior around the origin, as we see in Figs. 4.4.1 – 4.4.3. This makes the numerical evaluation of the integral more difficult, both because of the shape of the graph and because we are subtracting two almost equal areas.

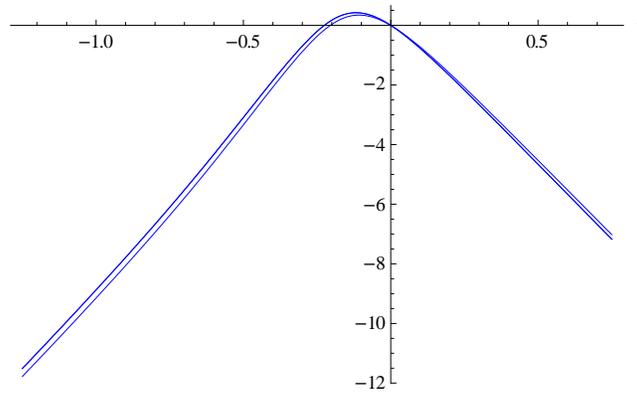

Figure 4.3. Graph of the part of the integrand that appears in braces in (4.4) for the third BQ and for the OP (0.25, 0.25, $h$). There are three curves corresponding to the values $h = 0.1$, $10^{-5}$ and $10^{-19}$.

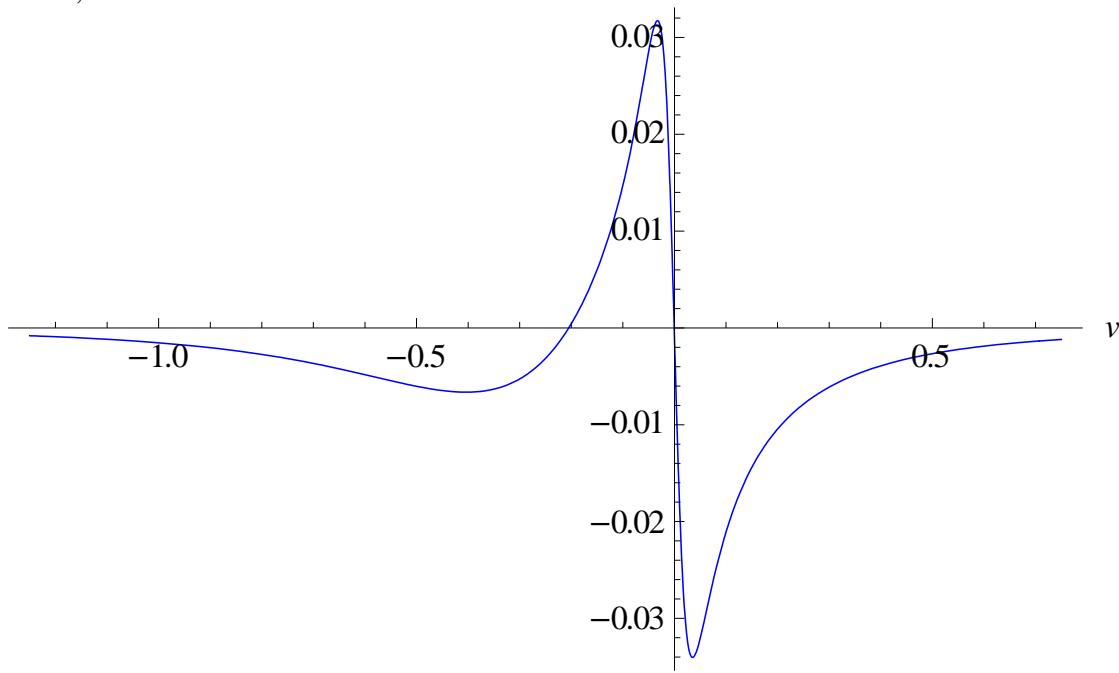

Figure 4.4.1. Graph of the entire integrand in (4.4) for the third BQ and for the OP at (0.25, 0.25, 0.1).

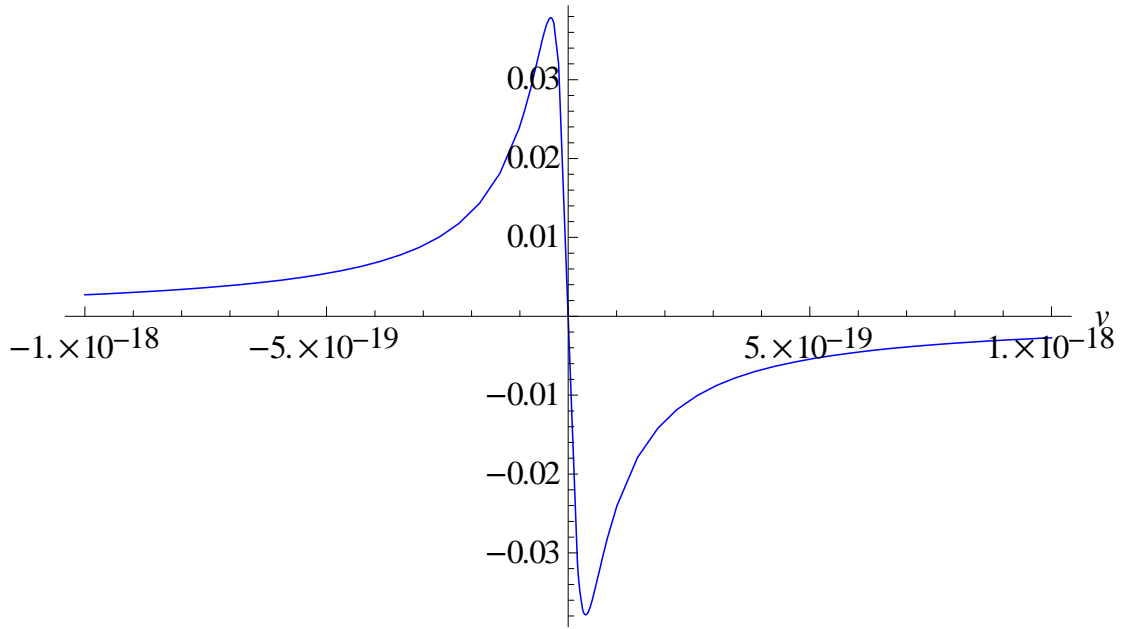

Figure 4.4.2. Graph of the entire integrand in (4.4) for the third BQ and for the OP at $(0.25, 0.25, 10^{-5})$.

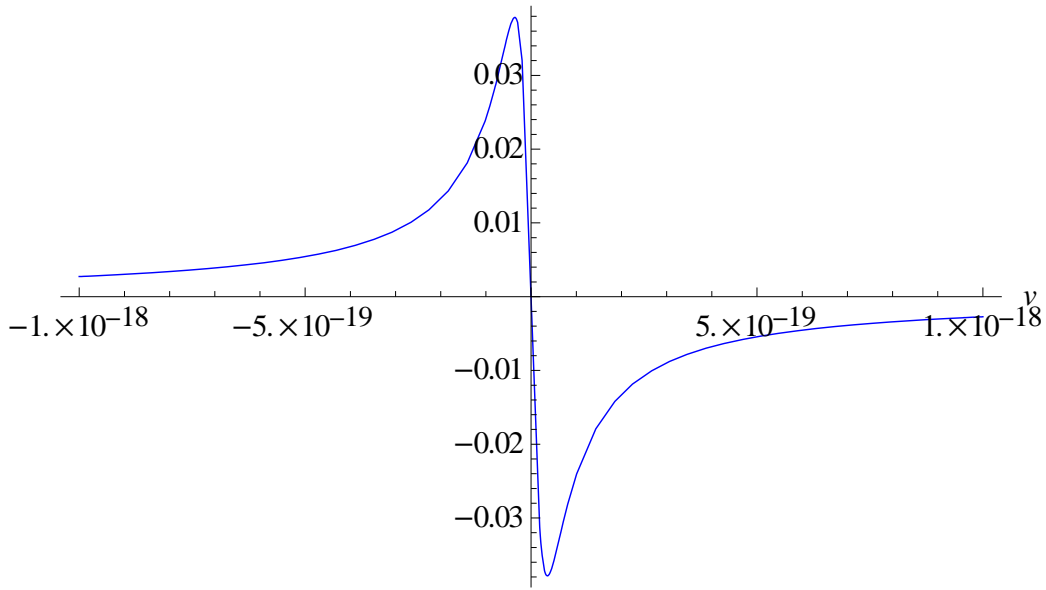

Figure 4.4.3. Graph of the entire integrand in (4.4) for the third BQ and for the OP at $(0.25, 0.25, 10^{-19})$.

We turn next to the integrand of (4.9). The graph of the term $hu / \Delta(u, h)$ is now resembling a doublet, as we see in Figs. 4.5.1 – 4.5.3. The graph of the remaining integrand, however, is very insensitive to changes in $h$, as we see in Fig. 4.6. The entire integrand is then dominated by the doublet and its graph is shown in Figs. 4.7.1 – 4.7.3. Again, this is a difficult integral to evaluate numerically because of the shape of the graph and because the doublet has a decreasing duration as *h* gets smaller.

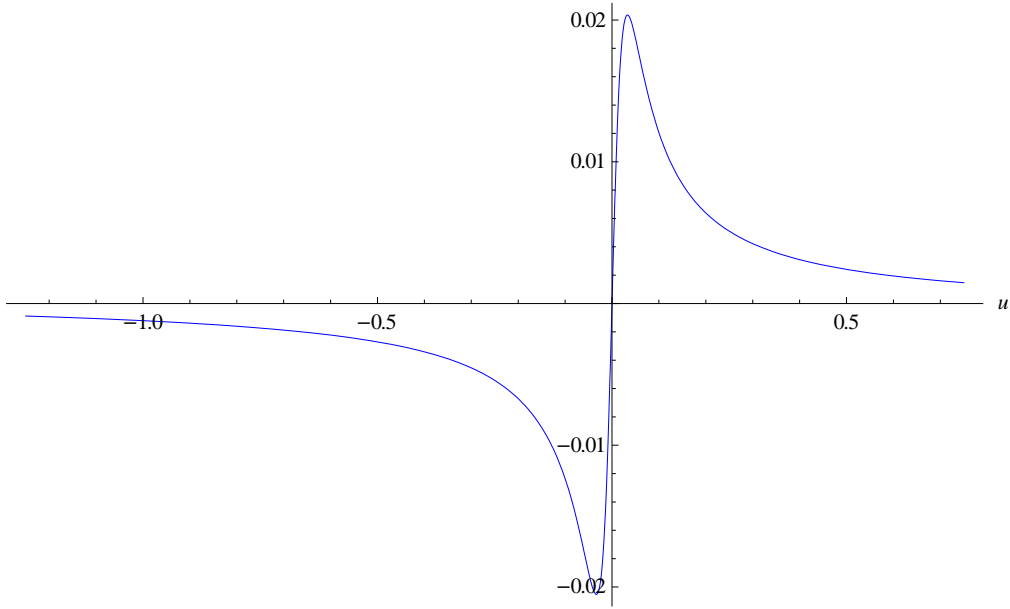

Figure 4.5.1. Graph of the entire integrand in (4.9) for the second BQ and for the OP at (0.25, 0.25, 0.1).

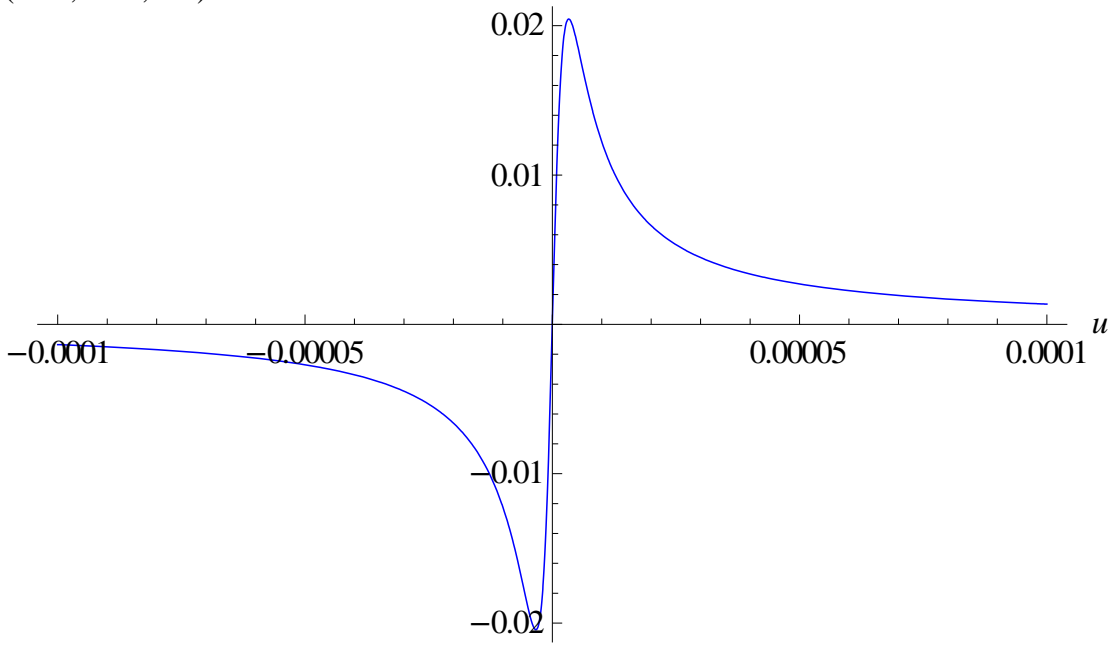

Figure 4.5.2. Graph of the entire integrand in (4.9) for the second BQ and for the OP at (0.25, 0.25, $10^{-5}$).

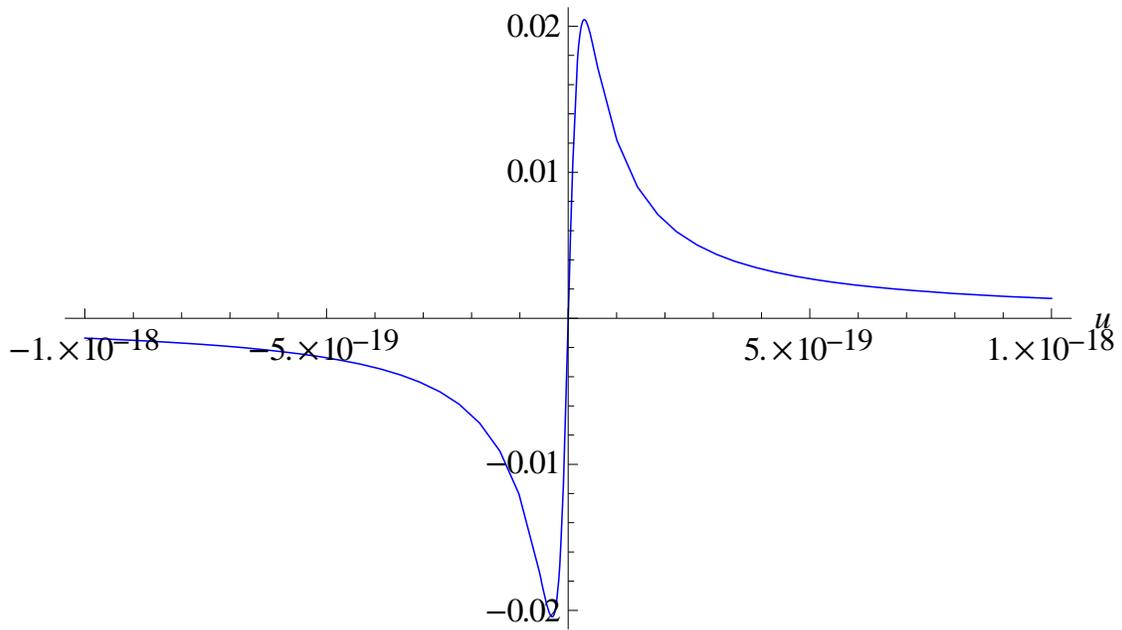

Figure 4.5.3. Graph of the entire integrand in (4.9) for the second BQ and for the OP at $(0.25, 0.25, 10^{-19})$.

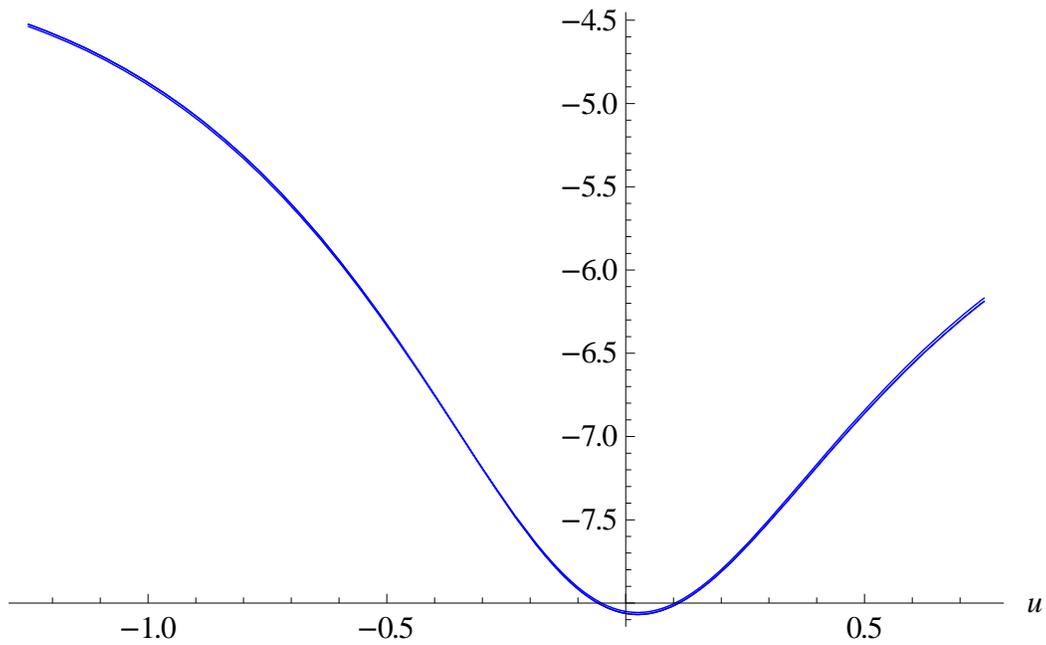

Figure 4.6. Graph of the part of the integrand that appears in braces in (4.9) for the second BQ and for the OP at $(0.25, 0.25, h)$. There are three curves corresponding to the values $h = 0.1$, $10^{-5}$ and $10^{-19}$.

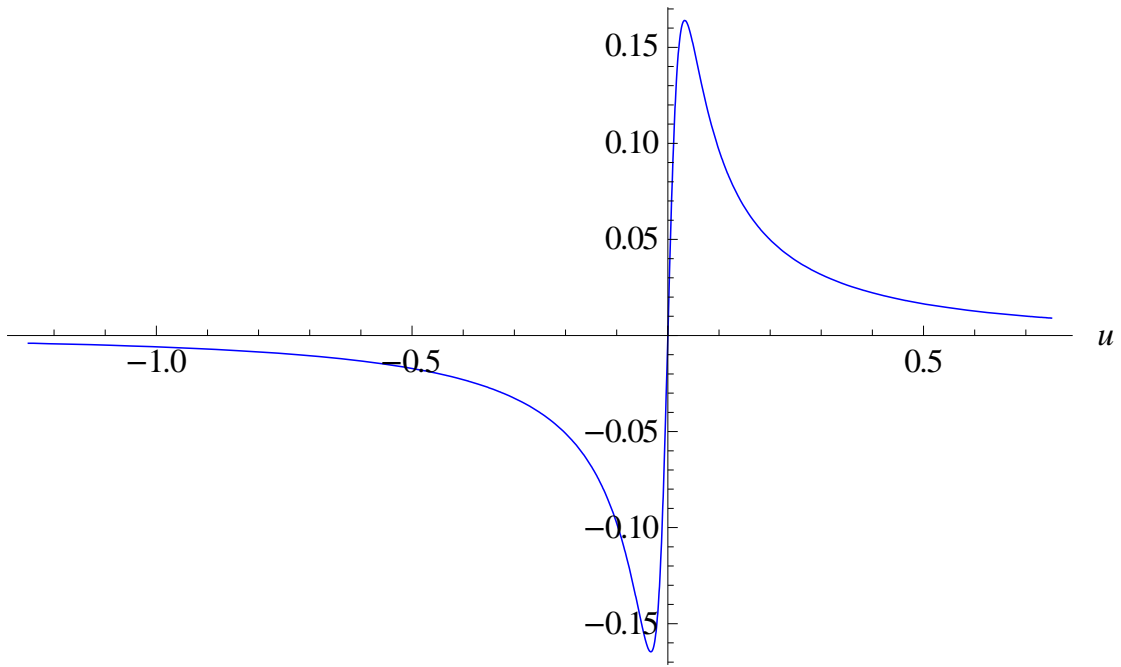

Figure 4.7.1. Graph of the entire integrand in (4.9) for the second BQ and for the OP at (0.25, 0.25, 0.1).

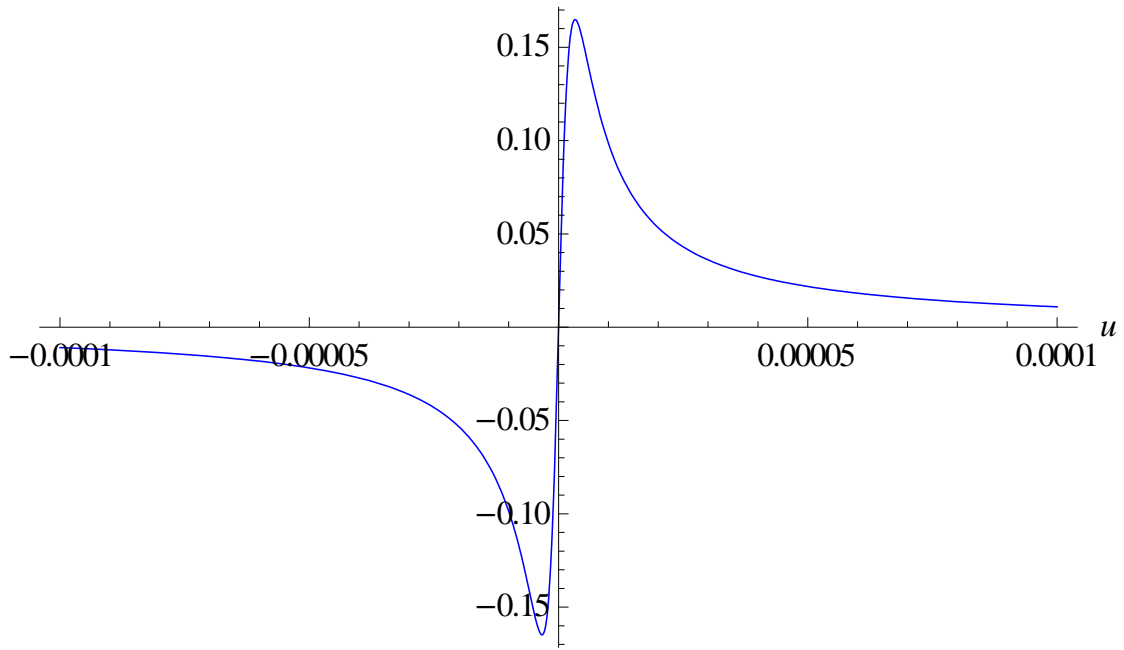

Figure 4.7.2. Graph of the entire integrand in (4.9) for the second BQ and for the OP at (0.25, 0.25, $10^{-5}$).

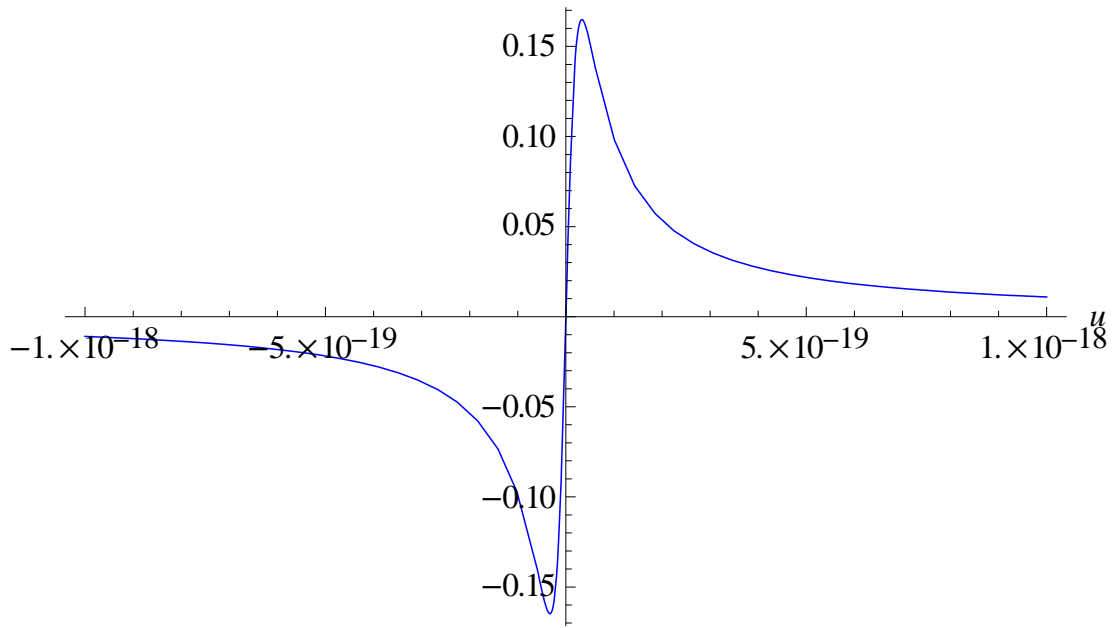

Figure 4.7.3. Graph of the entire integrand in (4.9) for the second BQ and for the OP at $(0.25, 0.25, 10^{-19})$.

In closing this section, we note that the third integral in (2.44) is exactly of the same kind as the integral in (4.1) except that the roles of $u$ and $v$ have been interchanged.

# 5. EVALUATION OF INTEGRALS IN $R^{-3}$. PART III

In this section we examine the fourth integral in (2.44). This integral is the same as the first integral in the two expressions in (2.45) except that the integrand in (2.44) contains the multiplicative factor $h$. When this factor is present, the integral is equal to zero when $h$ is equal to zero. We proceed to examine the integral without this factor. This integral is

$$I_1(p',q',h) = \int_S \frac{uv}{J(p,q)R^3} dS = \int_{-1-q'}^{1-q'} v\,dv \int_{-1-p'}^{1-p'} \frac{u\,du}{R^3(u,v,h)} \qquad (5.1)$$

and, clearly, it does not matter which integral we evaluate first. The graph of the integrand is smooth everywhere except near the origin. In Fig. 5.1, we display this graph for the second BQ with $h = 10^{-5}$ and $p' = q' = 0.25$. Along the directions $u$ and $v$, we have doublet-like curves, as in Fig. 5.2. When $h = 0.0$, the limit at the origin does not exist since the integrand goes to zero along the directions $u$ and $v$, while it goes to infinity along the direction $v = u$.

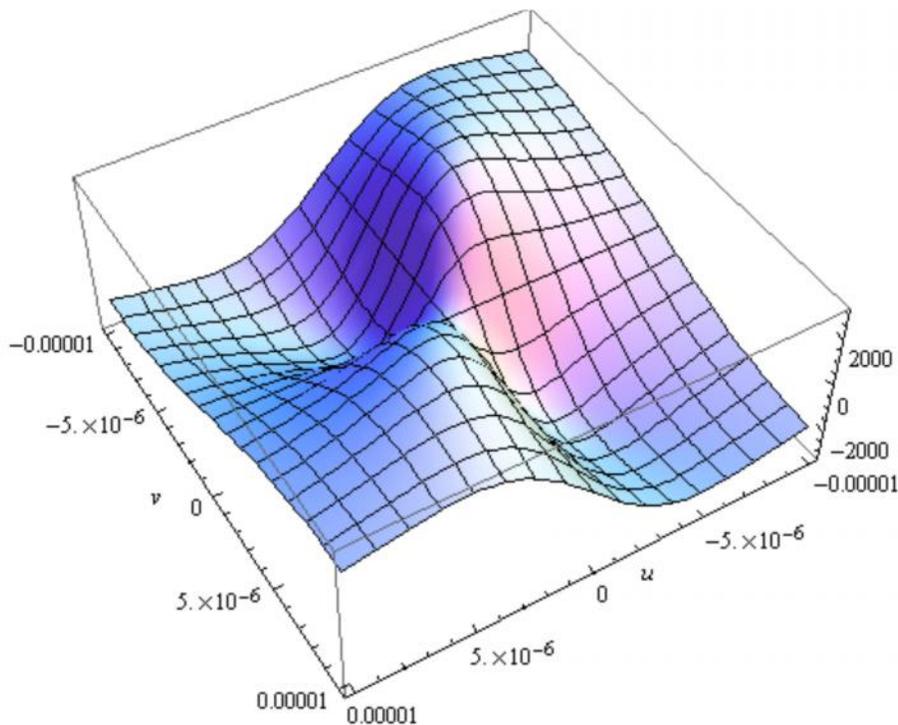

Figure 5.1. Graph of integrand of (5.1) for the second BQ with $h = 10^{-5}$ and $p' = q' = 0.25$.

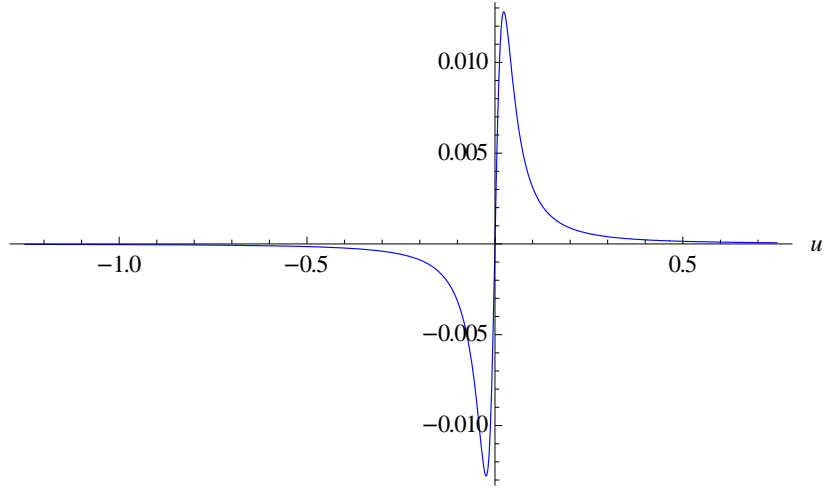

Figure 5.2. Graph of integrand of (5.1) for the second BQ with $v = 0.001$, $h = 10^{-5}$ and $p' = q' = 0.25$.

Using (4.3), we can write (5.1) in the form

$$I_1(p',q',h) = -2 \lim_{u \to 0^+} \left\{ \int_{-1-q'}^{-u} \frac{vdv}{\mathsf{U}(v,h)} \lim_{v \to 0^+} \left[ \frac{2A(v,h) + B(v,h)u}{R(u,v,h)} \bigg|_{-1-p'}^{-v} + \frac{2A(v,h) + B(v,h)u}{R(u,v,h)} \bigg|_{v}^{1-p'} \right] \right.$$

$$\left. + \int_{u}^{1-q'} \frac{vdv}{\mathsf{U}(v,h)} \lim_{v \to 0^+} \left[ \frac{2A(v,h) + B(v,h)u}{R(u,v,h)} \bigg|_{-1-p'}^{-v} + \frac{2A(v,h) + B(v,h)u}{R(u,v,h)} \bigg|_{v}^{1-p'} \right] \right\}$$

$$= -2 \lim_{u \to 0^+} \left\{ \int_{-1-q'}^{-u} \frac{vdv}{\mathsf{U}(v,h)} \frac{2A(v,h) + B(v,h)u}{R(u,v,h)} \bigg|_{-1-p'}^{1-p'} + \int_{u}^{1-q'} \frac{vdv}{\mathsf{U}(v,h)} \frac{2A(v,h) + B(v,h)u}{R(u,v,h)} \bigg|_{-1-p'}^{1-p'} \right\}$$

$$= 2 \lim_{u \to 0^+} \int_{-1-q'}^{-u} \frac{vdv}{\mathsf{U}(v,h)} \left\{ \frac{2A(v,h) - B(v,h)(1+p')}{R(-(1+p'),v,h)} - \frac{2A(v,h) + B(v,h)(1-p')}{R(1-p',v,h)} \right\}$$

$$+ 2 \lim_{u \to 0^+} \int_{u}^{1-q'} \frac{vdv}{\mathsf{U}(v,h)} \left\{ \frac{2A(v,h) - B(v,h)(1+p')}{R(-(1+p'),v,h)} - \frac{2A(v,h) + B(v,h)(1-p')}{R(1-p',v,h)} \right\}. \tag{5.2}$$

This is the same integral as in (4.4) except for the replacement of the multiplicative factor $h$ by $v$. As with that integral, the present one can be evaluated analytically but, for the same reasons, we avoid this path.

We have evaluated (5.2) numerically for the OP (0.25, 0.25, $h$) and the second BQ of Appendix B. We have specified the origin in Mathematica® as a singular point. We display the results in Table 5.1. In this table, we have multiplied the integrand in (5.2) by $h$; thus, this is the result for the fourth integral in (2.44). We observe that the two quadratures agree to at least 14 SD except for $h = 10^{-11}$ and $h = 10^{-13}$. Although these values are not in the range of practical interest, we investigate whether we can improve the agreement.

Table 5.1. Eq. (5.2) multiplied by *h* and evaluated by the two quadratures at the OP (0.25, 0.25, *h*).

| | SECOND BQ, EQ. (5.2) MULTIPLIED BY *h* | | | | | | | | |
|---|---|---|---|---|---|---|---|---|---|
| *h* | GKQ | SD | NOR | TIME* | DEQ | SD | NOR | TIME* | DOA |
| 1 | **1.49944916156393E-02** | 15 | 5 | 0.016 | **1.49944916156393E-02** | 15 | 3 | 0.015 | 15 |
| 1.E-01 | **7.40200806750922E-04** | 15 | 6 | 0.016 | **7.40200806750921E-04** | 15 | 4 | 0.016 | 14 |
| 1.E-03 | **4.75463049979300E-06** | 15 | 10 | 0.016 | **4.75463049979299E-06** | 15 | 5 | 0.032 | 14 |
| 1.E-05 | **4.72535082669790E-08** | 15 | 11 | 0.016 | **4.72535082669790E-08** | 15 | 5 | 0.031 | 15 |
| 1.E-07 | **4.72505773720324E-10** | 15 | 11 | 0.031 | **4.72505773720323E-10** | 15 | 5 | 0.031 | 14 |
| 1.E-09 | **4.72505480627900E-12** | 15 | 11 | 0.015 | **4.72505480627899E-12** | 15 | 5 | 0.016 | 14 |
| 1.E-11 | **4.72505477695218E-14** | 15 | 4 | 0.016 | **4.72505477696975E-14** | 15 | 5 | 0.031 | 11 |
| 1.E-13 | **4.72505477667649E-16** | 15 | 4 | 0.016 | **4.72505477667666E-16** | 15 | 4 | 0.000 | 13 |
| 1.E-15 | **4.72505477667373E-18** | 15 | 4 | 0.016 | **4.72505477667373E-18** | 15 | 3 | 0.016 | 15 |
| 1.E-17 | **4.72505477667370E-20** | 15 | 4 | 0.016 | **4.72505477667370E-20** | 15 | 3 | 0.000 | 15 |
| 1.E-19 | **4.72505477667371E-22** | 15 | 4 | 0.000 | **4.72505477667370E-22** | 15 | 3 | 0.015 | 14 |

*CPU time (in seconds) spent in Mathematica® kernel

If we do not multiply (5.2) by *h*, then we get the results displayed in Table 5.2. What we hoped to accomplish by multiplying the integrand of (5.2) by *h* (rather than multiplying the outcome of the integration by *h*) was to stabilize the quadratures. As we see in Fig. 5.3.1, the term $2v / \Delta(v, h)$ grows to large magnitudes as *h* gets smaller. When we multiply this term by *h*, we bring down this value to much smaller levels, as shown in Fig. 5.3.2. We thought this would help the numerical integration and alleviate the discrepancy between the two quadratures. A comparison, however, of the two tables shows that the effect is small. Specifically, we see no improvement for $h = 10^{-11}$ and $h = 10^{-13}$.

Table 5.2. Eq. (5.2) evaluated by the two quadratures at the OP (0.25, 0.25, *h*).

| | SECOND BQ, EQ. (5.2) | | | | | | | | |
|---|---|---|---|---|---|---|---|---|---|
| *h* | GKQ | SD | NOR | TIME* | DEQ | SD | NOR | TIME* | DOA |
| 1 | **1.49944916156393E-02** | 15 | 5 | 0.000 | **1.49944916156393E-02** | 15 | 3 | 0.000 | 15 |
| 1.E-01 | **7.40200806750922E-03** | 15 | 6 | 0.016 | **7.40200806750921E-03** | 15 | 4 | 0.016 | 14 |
| 1.E-03 | **4.75463049979300E-03** | 15 | 10 | 0.031 | **4.75463049979299E-03** | 15 | 5 | 0.015 | 14 |
| 1.E-05 | **4.72535082669790E-03** | 15 | 11 | 0.032 | **4.72535082669790E-03** | 15 | 5 | 0.031 | 15 |
| 1.E-07 | **4.72505773720324E-03** | 15 | 11 | 0.015 | **4.72505773720323E-03** | 15 | 5 | 0.015 | 14 |
| 1.E-09 | **4.72505480627900E-03** | 15 | 11 | 0.016 | **4.72505480627900E-03** | 15 | 5 | 0.032 | 15 |
| 1.E-11 | **4.72505477695218E-03** | 15 | 4 | 0.015 | **4.72505477696975E-03** | 15 | 5 | 0.015 | 11 |
| 1.E-13 | **4.72505477667649E-03** | 15 | 4 | 0.000 | **4.72505477667666E-03** | 15 | 4 | 0.015 | 13 |
| 1.E-15 | **4.72505477667373E-03** | 15 | 4 | 0.000 | **4.72505477667373E-03** | 15 | 3 | 0.000 | 15 |
| 1.E-17 | **4.72505477667370E-03** | 15 | 4 | 0.000 | **4.72505477667370E-03** | 15 | 3 | 0.016 | 15 |
| 1.E-19 | **4.72505477667370E-03** | 15 | 4 | 0.000 | **4.72505477667370E-03** | 15 | 3 | 0.000 | 15 |
| 0.E+00 | **4.72505477667370E-03** | 15 | 4 | 0.016 | **4.72505477667370E-03** | 15 | 3 | 0.000 | 15 |

*CPU time (in seconds) spent in Mathematica® kernel

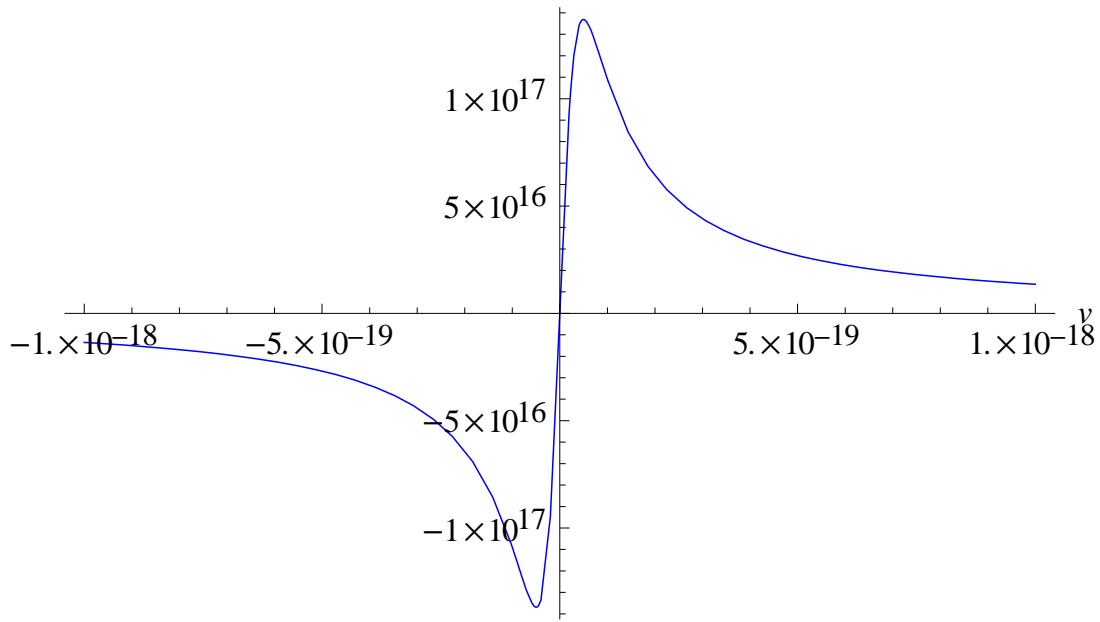

Figure 5.3.1. Graph of $2v / \Delta(v, h)$ near the origin for $h = 10^{-19}$.

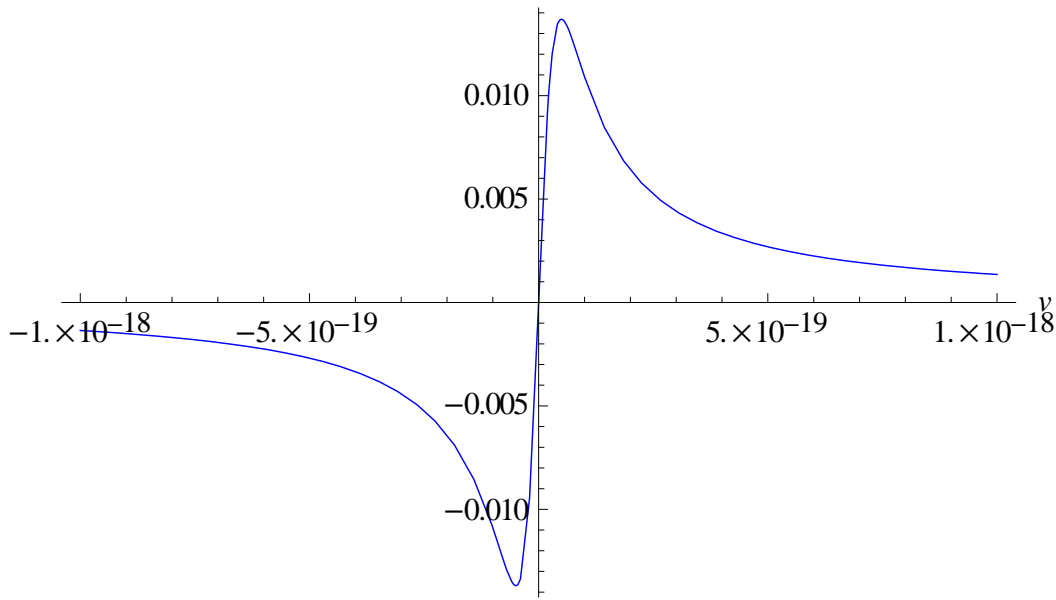

Figure 5.3.2. Graph of $2vh / \Delta(v, h)$ near the origin for $h = 10^{-19}$.

In an effort to pinpoint the cause of the discrepancy between the two quadratures at $h = 10^{-11}$ and $h = 10^{-13}$, we graphed the integrand of (5.2) for a number of values of $h$. We exhibit the graph for two values in Figs. 5.4.1 and 5.4.2. The graphs appear to be smooth. We know, however, from Fig. 5.1 that the behavior at the origin should not be. If we zoom to the origin, we see that, indeed, it is not. We display the impulse-like behavior in Figs. 5.5.1 and 5.5.2. It is this impulse that one or both of the quadratures may fail to capture.

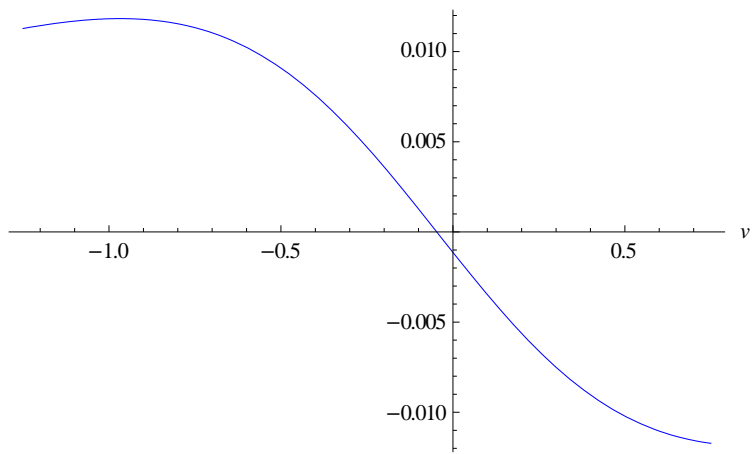

Figure 5.4.1. Graph of integrand of (5.2) for $h = 10^{-5}$.

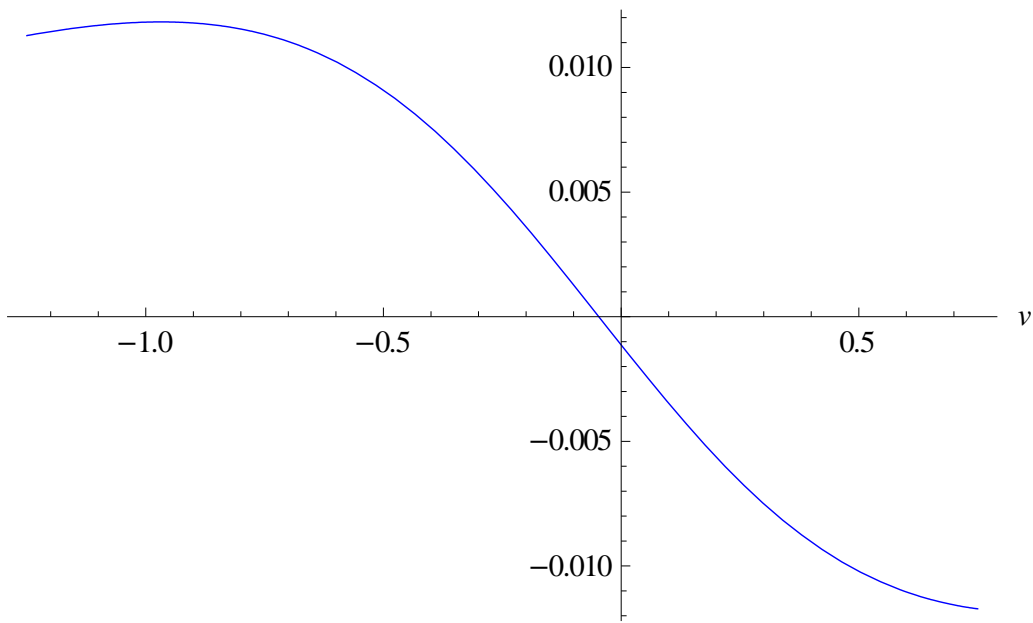

Figure 5.4.2. Graph of integrand of (5.2) for $h = 10^{-19}$.

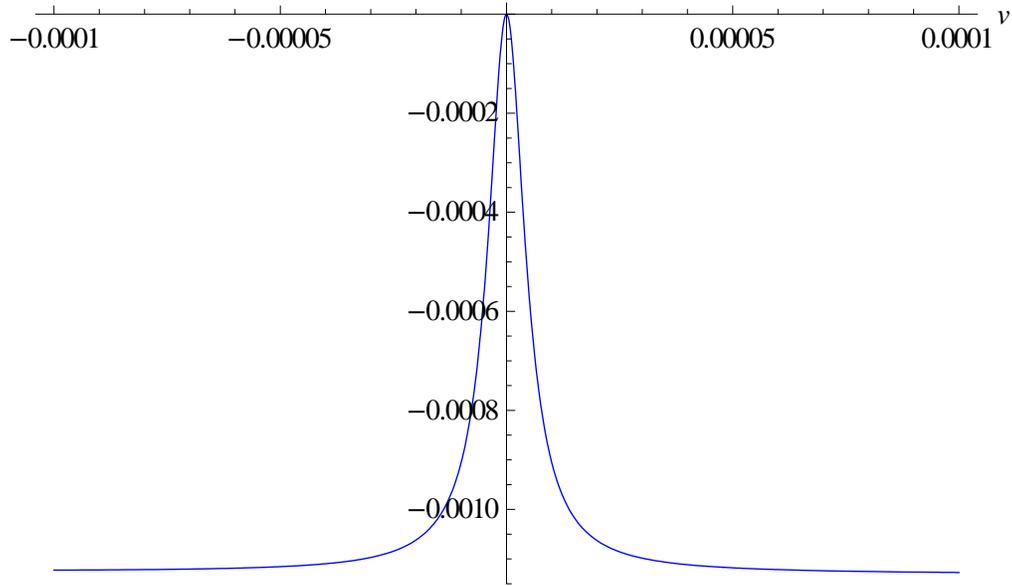

Figure 5.5.1. Graph of integrand of (5.2) about the origin for $h = 10^{-5}$.

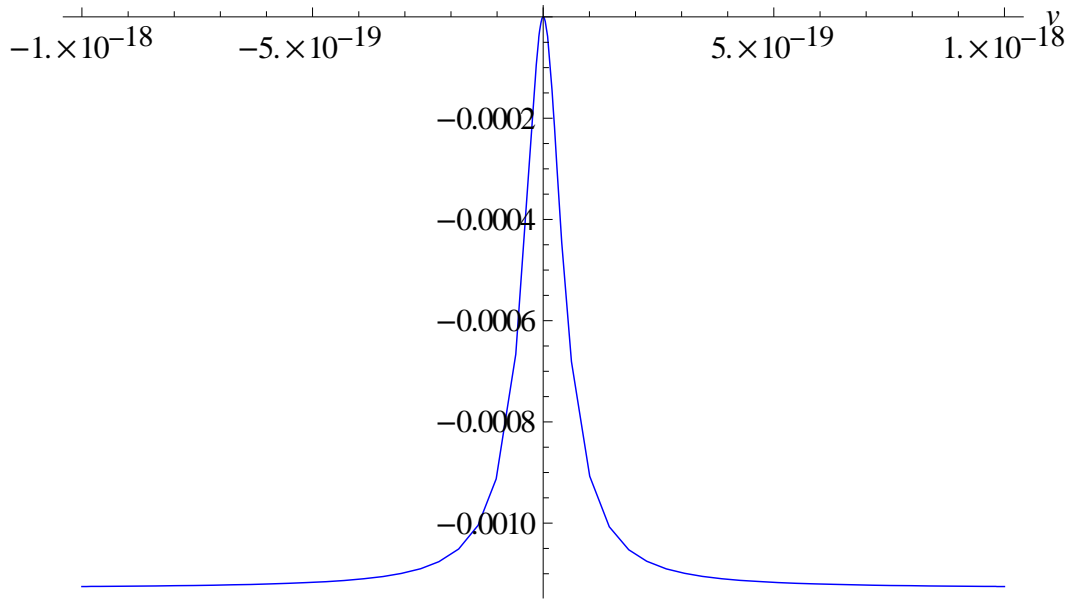

Figure 5.5.2. Graph of integrand of (5.2) about the origin for $h = 10^{-19}$.

We investigated this issue further by splitting the interval of integration $[-(1+q'), 1-q']$ into three intervals: $[-(1+q'), -h]$, $[-h, h]$ and $[h, 1-q']$, and by designating the origin as a singular point. We integrated over each subinterval and added up the results to obtain the value of (5.2). We display these results in Table 5.3. We see that the two quadratures are not in better agreement than before and that the center of disagreement has shifted to the values $h = 10^{-9}$ and $h = 10^{-11}$. In Table 5.4, we compare the one-integral to the three-integral approach for GKQ. In Table 5.5, we repeat using DEQ this time. We see that DOA for DEQ is better than that of GKQ. From Table 5.5, we also see that the one-integral DEQ is, in general, faster than the three-integral DEQ. This is the approach we recommend, *i.e.*, use DEQ as in Table 5.1. If we restrict the values of *h* to greater than $10^{-5}$, then either quadrature is acceptable.

Table 5.3. Eq. (5.2) multiplied by *h* and evaluated by the two quadratures at the OP (0.25, 0.25, *h*) as a sum of three integrals.

| | SECOND BQ, EQ. (5.2) AS SUM OF THREE INTEGRALS AND MULTIPLIED BY *h* | | | | | | |
|---|---|---|---|---|---|---|---|
| *h* | GKQ | SD | TIME* | DEQ | SD | TIME* | DOA |
| 1 | **1.49944916156393E-02** | 15 | 0.000 | **1.49944916156393E-02** | 15 | 0.032 | 15 |
| 1.E-01 | **7.40200806750922E-04** | 15 | 0.016 | **7.40200806750921E-04** | 15 | 0.015 | 14 |
| 1.E-03 | **4.75463049979300E-06** | 15 | 0.078 | **4.75463049979299E-06** | 15 | 0.031 | 14 |
| 1.E-05 | **4.72535082669790E-08** | 15 | 0.046 | **4.72535082669790E-08** | 15 | 0.032 | 15 |
| 1.E-07 | **4.72505773720324E-10** | 15 | 0.047 | **4.72505773720323E-10** | 15 | 0.031 | 14 |
| 1.E-09 | **4.72505480576348E-12** | 15 | 0.000 | **4.72505480627900E-12** | 15 | 0.031 | 9 |
| 1.E-11 | **4.72505477696460E-14** | 15 | 0.016 | **4.72505477696976E-14** | 15 | 0.031 | 12 |
| 1.E-13 | **4.72505477667662E-16** | 15 | 0.000 | **4.72505477667666E-16** | 15 | 0.016 | 14 |
| 1.E-15 | **4.72505477667373E-18** | 15 | 0.000 | **4.72505477667373E-18** | 15 | 0.031 | 15 |
| 1.E-17 | **4.72505477667371E-20** | 15 | 0.016 | **4.72505477667370E-20** | 15 | 0.016 | 14 |
| 1.E-19 | **4.72505477667370E-22** | 15 | 0.016 | **4.72505477667370E-22** | 15 | 0.015 | 15 |

*CPU time (in seconds) spent in Mathematica® kernel

Table 5.4. Eq. (5.2) multiplied by *h* and evaluated by GKQ at the OP (0.25, 0.25, *h*) as a single integral and as a sum of three integrals.

| | SECOND BQ, GKQ FOR EQ. (5.2) ALONE AND AS SUM OF THREE INTEGRALS , MULTIPLIED BY *h* | | | | | | |
|---|---|---|---|---|---|---|---|
| *h* | GKQ, ONE INTEGRAL | SD | TIME* | GKQ, THREE INTEGRALS | SD | TIME* | DOA |
| 1 | **1.49944916156393E-02** | 15 | 0.016 | **1.49944916156393E-02** | 15 | 0.000 | 15 |
| 1.E-01 | **7.40200806750922E-04** | 15 | 0.016 | **7.40200806750922E-04** | 15 | 0.016 | 15 |
| 1.E-03 | **4.75463049979300E-06** | 15 | 0.016 | **4.75463049979300E-06** | 15 | 0.078 | 15 |
| 1.E-05 | **4.72535082669790E-08** | 15 | 0.016 | **4.72535082669790E-08** | 15 | 0.046 | 15 |
| 1.E-07 | **4.72505773720324E-10** | 15 | 0.031 | **4.72505773720324E-10** | 15 | 0.047 | 15 |
| 1.E-09 | **4.72505480627900E-12** | 15 | 0.015 | **4.72505480576348E-12** | 15 | 0.000 | 9 |
| 1.E-11 | **4.72505477695218E-14** | 15 | 0.016 | **4.72505477696460E-14** | 15 | 0.016 | 11 |
| 1.E-13 | **4.72505477667649E-16** | 15 | 0.016 | **4.72505477667662E-16** | 15 | 0.000 | 13 |
| 1.E-15 | **4.72505477667373E-18** | 15 | 0.016 | **4.72505477667373E-18** | 15 | 0.000 | 15 |
| 1.E-17 | **4.72505477667370E-20** | 15 | 0.016 | **4.72505477667371E-20** | 15 | 0.016 | 14 |
| 1.E-19 | **4.72505477667371E-22** | 15 | 0.000 | **4.72505477667370E-22** | 15 | 0.016 | 14 |

*CPU time (in seconds) spent in Mathematica® kernel

Table 5.5. Eq. (5.2) multiplied by $h$ and evaluated by DEQ at the OP (0.25, 0.25, $h$) as a single integral and as a sum of three integrals.

| | SECOND BQ, DEQ FOR EQ. (5.2) ALONE AND AS SUM OF THREE INTEGRALS, MULTIPLIED BY $h$ | | | | | | |
|---|---|---|---|---|---|---|---|
| $h$ | DEQ, ONE INTEGRAL | SD | TIME* | DEQ, THREE INTEGRALS | SD | TIME* | DOA |
| 1 | 1.49944916156393E-02 | 15 | 0.015 | 1.49944916156393E-02 | 15 | 0.032 | 15 |
| 1.E-01 | 7.40200806750921E-04 | 15 | 0.016 | 7.40200806750921E-04 | 15 | 0.015 | 15 |
| 1.E-03 | 4.75463049979299E-06 | 15 | 0.032 | 4.75463049979299E-06 | 15 | 0.031 | 15 |
| 1.E-05 | 4.72535082669790E-08 | 15 | 0.031 | 4.72535082669790E-08 | 15 | 0.032 | 15 |
| 1.E-07 | 4.72505773720323E-10 | 15 | 0.031 | 4.72505773720323E-10 | 15 | 0.031 | 15 |
| 1.E-09 | 4.72505480627899E-12 | 15 | 0.016 | 4.72505480627900E-12 | 15 | 0.031 | 14 |
| 1.E-11 | 4.72505477696975E-14 | 15 | 0.031 | 4.72505477696976E-14 | 15 | 0.031 | 14 |
| 1.E-13 | 4.72505477667666E-16 | 15 | 0.000 | 4.72505477667666E-16 | 15 | 0.016 | 15 |
| 1.E-15 | 4.72505477667373E-18 | 15 | 0.016 | 4.72505477667373E-18 | 15 | 0.031 | 15 |
| 1.E-17 | 4.72505477667370E-20 | 15 | 0.000 | 4.72505477667370E-20 | 15 | 0.016 | 15 |
| 1.E-19 | 4.72505477667370E-22 | 15 | 0.015 | 4.72505477667370E-22 | 15 | 0.015 | 15 |

*CPU time (in seconds) spent in Mathematica® kernel

In Table 5.2, we also gave the value of the integral for $h = 0$. We conclude this section by computing the value of the integral in (5.2) when $h$ is equal to zero. We first show that the singularity at the origin is removable. When $h$ is equal to zero, we have from (3.4) that

$$A(v,0) = |\mathbf{q}'|^2 v^2, \quad B(v,0) = 2\left[\mathbf{q}' \cdot (\mathbf{p}' + \mathbf{r}_{pq} v)\right] v, \quad C(v) = |\mathbf{p}' + \mathbf{r}_{pq} v| \qquad (5.3)$$

while, from (3.6),

$$\begin{aligned}
\Delta(v,0) &= 4C^2(v) A(v,0) - B^2(v,0) \\
&= 4\left|\mathbf{p}' + \mathbf{r}_{pq} v\right|^2 |\mathbf{q}'|^2 v^2 - 4\left[\mathbf{q}' \cdot (\mathbf{p}' + \mathbf{r}_{pq} v)\right]^2 v^2 \\
&= 4v^2 \left\{\left|\mathbf{p}' + \mathbf{r}_{pq} v\right|^2 |\mathbf{q}'|^2 - \left[\mathbf{q}' \cdot (\mathbf{p}' + \mathbf{r}_{pq} v)\right]^2\right\} \\
&= 4v^2 \left|\mathbf{q}' \times (\mathbf{p}' + \mathbf{r}_{pq} v)\right|^2.
\end{aligned} \qquad (5.4)$$

Also,

$$2A(v,0) + B(v,0)u = 2v\left\{|\mathbf{q}'|^2 v + \left[\mathbf{q}' \cdot (\mathbf{p}' + \mathbf{r}_{pq} v)\right] u\right\} \qquad (5.5)$$

so that

$$\frac{2v}{\Delta(v,0)}\left[2A(v,0) + B(v,0)u\right] = \frac{|\mathbf{q}'|^2 v + \left[\mathbf{q}' \cdot (\mathbf{p}' + \mathbf{r}_{pq} v) u\right]}{\left|\mathbf{q}' \times (\mathbf{p}' + \mathbf{r}_{pq} v)\right|^2}. \qquad (5.6)$$

From (3.3)

$$R(u,v,0) = \sqrt{C(v)^2 u^2 + B(v,0)u + A(v,0)}$$
$$= \left\{ \left|\mathbf{p}' + \mathbf{r}_{pq}v\right|^2 u^2 + 2\left[\mathbf{q}'\cdot\left(\mathbf{p}' + \mathbf{r}_{pq}v\right)\right]vu + \left|\mathbf{q}'\right|^2 v^2 \right\}^{1/2}$$
$$= \left|\mathbf{q}'v + \left(\mathbf{p}' + \mathbf{r}_{pq}v\right)u\right|. \tag{5.7}$$

When we substitute these expressions in (5.2), we get that

$$I_1(p',q',0) = 2\int_{-1-q'}^{1-q'} \frac{v\,dv}{\Delta(v,0)} \left\{ \frac{2A(v,0) - B(v,0)(1+p')}{R(-(1+p'),v,0)} - \frac{2A(v,0) + B(v,0)(1-p')}{R(1-p',v,0)} \right\}$$
$$= \int_{-1-q'}^{1-q'} \frac{dv}{\left|\mathbf{q}'\times(\mathbf{p}'+\mathbf{r}_{pq}v)\right|^2} \left\{ \frac{|\mathbf{q}'|^2 v - (1+p')\left[\mathbf{q}'\cdot(\mathbf{p}'+\mathbf{r}_{pq}v)\right]}{\left|\mathbf{q}'v - (1+p')(\mathbf{p}'+\mathbf{r}_{pq}v)\right|} - \frac{|\mathbf{q}'|^2 v + (1-p')\left[\mathbf{q}'\cdot(\mathbf{p}'+\mathbf{r}_{pq}v)\right]}{\left|\mathbf{q}'v + (1-p')(\mathbf{p}'+\mathbf{r}_{pq}v)\right|} \right\}.$$
$$\tag{5.8}$$

We see that, for any interior OP, the integrand attains a finite value. This is the expression we should use when $h$ is equal to zero. We exhibit the graph of the integrand for the second BQ of Appendix B and for $p' = q' = 0.25$ in Figs. 5.6.1 and 5.6.2. The value of the integrand at the origin is -0.00112795 and, by either quadrature, the value of the integral is $4.7250547766737 \times 10^{-3}$. This agrees with the result in Table 5.2 were the result was obtained by setting $h$ to zero in (5.2).

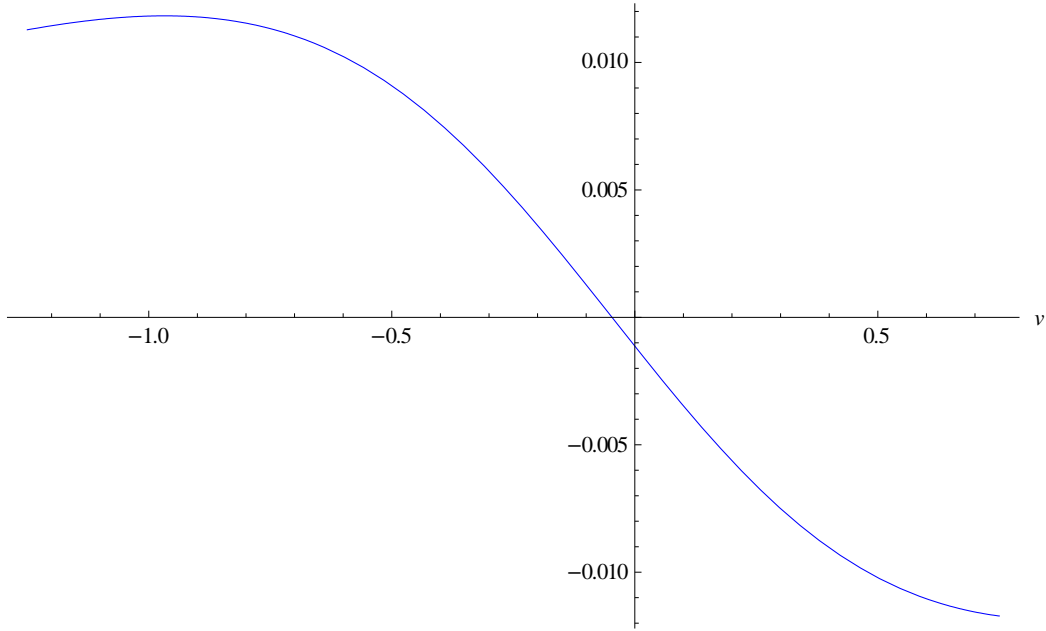

Figure 5.6.1. Graph of integrand of (5.8) for the second BQ of Appendix B and for $p' = q' = 0.25$.

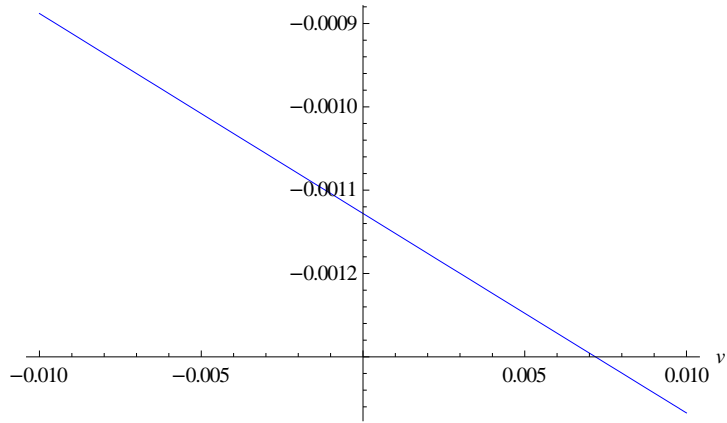

Figure 5.6.2. Graph of integrand of (5.8) about the origin. Same conditions as for Fig. 5.6.1.

For OPs on the boundary of the BQ, we still have a well-defined integral. For example,

$$I_1(-1,-1,0) = \int_0^2 \frac{dv}{\left|\mathbf{q}'\times(\mathbf{p}'+\mathbf{r}_{pq}v)\right|^2}\left\{|\mathbf{q}'| - \frac{|\mathbf{q}'|^2 v + 2\left[\mathbf{q}'\cdot(\mathbf{p}'+\mathbf{r}_{pq}v)\right]}{\left|\mathbf{q}'v + 2(\mathbf{p}'+\mathbf{r}_{pq}v)\right|}\right\}. \tag{5.9}$$

# 6. EVALUATION OF INTEGRALS IN $R^{-3}$. PART IV

In this section, we examine the third integral of the second expression in (2.45), namely,

$$I_1(p',q',h) = \int_{-1-q'}^{1-q'} v^2 dv \int_{-1-p'}^{1-p'} \frac{du}{R^3(u,v,h)}. \qquad (6.1)$$

From the graph of the integrand in Fig. 6.1, we see that we have an impulse-like behavior along $u$, and a double-hump behavior along $v$. For this reason and, also, in order to take advantage of the results of previous sections, we proceed with the $u$-integration first.

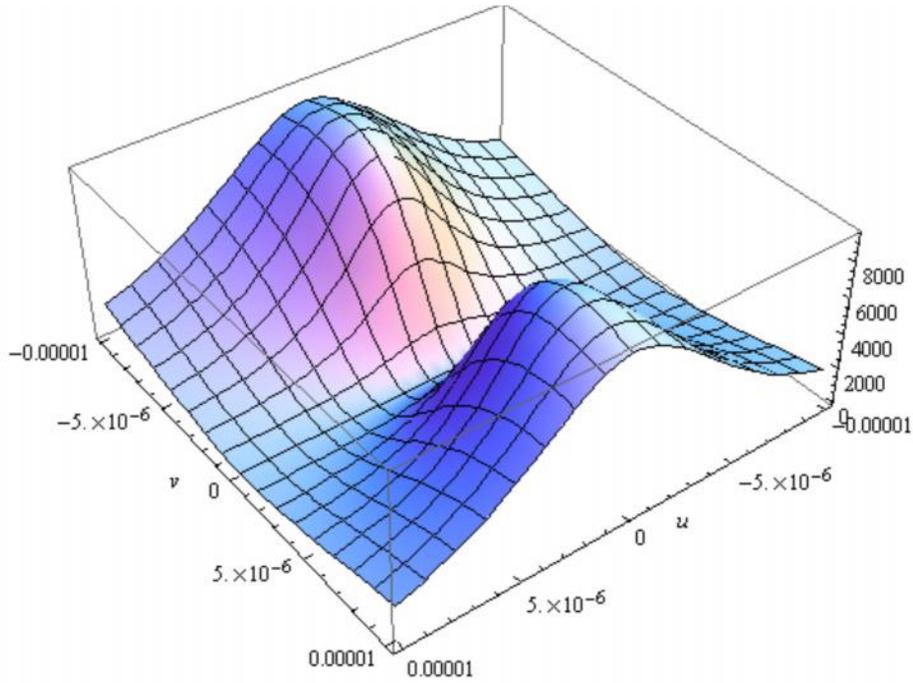

Figure 6.1. Graph of the integrand in (6.1) near the origin for the second BQ of Appendix B and for $h = 10^{-5}$. The OP is (0.25, 0.25, $h$).

From (3.5) and (3.6)

$$I_1(p',q',h) = 2\int_{-1-q'}^{1-q'} \frac{v^2 dv}{\Delta(v,h)} \left\{ \frac{[2C^2(v)(1-p') + B(v,h)]}{R(1-p',v,h)} - \frac{[-2C^2(v)(1+p') + B(v,h)]}{R(-(1+p'),v,h)} \right\}$$

(6.2)

This integral is well defined when $h$ is equal to zero. Using (5.3), (5.4) and (5.7), we have

$$I_1(p', q', 0) = \int_{-1-q'}^{1-q'} \frac{dv}{\left|\mathbf{q}' \times (\mathbf{p}' + \mathbf{r}_{pq}v)\right|^2}$$

$$\cdot \left\{ \frac{\left|\mathbf{p}' + \mathbf{r}_{pq}v\right|^2 (1 - p') + \left[\mathbf{q}' \cdot (\mathbf{p}' + \mathbf{r}_{pq}v)\right] v}{\left|\mathbf{q}'v + (1 - p')(\mathbf{p}' + \mathbf{r}_{pq}v)\right|} - \frac{-\left|\mathbf{p}' + \mathbf{r}_{pq}v\right|^2 (1 + p') + \left[\mathbf{q}' \cdot (\mathbf{p}' + \mathbf{r}_{pq}v)\right] v}{\left|\mathbf{q}'v - (1 + p')(\mathbf{p}' + \mathbf{r}_{pq}v)\right|} \right\}. \quad (6.3)$$

We are then concerned with the computability of (6.2) for small values of $h$. We conduct numerical experiments as in the previous sections using the second BQ in Appendix B and the OP (0.25, 0.25, $h$). In Fig. 6.2, we present the graph of the integrand for three values of $h$. It appears that for $h = 0.1$, the graph makes a valley about the origin while, for the other two values ($10^{-5}$ and $10^{-19}$), it does not. If we zoom, however, toward the origin, we find that the other two graphs exhibit the same behavior there (Figs. 6.3 and 6.4).

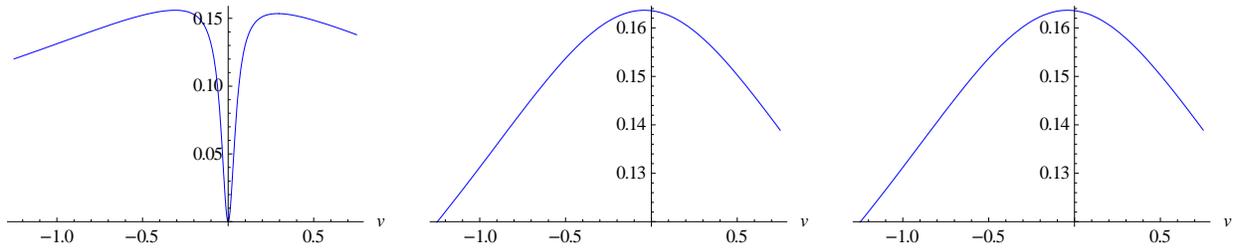

Figure 6.2. Graph of the integrand of (6.2) for the second BQ of Appendix B and $p' = q' = 0.25$. From left to right: $h = 10^{-1}$, $10^{-5}$ and $10^{-19}$.

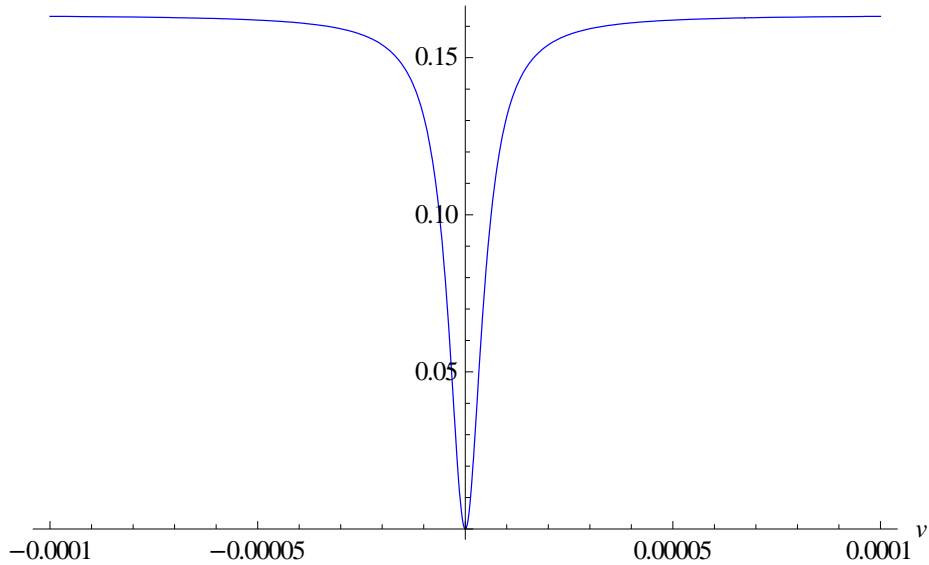

Figure 6.3. Middle graph in Fig. 6.2 near the origin.

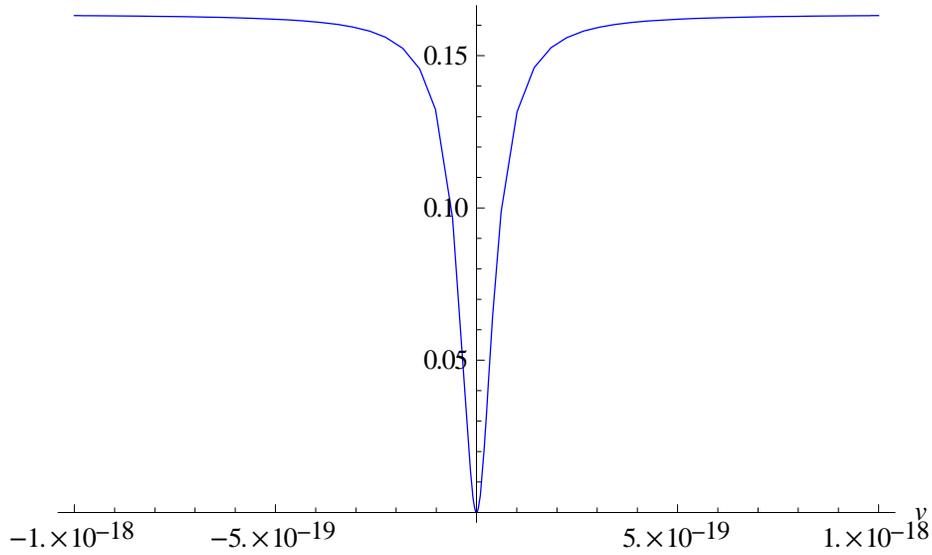

Figure 6.4. Right graph in Fig. 6.2 near the origin.

We exhibit the results of our calculations in Mathematica® in Table 6.1 where we see that the two quadratures are not in total agreement. We note that the DEQ result is monotonically increasing. The GKQ result does the same except in going from $10^{-11}$ to $10^{-13}$ where we notice a drop. This is not a proof of anything but we tend to trust the DEQ result more.

We have also computed the value when $h$ is zero. Both quadratures returned the value 0.298429854494576. This value is in agreement with the GKQ and DEQ values using (6.2). In Fig. 6.5.1, we graph the integrand of (6.3) while in Fig. 6.5.2 we zoom toward the origin to demonstrate that the integrand is continuous there.

We conclude by noting that the third integral in the first expression in (2.45) may be computed exactly as above by interchanging the roles of $u$ and $v$.

Table 6.1. Eq. (6.2) evaluated by the two quadratures at the OP (0.25, 0.25, $h$).

| | SECOND BQ, EQ. (6.2) | | | | | | |
|---|---|---|---|---|---|---|---|
| $h$ | GKQ | SD | TIME* | DEQ | SD | TIME* | DOA |
| 1 | **1.28025716887159E-01** | 15 | 0.015 | **1.28025716887159E-01** | 15 | 0.016 | 15 |
| 1.E-01 | **2.74030347929187E-01** | 15 | 0.016 | **2.74030347929187E-01** | 15 | 0.015 | 15 |
| 1.E-03 | **2.98176063303905E-01** | 15 | 0.015 | **2.98176063303904E-01** | 15 | 0.031 | 14 |
| 1.E-05 | **2.98427315588154E-01** | 15 | 0.015 | **2.98427315588154E-01** | 15 | 0.047 | 15 |
| 1.E-07 | **2.98429829105412E-01** | 15 | 0.047 | **2.98429829105412E-01** | 15 | 0.031 | 15 |
| 1.E-09 | **2.98429854240685E-01** | 15 | 0.031 | **2.98429854240684E-01** | 15 | 0.031 | 14 |
| 1.E-11 | **2.98429854494587E-01** | 15 | 0.015 | **2.98429854492037E-01** | 15 | 0.032 | 11 |
| 1.E-13 | **2.98429854494576E-01** | 15 | 0.015 | **2.98429854494550E-01** | 15 | 0.031 | 13 |
| 1.E-15 | **2.98429854494576E-01** | 15 | 0.000 | **2.98429854494576E-01** | 15 | 0.000 | 15 |
| 1.E-17 | **2.98429854494576E-01** | 15 | 0.000 | **2.98429854494576E-01** | 15 | 0.000 | 15 |
| 1.E-19 | **2.98429854494576E-01** | 15 | 0.016 | **2.98429854494576E-01** | 15 | 0.000 | 15 |

*CPU time (in seconds) spent in Mathematica® kernel

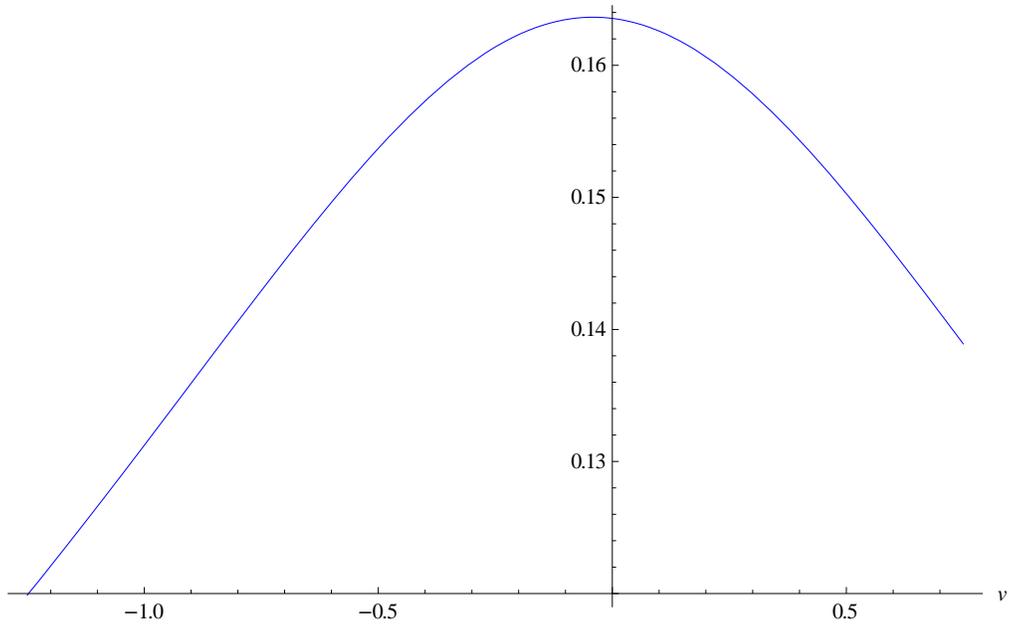

Figure 6.5.1. Graph of integrand of (6.3) for the second BQ of Appendix B and $p' = q' = 0.25$.

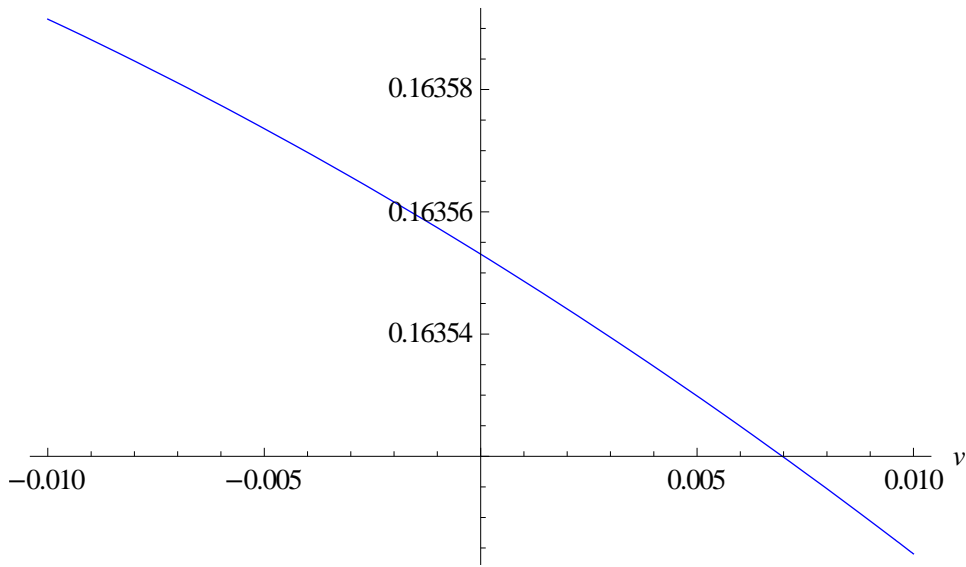

Figure 6.5.2. Graph of integrand of (6.3) about the origin for the second BQ of Appendix B and $p' = q' = 0.25$.

# 7. EVALUATION OF INTEGRALS IN $R^{-3}$. PART V

In this section, we examine the second integral of the first expression in (2.45), namely,

$$I_1(p',q',h) = \int_S \frac{uv^2}{J(p,q)R^3} dS = \int_{-1-q'}^{1-q'} v^2 dv \int_{-1-p'}^{1-p'} \frac{udu}{R^3(u,v,h)}. \tag{7.1}$$

We have graphed the integrand in Fig. 7.1 for the second BQ of Appendix B and for $h = 10^{-5}$. We see that in the neighborhood of the origin ($p' = q' = 0.25$) the graph changes rapidly. The integration along $u$ appears to be the simpler one and that is how we proceed.

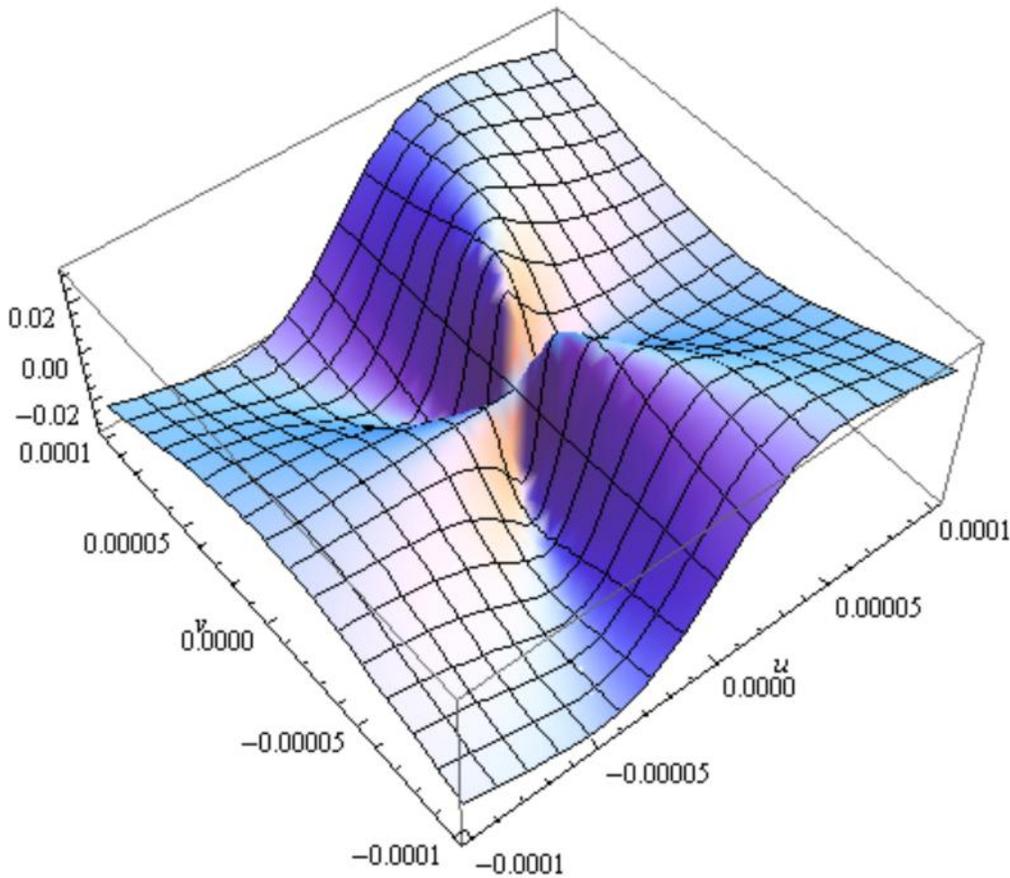

Figure 7.1. Graph of the integrand in (7.1) about the origin for the second BQ and with $h = 10^{-5}$.

From (5.1) and (5.2), we can write

$$I_1(p',q',h) = -2 \int_{-1-q'}^{1-q'} v^2 dv \frac{2A(v,h) + B(v,h)u}{U(v,h)R(u,v,h)}\bigg|_{-1-p'}^{1-p'}$$

$$= 2\int_{-1-q'}^{1-q'} \frac{v^2 dv}{\mathsf{U}(v,h)} \left\{ \frac{2A(v,h)-B(v,h)(1+p')}{R(-(1+p'),v,h)} - \frac{2A(v,h)+B(v,h)(1-p')}{R(1-p',v,h)} \right\}. \qquad (7.2)$$

Also, as with (5.8), we can write

$$I_1(p',q',0) = 2\int_{-1-q'}^{1-q'} \frac{v^2 dv}{\Delta(v,0)} \left\{ \frac{2A(v,0)-B(v,0)(1+p')}{R(-(1+p'),v,0)} - \frac{2A(v,0)+B(v,0)(1-p')}{R(1-p',v,0)} \right\}$$

$$= \int_{-1-q'}^{1-q'} \frac{v\,dv}{|\mathbf{q}'\times(\mathbf{p}'+\mathbf{r}_{pq}v)|^2} \left\{ \frac{|\mathbf{q}'|^2 v - (1+p')[\mathbf{q}'\cdot(\mathbf{p}'+\mathbf{r}_{pq}v)]}{|\mathbf{q}'v - (1+p')(\mathbf{p}'+\mathbf{r}_{pq}v)|} - \frac{|\mathbf{q}'|^2 v + (1-p')[\mathbf{q}'\cdot(\mathbf{p}'+\mathbf{r}_{pq}v)]}{|\mathbf{q}'v + (1-p')(\mathbf{p}'+\mathbf{r}_{pq}v)|} \right\}.$$

(7.3)

In Fig. 7.2, we display the graph of the integrand of (7.2) for the second BQ of Appendix B and for the OP (0.25, 0.25, $h$). This time, the origin is a regular point, as we see in Fig. 7.3.

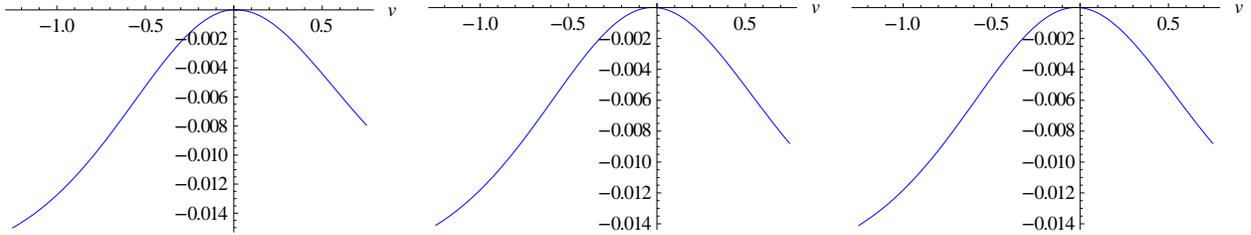

Figure 7.2. Graph of the integrand of (7.2) for the second BQ of Appendix B and $p' = q' = 0.25$. From left to right: $h = 10^{-1}$, $10^{-5}$ and $10^{-19}$.

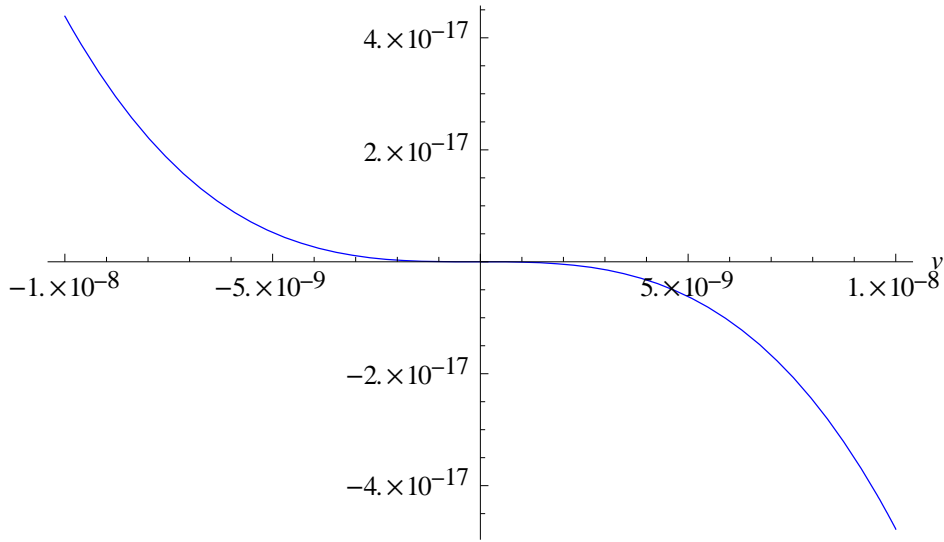

Figure 7.3. Middle graph in Fig. 7.2 near the origin.

In Table 7.1, we display the results of the evaluation of (7.2). We see that the two quadratures are in quite close agreement. We have also computed the value when $h$ is zero using (7.3). Both quadratures returned the value -0.0109529831506828. This value is in agreement with the value obtained using (7.2). In Fig. 7.4.1, we graph the integrand of (7.3) while in Fig. 7.4.2 we enlarge the region about the origin to demonstrate that the integrand is continuous there.

We conclude this section by noting that the second integral in the second expression in (2.45) can be evaluated in the same fashion by interchanging the roles of $u$ and $v$.

Table 7.1. Eq. (7.2) evaluated by the two quadratures at the OP (0.25, 0.25, $h$).

| | SECOND BQ, EQ. (7.2) | | | | | | |
|---|---|---|---|---|---|---|---|
| $h$ | GKQ | SD | TIME* | DEQ | SD | TIME* | DOA |
| **1** | **-1.29266108527094E-02** | **15** | **0.015** | **-1.29266108527093E-02** | **15** | **0.015** | **14** |
| 1.E-01 | -1.14180359320254E-02 | 15 | 0.000 | -1.14180359320253E-02 | 15 | 0.016 | 14 |
| 1.E-03 | -1.09578998125534E-02 | 15 | 0.000 | -1.09578998125534E-02 | 15 | 0.016 | 15 |
| 1.E-05 | -1.09530323435366E-02 | 15 | 0.000 | -1.09530323435365E-02 | 15 | 0.016 | 14 |
| 1.E-07 | -1.09529836426140E-02 | 15 | 0.031 | -1.09529836426139E-02 | 15 | 0.000 | 14 |
| 1.E-09 | -1.09529831556021E-02 | 15 | 0.000 | -1.09529831556021E-02 | 15 | 0.000 | 15 |
| 1.E-11 | -1.09529831507320E-02 | 15 | 0.016 | -1.09529831507320E-02 | 15 | 0.015 | 15 |
| 1.E-13 | -1.09529831506833E-02 | 15 | 0.000 | -1.09529831506833E-02 | 15 | 0.016 | 15 |
| 1.E-15 | -1.09529831506828E-02 | 15 | 0.000 | -1.09529831506828E-02 | 15 | 0.016 | 15 |
| 1.E-17 | -1.09529831506828E-02 | 15 | 0.016 | -1.09529831506828E-02 | 15 | 0.000 | 15 |
| 1.E-19 | -1.09529831506828E-02 | 15 | 0.016 | -1.09529831506828E-02 | 15 | 0.000 | 15 |

*CPU time (in seconds) spent in Mathematica® kernel

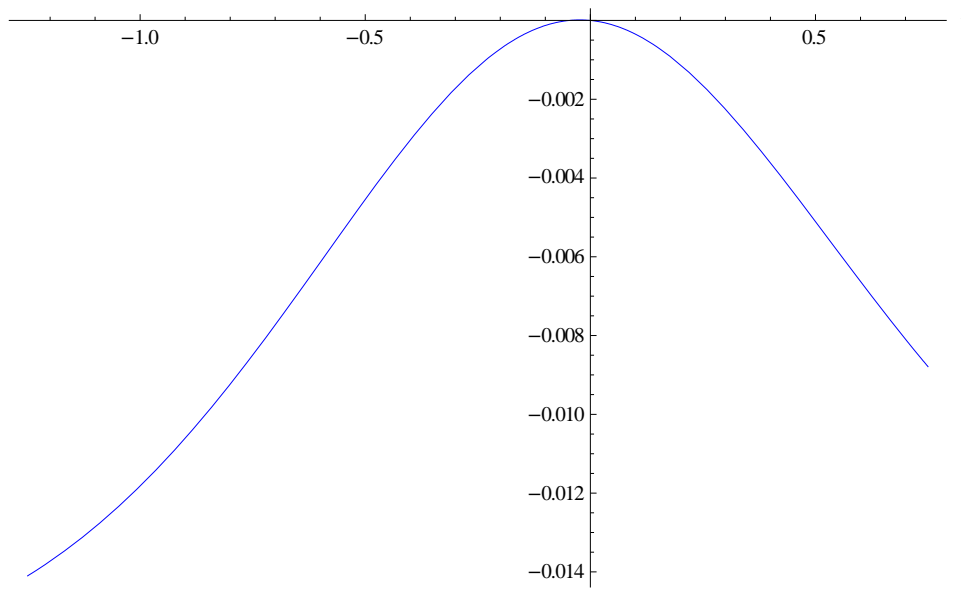

Figure 7.4.1. Graph of integrand of (7.3) for the second BQ of Appendix B and $p' = q' = 0.25$.

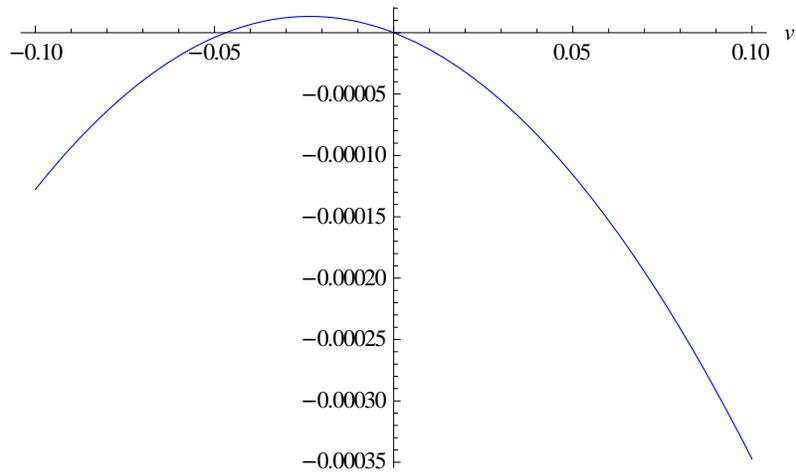

Figure 7.4.2. Graph of integrand of (7.3) about the origin for the second BQ of Appendix B and $p' = q' = 0.25$.

# 8. EVALUATION OF INTEGRALS IN $R^{-3}$. PART VI

In this section, we examine the fourth integral of either expression in (2.45), namely,

$$I_1(p',q',h) = \lim_{u \to 0} \int_{-1-p'}^{-u} u^2 du \lim_{v \to 0} \left\{ \int_{-1-q'}^{-v} \frac{v^2 dv}{R^3(u,v,h)} + \int_{v}^{1-q'} \frac{v^2 dv}{R^3(u,v,h)} \right\}$$

$$+ \lim_{u \to 0} \int_{u}^{1-p'} u^2 du \lim_{v \to 0} \left\{ \int_{-1-q'}^{-v} \frac{v^2 dv}{R^3(u,v,h)} + \int_{v}^{1-q'} \frac{v^2 dv}{R^3(u,v,h)} \right\}. \tag{8.1}$$

We have reversed the order of integration so that we have an example where we integrate analytically with respect to $v$ rather than $u$. We present the graph of the integrand of (8.1) in Fig. 8.1. We have used the second BQ of Appendix B, and the OP (0.25, 0.25, $10^{-5}$). In this case, the influence of $h$ is minimal as we see from Fig. 8.2. We might also expect the graph to be symmetric along the principal axes. We must remember, however, that, for the particular BQ, $R$ is not symmetric in $u$ and $v$.

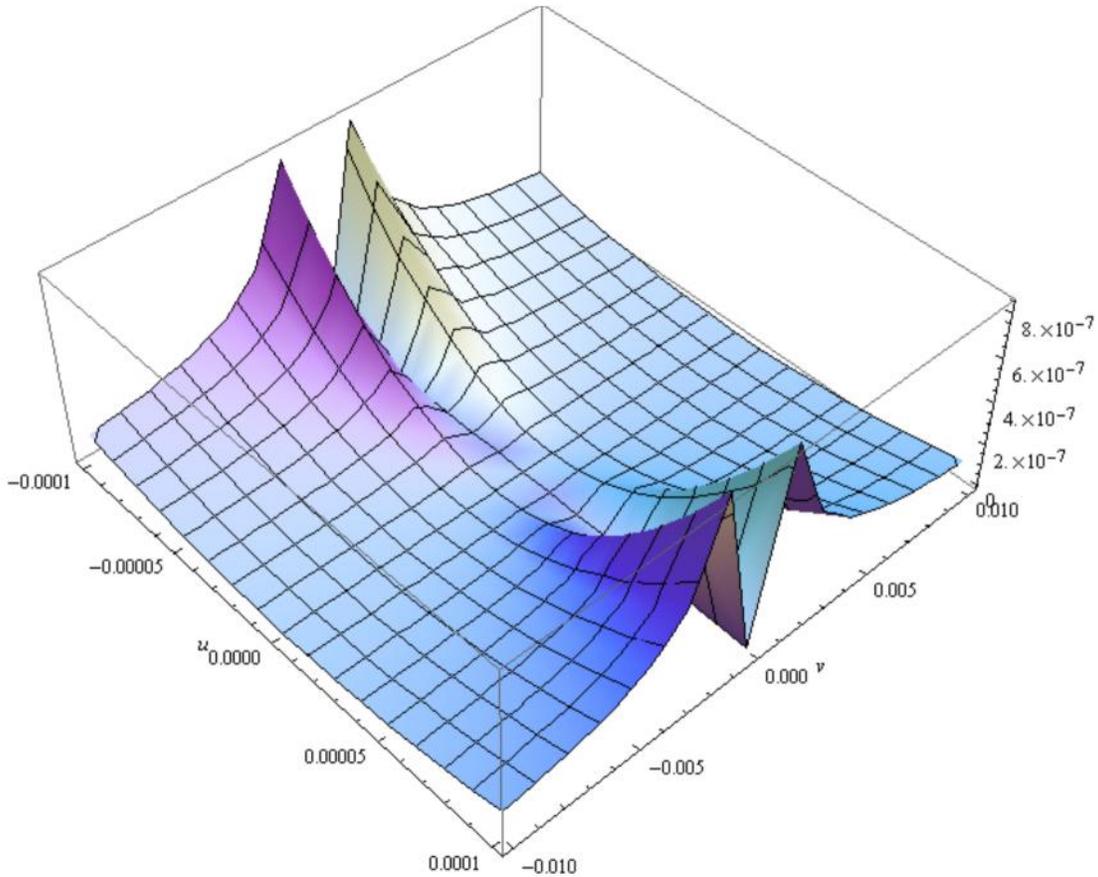

Figure 8.1. Graph of integrand in (8.1) for the second BQ of Appendix B, and the OP (0.25, 0.25, $10^{-5}$).

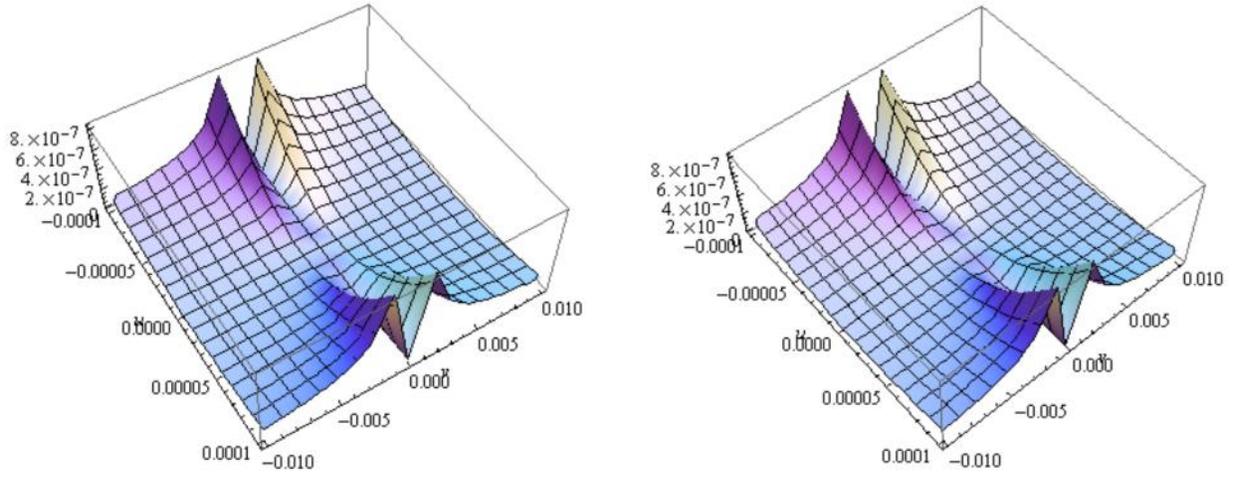

Figure 8.2. Graph of integrand in (8.1) for the second BQ of Appendix B, and the OP (0.25, 0.25, $h$). Left: $h = 10^{-5}$. Right: $h = 10^{-19}$.

From (2.22), we can write for the distance function

$$R = \sqrt{F^2 v^2 + Ev + D} \tag{8.2}$$

with

$$D(u,h) = |\mathbf{p}'|^2 u^2 + h^2, \quad E(u,h) = 2\left[\mathbf{p}' \cdot (\mathbf{q}' + \mathbf{r}_{pq} u) - h\hat{n}' \cdot \mathbf{r}_{pq}\right] u, \quad F(u) = |\mathbf{q}' + \mathbf{r}_{pq} u|. \tag{8.3}$$

From [11], p. 83, Formula 2.264.7, we have that

$$\int \frac{v^2 dv}{R^3(u,v,h)} = -\frac{\left[\Delta(u,h) - E^2(u,h)\right]v - 2D(u,h)E(u,h)}{F^2(u,h)\Delta(u,h)R(u,v,h)} + \frac{1}{F^2(u,h)} \int \frac{dv}{R(u,v,h)}. \tag{8.4}$$

where

$$\mathsf{U}(u,h) = 4F^2(u)D(u,h) - E^2(u,h). \tag{8.5}$$

Applying it to (8.1), we get

$$I_1 = \lim_{u \to 0} \int_{-1-p'}^{-u} u^2 du \lim_{v \to 0} \left\{ -\frac{\left[\Delta(u,h) - E^2(u,h)\right]v - 2D(u,h)E(u,h)}{F^2(u,h)\Delta(u,h)R(u,v,h)} \bigg|_{-1-q'}^{-v} + \frac{1}{F^2(u,h)} \int_{-1-q'}^{-v} \frac{dv}{R(u,v,h)} \right.$$

$$\left. - \frac{\left[\Delta(u,h) - E^2(u,h)\right]v - 2D(u,h)E(u,h)}{F^2(u,h)\Delta(u,h)R(u,v,h)} \bigg|_{v}^{1-q'} + \frac{1}{F^2(u,h)} \int_{v}^{1-q'} \frac{dv}{R(u,v,h)} \right\}$$

$$+ \lim_{u \to 0} \int_{u}^{1-p'} u^2 du \lim_{v \to 0} \left\{ -\frac{\left[\Delta(u,h) - E^2(u,h)\right]v - 2D(u,h)E(u,h)}{F^2(u,h)\Delta(u,h)R(u,v,h)} \bigg|_{-1-q'}^{-v} + \frac{1}{F^2(u,h)} \int_{-1-q'}^{-v} \frac{dv}{R(u,v,h)} \right.$$

$$\left. - \frac{\left[\Delta(u,h) - E^2(u,h)\right]v - 2D(u,h)E(u,h)}{F^2(u,h)\Delta(u,h)R(u,v,h)} \bigg|_{v}^{1-q'} + \frac{1}{F^2(u,h)} \int_{v}^{1-q'} \frac{dv}{R(u,v,h)} \right\}$$

$$= \lim_{u \to 0} \int_{-1-p'}^{-u} u^2 du \left\{ -\frac{\left[\Delta(u,h) - E^2(u,h)\right](1+q') + 2D(u,h)E(u,h)}{F^2(u,h)\Delta(u,h)R(u,-(1+q'),h)} \right.$$

$$\left. -\frac{\left[\Delta(u,h) - E^2(u,h)\right](1-q') - 2D(u,h)E(u,h)}{F^2(u,h)\Delta(u,h)R(u,1-q',h)} + \frac{1}{F^2(u,h)} \lim_{v \to 0} \left[ \int_{-1-q'}^{-v} \frac{dv}{R(u,v,h)} + \int_{v}^{1-q'} \frac{dv}{R(u,v,h)} \right] \right\}$$

$$+ \lim_{u \to 0} \int_{u}^{1-p'} u^2 du \left\{ -\frac{\left[\Delta(u,h) - E^2(u,h)\right](1+q') + 2D(u,h)E(u,h)}{F^2(u,h)\Delta(u,h)R(u,-(1+q'),h)} \right.$$

$$\left. -\frac{\left[\Delta(u,h) - E^2(u,h)\right](1-q') - 2D(u,h)E(u,h)}{F^2(u,h)\Delta(u,h)R(u,1-q',h)} + \frac{1}{F^2(u,h)} \lim_{v \to 0} \left[ \int_{-1-q'}^{-v} \frac{dv}{R(u,v,h)} + \int_{v}^{1-q'} \frac{dv}{R(u,v,h)} \right] \right\}$$

(8.6)

We have dealt with the remaining integral in *v* in [5]. This integral exists and can be evaluated analytically. The integral, however, with respect to *u* requires numerical integration. In order to proceed, we express the original integral as the sum of two integrals

$$I_1(p',q',h) = I_{11}(p',q',h) + I_{12}(p',q',h) \qquad (8.7)$$

where

$$I_{11}(p',q',h) = -\lim_{u \to 0} \int_{-1-p'}^{-u} u^2 du \left\{ \frac{\left[\Delta(u,h) - E^2(u,h)\right](1+q') + 2D(u,h)E(u,h)}{F^2(u,h)\Delta(u,h)R(u,-(1+q'),h)} \right.$$

$$\left. + \frac{\left[\Delta(u,h) - E^2(u,h)\right](1-q') - 2D(u,h)E(u,h)}{F^2(u,h)\Delta(u,h)R(u,1-q',h)} \right\}$$

$$-\lim_{u \to 0} \int_u^{1-p'} u^2 du \left\{ \frac{\left[\Delta(u,h) - E^2(u,h)\right](1+q') + 2D(u,h)E(u,h)}{F^2(u,h)\Delta(u,h)R(u,-(1+q'),h)} \right.$$

$$\left. + \frac{\left[\Delta(u,h) - E^2(u,h)\right](1-q') - 2D(u,h)E(u,h)}{F^2(u,h)\Delta(u,h)R(u,1-q',h)} \right\} \quad (8.8)$$

and

$$I_{12}(p',q',h) = \lim_{u \to 0} \int_{-1-p'}^{-u} \frac{u^2 du}{F^2(u,h)} \lim_{v \to 0} \left[ \int_{-1-q'}^{-v} \frac{dv}{R(u,v,h)} + \int_v^{1-q'} \frac{dv}{R(u,v,h)} \right]$$

$$+ \lim_{u \to 0} \int_u^{1-p'} \frac{u^2 du}{F^2(u,h)} \lim_{v \to 0} \left[ \int_{-1-q'}^{-v} \frac{dv}{R(u,v,h)} + \int_v^{1-q'} \frac{dv}{R(u,v,h)} \right]. \quad (8.9)$$

We have performed numerical calculations for (8.8) using the second BQ in Appendix B and the OP ($p' = 0.25$, $q' = 0.25$, $h$). The results of the two quadratures agree to at least 14 SD. The execution times are also very comparable. We also note that the smaller $h$ gets, the less it influences the value of the integral, something we noticed in Fig. 8.2 also. The integrand in this case is quite smooth, as we see from Fig. 8.3.

Table 8.1. Integral in (8.8) evaluated for the second BQ using GKQ and DEQ ($p' = q' = 0.25$).

| | SECOND BQ, INTEGRAL IN (8.8) | | | | | |
|---|---|---|---|---|---|---|
| h | GKQ | GKQ SD | GKQ TIME* | DEQ | DEQ SD | DEQ TIME* |
| 1 | **-9.92920282567043E-02** | 15 | 0.000 | **-9.92920282567042E-02** | 15 | 0.016 |
| 1.E-01 | **-1.04431229587766E-01** | 15 | 0.015 | **-1.04431229587766E-01** | 15 | 0.016 |
| 1.E-03 | **-1.04493450447920E-01** | 15 | 0.016 | **-1.04493450447920E-01** | 15 | 0.016 |
| 1.E-05 | **-1.04493479244954E-01** | 15 | 0.015 | **-1.04493479244954E-01** | 15 | 0.015 |
| 1.E-07 | **-1.04493479473423E-01** | 15 | 0.016 | **-1.04493479473422E-01** | 15 | 0.016 |
| 1.E-09 | **-1.04493479475701E-01** | 15 | 0.000 | **-1.04493479475701E-01** | 15 | 0.015 |
| 1.E-11 | **-1.04493479475724E-01** | 15 | 0.016 | **-1.04493479475724E-01** | 15 | 0.015 |
| 1.E-13 | **-1.04493479475724E-01** | 15 | 0.015 | **-1.04493479475724E-01** | 15 | 0.016 |
| 1.E-15 | **-1.04493479475724E-01** | 15 | 0.000 | **-1.04493479475724E-01** | 15 | 0.015 |
| 1.E-17 | **-1.04493479475724E-01** | 15 | 0.000 | **-1.04493479475724E-01** | 15 | 0.000 |
| 1.E-19 | **-1.04493479475724E-01** | 15 | 0.015 | **-1.04493479475724E-01** | 15 | 0.016 |
| 0.E+00 | **-1.04493479475724E-01** | 15 | 0.000 | **-1.04493479475724E-01** | 15 | 0.016 |

*CPU time (in seconds) spent in Mathematica® kernel

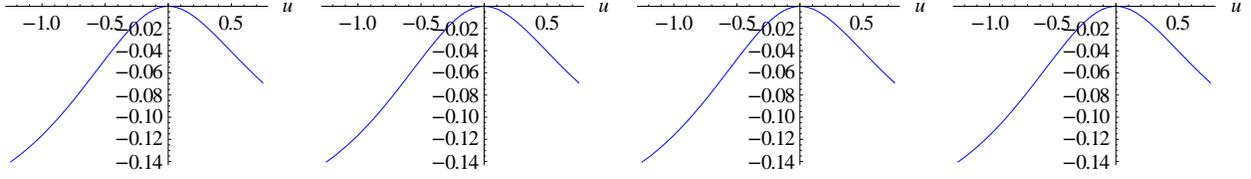

Figure 8.3. Graph of integrand in (8.8) for the second BQ of Appendix B, and the OP (0.25, 0.25, $h$). From left to right: $h = 0.1$, $10^{-5}$, $10^{-19}$, 0.0.

We proceed to evaluate (8.8) when $h$ is equal to zero. From (8.3)

$$D(u,0) = |\mathbf{p}'|^2 u^2, \quad E(u,h) = 2\left[\mathbf{p}' \cdot (\mathbf{q}' + \mathbf{r}_{pq}u)\right]u, \quad F(u) = |\mathbf{q}' + \mathbf{r}_{pq}u| \tag{8.10}$$

while, from (8.4),

$$\Delta(u,0) = 4F^2(u)D(u,0) - E^2(u,0)$$
$$= 4|\mathbf{q}' + \mathbf{r}_{pq}u|^2 |\mathbf{p}'|^2 u^2 - 4u^2 \left[\mathbf{p}' \cdot (\mathbf{q}' + \mathbf{r}_{pq}u)\right]^2$$
$$= 4u^2 \left\{|\mathbf{q}' + \mathbf{r}_{pq}u|^2 |\mathbf{p}'|^2 - \left[\mathbf{p}' \cdot (\mathbf{q}' + \mathbf{r}_{pq}u)\right]^2\right\}$$
$$= 4u^2 |\mathbf{p}' \times (\mathbf{q}' + \mathbf{r}_{pq}u)|^2. \tag{8.11}$$

Similarly,

$$\left[\Delta(u,0) - E^2(u,0)\right]v - 2D(u,0)E(u,0)$$
$$= \left[4u^2 |\mathbf{p}' \times (\mathbf{q}' + \mathbf{r}_{pq}u)|^2 - 4u^2\left[\mathbf{p}' \cdot (\mathbf{q}' + \mathbf{r}_{pq}u)\right]^2\right]v - 4|\mathbf{p}'|^2 u^3 \left[\mathbf{p}' \cdot (\mathbf{q}' + \mathbf{r}_{pq}u)\right]$$
$$= 4u^2 \left\{\left[|\mathbf{p}' \times (\mathbf{q}' + \mathbf{r}_{pq}u)|^2 - \left[\mathbf{p}' \cdot (\mathbf{q}' + \mathbf{r}_{pq}u)\right]^2\right]v - |\mathbf{p}'|^2 u\left[\mathbf{p}' \cdot (\mathbf{q}' + \mathbf{r}_{pq}u)\right]\right\}. \tag{8.12}$$

Substituting these results and (5.7) in (8.8), we have

$$I_{11}(p', q', 0) = -\int_{-1-p'}^{1-p'} u^2 du \left\{\frac{\left[|\mathbf{p}' \times (\mathbf{q}' + \mathbf{r}_{pq}u)|^2 - \left[\mathbf{p}' \cdot (\mathbf{q}' + \mathbf{r}_{pq}u)\right]^2\right](1+q') + |\mathbf{p}'|^2 u\left[\mathbf{p}' \cdot (\mathbf{q}' + \mathbf{r}_{pq}u)\right]}{|\mathbf{q}' + \mathbf{r}_{pq}u|^2 |\mathbf{p}' \times (\mathbf{q}' + \mathbf{r}_{pq}u)|^2 |\mathbf{p}'u - (\mathbf{q}' + \mathbf{r}_{pq}u)(1+q')|}\right.$$

$$+\frac{\left[\left|\mathbf{p}'\times(\mathbf{q}'+\mathbf{r}_{pq}u)\right|^2-\left[\mathbf{p}'\cdot(\mathbf{q}'+\mathbf{r}_{pq}u)\right]^2\right](1-q')-|\mathbf{p}'|^2 u\left[\mathbf{p}'\cdot(\mathbf{q}'+\mathbf{r}_{pq}u)\right]}{\left|\mathbf{q}'+\mathbf{r}_{pq}u\right|^2\left|\mathbf{p}'\times(\mathbf{q}'+\mathbf{r}_{pq}u)\right|^2\left|\mathbf{p}'u+(\mathbf{q}'+\mathbf{r}_{pq}u)(1-q')\right|}\Bigg\}. \tag{8.13}$$

If the OP is on the boundary, special formulas can be used as, for example,

$$I_{11}(p',1,0)=\int_{-1-p'}^{1-p'}u^2 du\Bigg\{\frac{2\left[\left|\mathbf{p}'\times(\mathbf{q}'+\mathbf{r}_{pq}u)\right|^2-\left[\mathbf{p}'\cdot(\mathbf{q}'+\mathbf{r}_{pq}u)\right]^2\right]+|\mathbf{p}'|^2 u\left[\mathbf{p}'\cdot(\mathbf{q}'+\mathbf{r}_{pq}u)\right]}{\left|\mathbf{q}'+\mathbf{r}_{pq}u\right|^2\left|\mathbf{p}'\times(\mathbf{q}'+\mathbf{r}_{pq}u)\right|^2\left|\mathbf{p}'u-2(\mathbf{q}'+\mathbf{r}_{pq})\right|}$$

$$-\frac{|\mathbf{p}'|u\left[\mathbf{p}'\cdot(\mathbf{q}'+\mathbf{r}_{pq}u)\right]}{\left|\mathbf{q}'+\mathbf{r}_{pq}u\right|^2\left|\mathbf{p}'\times(\mathbf{q}'+\mathbf{r}_{pq}u)\right|^2|u|}\Bigg\}. \tag{8.14}$$

When we compute (8.13) for the second BQ and the OP of Table 8.1, we get exactly the value that appears in the last row of that table.

We turn next to (8.9). From Formula 2.261, p. 81 of [11], we have

$$\lim_{v\to 0}\left[\int_{-1-q'}^{-v}\frac{dv}{R(u,v,h)}+\int_{v}^{1-q'}\frac{dv}{R(u,v,h)}\right]$$
$$=\frac{1}{F(u)}\ln\left|\frac{2F(u)R(u,1-q',h)+2F^2(u)(1-q')+E(u,h)}{2F(u)R(u,-(1+q'),h)-2F^2(u)(1+q')+E(u,h)}\right| \tag{8.15}$$

We substitute in (8.9)

$$I_{12}(p',q',h)=\lim_{u\to 0}\int_{-1-p'}^{-u}\frac{u^2 du}{F^3(u,h)}\ln\left|\frac{2F(u)R(u,1-q',h)+2F^2(u)(1-q')+E(u,h)}{2F(u)\left[R(u,-(1+q'),h)-F(u)(1+q')\right]+E(u,h)}\right|$$
$$+\lim_{u\to 0}\int_{u}^{1-p'}\frac{u^2 du}{F^3(u,h)}\ln\left|\frac{2F(u)\left[R(u,1-q',h)+F(u)(1-q')\right]+E(u,h)}{2F(u)\left[R(u,-(1+q'),h)-F(u)(1+q')\right]+E(u,h)}\right|. \tag{8.16}$$

As we have discussed in detail in Section 7 of [5], the term in the square brackets in the denominator of the logarithmic argument involves the subtraction of two terms that are almost identical when $h$ and $u$ are small. This leads to a corruption of the numerical results. For this reason, we rewrite the argument of the logarithm as follows

$$\frac{2F(u)R(u,1-q',h)+2F^2(u)(1-q')+E(u,h)}{2F(u)\left[R(u,-(1+q'),h)-F(u)(1+q')\right]+E(u,h)}$$

$$=\frac{2F(u)R(u,1-q',h)+2F^2(u)(1-q')+E(u,h)}{2F(u)\left[\dfrac{R^2(u,-(1+q'),h)-F^2(u)(1+q')^2}{R(u,-(1+q'),h)+F(u)(1+q')}\right]+E(u,h)}$$

$$=\frac{\left[2F(u)R(u,1-q',h)+2F^2(u)(1-q')+E(u,h)\right]\left[R(u,-(1+q'),h)+F(u)(1+q')\right]}{2F(u)\left[R^2(u,-(1+q'),h)-F^2(u)(1+q')^2\right]+E(u,h)\left[R(u,-(1+q'),h)+F(u)(1+q')\right]}$$

(8.17)

For $R^2$, we take into consideration (8.2) and (8.3) to re-write this as

$$\frac{2F(u)R(u,1-q',h)+2F^2(u)(1-q')+E(u,h)}{2F(u)\left[R(u,-(1+q'),h)-F(u)(1+q')\right]+E(u,h)}$$

$$=\frac{\left[2F(u)R(u,1-q',h)+2F^2(u)(1-q')+E(u,h)\right]\left[R(u,-(1+q'),h)+F(u)(1+q')\right]}{2F(u)\left[-(1+q')E(u,h)+D(u,h)\right]+E(u,h)\left[R(u,-(1+q'),h)+F(u)(1+q')\right]}$$

$$=\frac{\left\{2F(u)\left[R(u,1-q',h)+F(u)(1-q')\right]+E(u,h)\right\}\left[R(u,-(1+q'),h)+F(u)(1+q')\right]}{2F(u)D(u,h)+E(u,h)\left[R(u,-(1+q'),h)-F(u)(1+q')\right]}. \quad (8.18)$$

We substitute this result in (8.16)

$$I_{12}(p',q',h)$$
$$=\lim_{u\to 0}\int_{-1-p'}^{-u}\frac{u^2 du}{F^3(u,h)}\ln\left|\frac{\left\{2F(u)\left[R(u,1-q',h)+F(u)(1-q')\right]+E(u,h)\right\}\left[R(u,-(1+q'),h)+F(u)(1+q')\right]}{2F(u)D(u,h)+E(u,h)\left[R(u,-(1+q'),h)-F(u)(1+q')\right]}\right|$$
$$+\lim_{u\to 0}\int_{u}^{1-p'}\frac{u^2 du}{F^3(u,h)}\ln\left|\frac{\left\{2F(u)\left[R(u,1-q',h)+F(u)(1-q')\right]+E(u,h)\right\}\left[R(u,-(1+q'),h)+F(u)(1+q')\right]}{2F(u)D(u,h)+E(u,h)\left[R(u,-(1+q'),h)-F(u)(1+q')\right]}\right|$$

(8.19)

We have computed this expression for the same BQ and OP as in Table 8.1, and we present the results in Table 8.2. The same comments as for Table 8.1 hold here. We also present the graph of the integrand in Fig. 8.4.

Table 8.2. Integral in (8.19) evaluated for the second BQ using GKQ and DEQ ($p' = q' = 0.25$).

| | SECOND BQ, INTEGRAL IN (8.19) | | | | | |
|---|---|---|---|---|---|---|
| h | GKQ | GKQ SD | GKQ TIME* | DEQ | DEQ SD | DEQ TIME* |
| 1 | **1.22541444754356E-01** | 15 | 0.016 | **1.22541444754356E-01** | 15 | 0.016 |
| 1.E-01 | **1.33304710210483E-01** | 15 | 0.016 | **1.38555779969812E-01** | 15 | 0.015 |
| 1.E-03 | **1.33367634066845E-01** | 15 | 0.015 | **1.33367634066844E-01** | 15 | 0.016 |
| 1.E-05 | **1.33366468479235E-01** | 15 | 0.016 | **1.33366468479235E-01** | 15 | 0.016 |
| 1.E-07 | **1.33366456634967E-01** | 15 | 0.015 | **1.33366456634967E-01** | 15 | 0.016 |
| 1.E-09 | **1.33366456516505E-01** | 15 | 0.000 | **1.33366456516505E-01** | 15 | 0.000 |
| 1.E-11 | **1.33366456515321E-01** | 15 | 0.000 | **1.33366456515321E-01** | 15 | 0.000 |
| 1.E-13 | **1.33366456515309E-01** | 15 | 0.016 | **1.33366456515309E-01** | 15 | 0.000 |
| 1.E-15 | **1.33366456515309E-01** | 15 | 0.016 | **1.33366456515309E-01** | 15 | 0.015 |
| 1.E-17 | 1.33366456515309E-01 | 15 | 0.015 | 1.33366456515309E-01 | 15 | 0.000 |
| 1.E-19 | **1.33366456515309E-01** | 15 | 0.016 | **1.33366456515309E-01** | 15 | 0.016 |
| 0.E+00 | **1.33366456515309E-01** | 15 | 0.031 | **1.33366456515309E-01** | 15 | 0.015 |

*CPU time (in seconds) spent in Mathematica® kernel

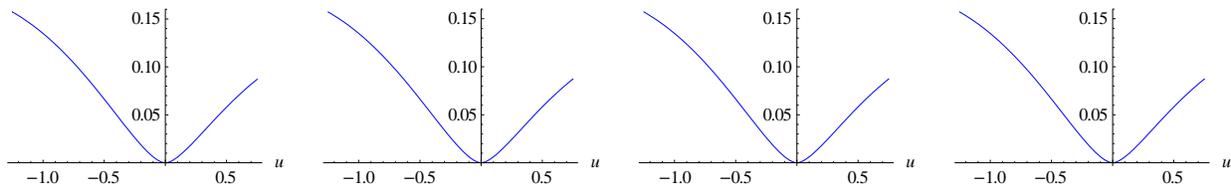

Figure 8.4. Graph of integrand in (8.19) for the second BQ of Appendix B, and the OP (0.25, 0.25, h). From left to right: $h = 0.1$, $10^{-5}$, $10^{-19}$, 0.0.

# 9. EVALUATION OF INTEGRALS IN $R^{-1}$

In this section, we return to (2.44) and (2.45) and consider the integrals that contain the factor $R^{-1}$ in their integrands. The integrals in (2.44), as principal-value integrals, are equal to zero when $h$ is equal to zero. The first integral in (2.44) has been extensively studied in [5]. The second integral is

$$I_1 = h \int_{-(1+p')}^{1-p'} u\,du \int_{-(1+q')}^{1-q'} \frac{dv}{R(u,v,h)}. \tag{9.1}$$

We display the graph of the integrand (multiplied by $h$) for the second BQ of Appendix B and for the OP (0.25, 0.25, $h$).

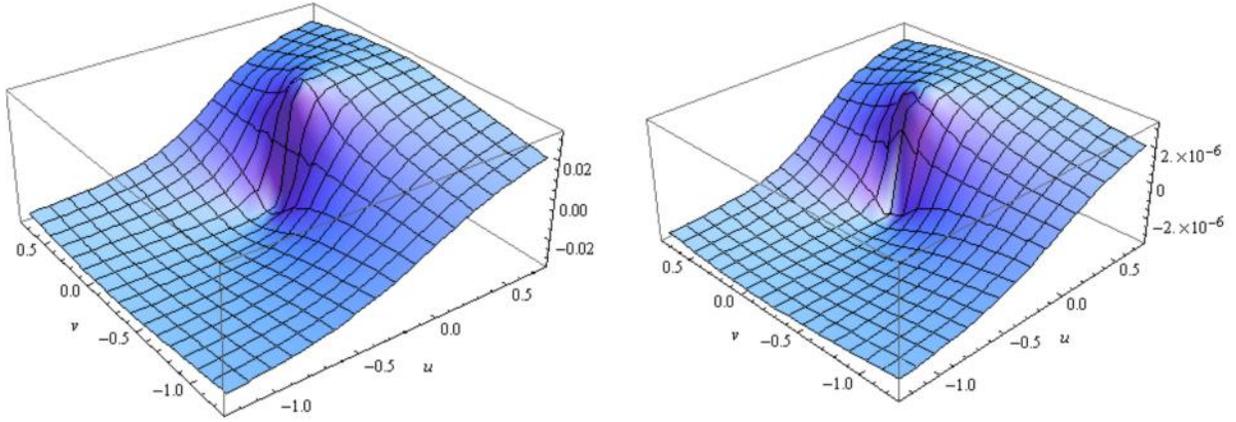

Figure 9.1. Integrand of (9.1) multiplies by $h$ for the second BQ of Appendix B and for the OP (0.25, 0.25, $h$). Left: $h = 0.1$. Right: $h = 10^{-5}$.

From (8.15)

$$\int_{-(1+q')}^{1-q'} \frac{dv}{R(u,v,h)} = \frac{1}{F(u)} \ln\left|\frac{2F(u)R(u,1-q',h)+2F^2(u)(1-q')+E(u,h)}{2F(u)R(u,-(1+q'),h)-2F^2(u)(1+q')+E(u,h)}\right| \tag{9.2}$$

so that

$$I_1 = h \int_{-(1+p')}^{1-p'} \frac{u\,du}{F(u)} \ln\left|\frac{2F(u)R(u,1-q',h)+2F^2(u)(1-q')+E(u,h)}{2F(u)[R(u,-(1+q'),h)-F(u)(1+q')]+E(u,h)}\right|. \tag{9.3}$$

As we did in Section 8 with $I_{12}$, we re-write the integrand so as to stabilize its computation for small values of $u$ and $h$. As with (8.19), we write

$$\begin{aligned}&I_1(p',q',h)\\ &= h \int_{-1-p'}^{1-p'} \frac{u\,du}{F(u)} \ln\left|\frac{\{2F(u)[R(u,1-q',h)+F(u)(1-q')]+E(u,h)\}[R(u,-(1+q'),h)+F(u)(1+q')]}{2F(u)D(u,h)+E(u,h)[R(u,-(1+q'),h)-F(u)(1+q')]}\right|\end{aligned} \tag{9.4}$$

The graph of the integrand (multiplied by $h$) has the form shown in Fig. 9.2.

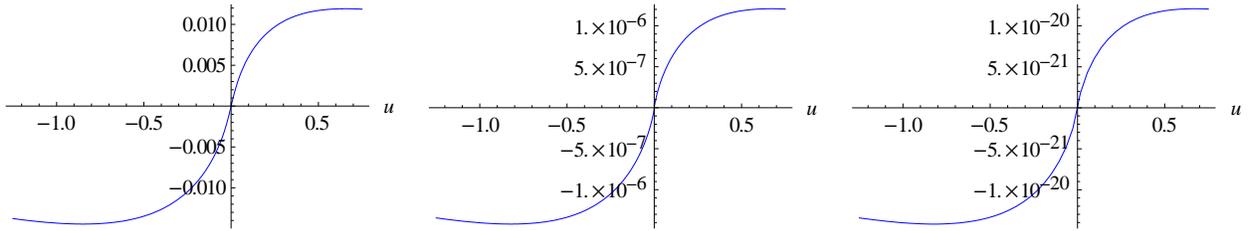

Figure 9.2. Integrand of (9.4), multiplied by $h$, for the same BQ and OP as in Fig. 9.1. Left: $h = 0.1$. Center: $h = 10^{-5}$. Right: $h = 10^{-19}$.

In Table 9.1, we display the calculations we performed using (9.4) on the second BQ of Appendix B and the OP ($p' = 0.25$, $q' = 0.25$, $h$). The results of the two quadratures agree to at least 14 SD. The execution times are also very comparable.

The third integral in (2.44) is of the same type as the integral we analyzed above and, hence, we omit it. The fourth integral in (2.44) is identical to the first integral in each of (2.45) except for the multiplicative factor $h$. We will consider the integral without this factor:

$$I_2 = \int_{-(1+p')}^{1-p'} u\,du \int_{-(1+q')}^{1-q'} \frac{v\,dv}{R(u,v,h)}. \tag{9.5}$$

The limit of the integrand at the origin exists for any value of $h$ and it is zero. From [11], p. 83, Formula 2.264.2

$$\int_{-(1+q')}^{1-q'} \frac{v\,dv}{R(u,v,h)} = \frac{R(u,1-q',h)-R(u,-(1+q'),h)}{F^2(u)} - \frac{E(u,h)}{2F^2(u)} \int_{-(1+q')}^{1-q'} \frac{dv}{R(u,v,h)} \tag{9.6}$$

so that we can write for (9.5) that

$$I_2 = \int_{-(1+p')}^{1-p'} \frac{u\,du}{F^2(u)}\left[R(u,1-q',h) - R(u,-(1+q'),h)\right] - \frac{1}{2}\int_{-(1+p')}^{1-p'} \frac{E(u,h)u\,du}{F^2(u)} \int_{-(1+q')}^{1-q'} \frac{dv}{R(u,v,h)}. \quad (9.7)$$

Table 9.1. Integral in (9.4) evaluated for the second BQ using GKQ and DEQ ($p' = q' = 0.25$).

| | SECOND BQ, EQ. (9.4) | | | | | |
|---|---|---|---|---|---|---|
| h | GKQ | GKQ SD | GKQ TIME* | DEQ | DEQ SD | DEQ TIME* |
| 1 | -9.76586089041958E-02 | 15 | 0.016 | -9.76586089041957E-02 | 15 | 0.016 |
| 1.E-01 | -8.06206633518322E-03 | 15 | 0.000 | -8.06206633518322E-03 | 15 | 0.016 |
| 1.E-03 | -7.79300181107822E-05 | 15 | 0.031 | -7.79300181107821E-05 | 15 | 0.031 |
| 1.E-05 | -7.79021513103433E-07 | 15 | 0.031 | -7.79021513103432E-07 | 15 | 0.016 |
| 1.E-07 | -7.79018725447172E-09 | 15 | 0.016 | -7.79018725447171E-09 | 15 | 0.016 |
| 1.E-09 | -7.79018697570512E-11 | 15 | 0.015 | -7.79018697570511E-11 | 15 | 0.015 |
| 1.E-11 | -7.79018697291745E-13 | 15 | 0.016 | -7.79018697291744E-13 | 15 | 0.000 |
| 1.E-13 | -7.79018697288957E-15 | 15 | 0.016 | -7.79018697288956E-15 | 15 | 0.015 |
| 1.E-15 | -7.79018697288930E-17 | 15 | 0.015 | -7.79018697288929E-17 | 15 | 0.016 |
| 1.E-17 | -7.79018697288929E-19 | 15 | 0.016 | -7.79018697288928E-19 | 15 | 0.016 |
| 1.E-19 | -7.79018697288929E-21 | 15 | 0.015 | -7.79018697288929E-21 | 15 | 0.015 |

*CPU time (in seconds) spent in Mathematica® kernel

The first integral on the right, namely,

$$I_{21} = \int_{-(1+p')}^{1-p'} \frac{u\,du}{F^2(u)}\left[R(u,1-q',h) - R(u,-(1+q'),h)\right] \quad (9.8)$$

is a regular integral. We display the integrand in Fig. 9.3. We have computed this integral for the same conditions as for Table 9.1 and we encountered no difficulties. We display the results in Table 9.2. There is agreement between the two quadratures to at least 14 SD. We might wish to further stabilize the integrand by multiplying and dividing by the conjugate of the factor in brackets but we do not believe that this is necessary.

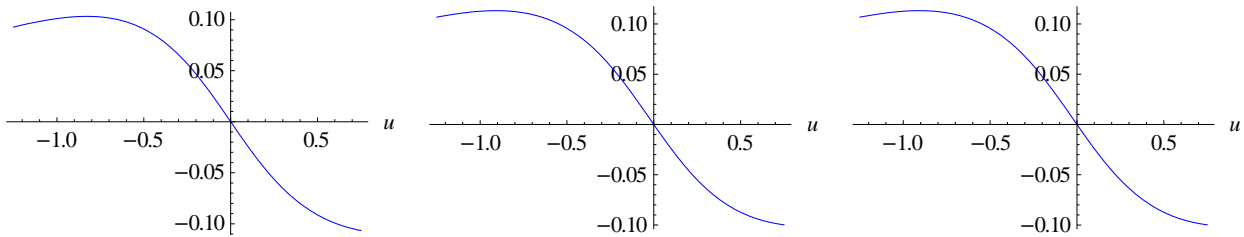

Figure 9.3. Integrand of (9.8) for the same BQ and OP as in Fig. 9.1. Left: $h = 0.1$. Center: $h = 10^{-5}$. Right: $h = 10^{-19}$.

Table 9.2. Integral in (9.8) evaluated for the second BQ using GKQ and DEQ ($p' = q' = 0.25$).

| | SECOND BQ, INTEGRAL IN (9.8) | | | | | |
|---|---|---|---|---|---|---|
| $h$ | GKQ | GKQ SD | GKQ TIME* | DEQ | DEQ SD | DEQ TIME* |
| 1 | -4.19172843647591E-02 | 15 | 0.000 | -4.19172843647591E-02 | 15 | 0.015 |
| 1.E-01 | 4.95995036294927E-02 | 15 | 0.015 | 4.95995036294927E-02 | 15 | 0.000 |
| 1.E-03 | 5.98914737934455E-02 | 15 | 0.000 | 5.98914737934454E-02 | 15 | 0.000 |
| 1.E-05 | 5.99941196996292E-02 | 15 | 0.000 | 5.99941196996291E-02 | 15 | 0.016 |
| 1.E-07 | 5.99951461260395E-02 | 15 | 0.000 | 5.99951461260395E-02 | 15 | 0.000 |
| 1.E-09 | 5.99951563903003E-02 | 15 | 0.000 | 5.99951563903003E-02 | 15 | 0.000 |
| 1.E-11 | 5.99951564929430E-02 | 15 | 0.016 | 5.99951564929429E-02 | 15 | 0.000 |
| 1.E-13 | 5.99951564939694E-02 | 15 | 0.000 | 5.99951564939694E-02 | 15 | 0.016 |
| 1.E-15 | 5.99951564939797E-02 | 15 | 0.000 | 5.99951564939796E-02 | 15 | 0.015 |
| 1.E-17 | 5.99951564939798E-02 | 15 | 0.000 | 5.99951564939797E-02 | 15 | 0.000 |
| 1.E-19 | 5.99951564939798E-02 | 15 | 0.000 | 5.99951564939797E-02 | 15 | 0.000 |
| 0.E+00 | 5.99951564939798E-02 | 15 | 0.016 | 5.99951564939797E-02 | 15 | 0.016 |

*CPU time (in seconds) spent in Mathematica® kernel

We can evaluate the second integral in (9.7), namely,

$$I_{22} = \int_{-(1+p')}^{1-p'} \frac{E(u,h)u\,du}{F^2(u)} \int_{-(1+q')}^{1-q'} \frac{dv}{R(u,v,h)} \tag{9.9}$$

by taking advantage of (9.2)

$$I_{22} = \int_{-(1+p')}^{1-p'} \left\{ \frac{E(u,h)u\,du}{F^3(u)} \right.$$
$$\left. \cdot \ln\left| \frac{\{2F(u)[R(u,1-q',h)+F(u)(1-q')]+E(u,h)\}[R(u,-(1+q'),h)+F(u)(1+q')]}{2F(u)D(u,h)+E(u,h)[R(u,-(1+q'),h)-F(u)(1+q')]} \right| \right\} \tag{9.10}$$

In Fig. 9.4 we present the graph of the integrand while, in Table 9.3, we calculate the integral for the same conditions as in Table 9.2. The observations are the same as for the previous two tables.

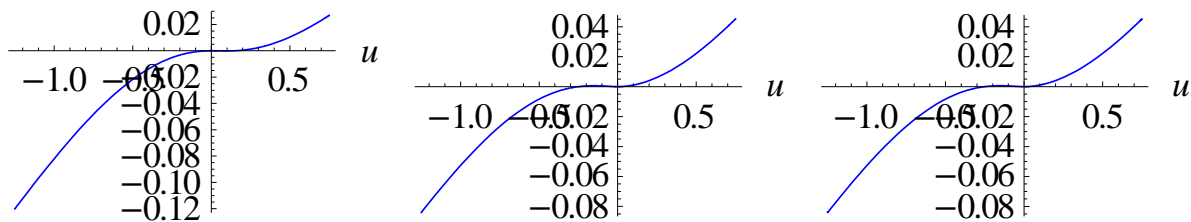

Figure 9.4. Integrand of (9.10) for the same BQ and OP as in Fig. 9.1. Left: $h = 0.1$. Center: $h = 10^{-5}$. Right: $h = 10^{-19}$.

Table 9.3. Integral in (9.10) evaluated for the second BQ using GKQ and DEQ ($p' = q' = 0.25$).

| | SECOND BQ, INTEGRAL IN (9.10) | | | | | |
|---|---|---|---|---|---|---|
| h | GKQ | GKQ SD | GKQ TIME* | DEQ | DEQ SD | DEQ TIME* |
| 1 | **-2.96427329188867E-01** | 15 | 0.015 | **-2.96427329188867E-01** | 15 | 0.016 |
| 1.E-01 | **-4.76427586157066E-02** | 15 | 0.000 | **-4.76427586157066E-02** | 15 | 0.016 |
| 1.E-03 | **-1.97396132066286E-02** | 15 | 0.016 | **-1.97396132066286E-02** | 15 | 0.015 |
| 1.E-05 | **-1.94631735454601E-02** | 15 | 0.015 | **-1.94631735454601E-02** | 15 | 0.016 |
| 1.E-07 | **-1.94604094544073E-02** | 15 | 0.016 | **-1.94604094544073E-02** | 15 | 0.015 |
| 1.E-09 | **-1.94603818135274E-02** | 15 | 0.000 | **-1.94603818135273E-02** | 15 | 0.016 |
| 1.E-11 | **-1.94603815371186E-02** | 15 | 0.000 | **-1.94603815371185E-02** | 15 | 0.015 |
| 1.E-13 | **-1.94603815343545E-02** | 15 | 0.015 | **-1.94603815343545E-02** | 15 | 0.016 |
| 1.E-15 | **-1.94603815343268E-02** | 15 | 0.016 | **-1.94603815343268E-02** | 15 | 0.016 |
| 1.E-17 | **-1.94603815343266E-02** | 15 | 0.015 | **-1.94603815343265E-02** | 15 | 0.015 |
| 1.E-19 | **-1.94603815343266E-02** | 15 | 0.016 | **-1.94603815343265E-02** | 15 | 0.016 |
| 0.E+00 | **-1.94603815343266E-02** | 15 | 0.016 | **-1.94603815343265E-02** | 15 | 0.015 |

*CPU time (in seconds) spent in Mathematica® kernel

The next integral in (2.45) involves the second power of one of the variables and the zeroth power of the other. We consider the integral

$$I_3 = \int_{-(1+p')}^{1-p'} u^2 du \int_{-(1+q')}^{1-q'} \frac{dv}{R(u,v,h)}. \qquad (9.11)$$

Using (9.2), after stabilization, we have

$$I_3 = \int_{-(1+p')}^{1-p'} \frac{u^2 du}{F(u)} \ln\left|\frac{\{2F(u)[R(u,1-q',h)+F(u)(1-q')]+E(u,h)\}[R(u,-(1+q'),h)+F(u)(1+q')]}{2F(u)D(u,h)+E(u,h)[R(u,-(1+q'),h)-F(u)(1+q')]}\right| \qquad (9.12)$$

We present the results of the computation of the integrand and integral (under the same conditions as for the rest of this section) in Fig. 9.5 and Table 9.4, respectively. The two quadratures agree to at least 14 SD and GKQ appears to have the time advantage.

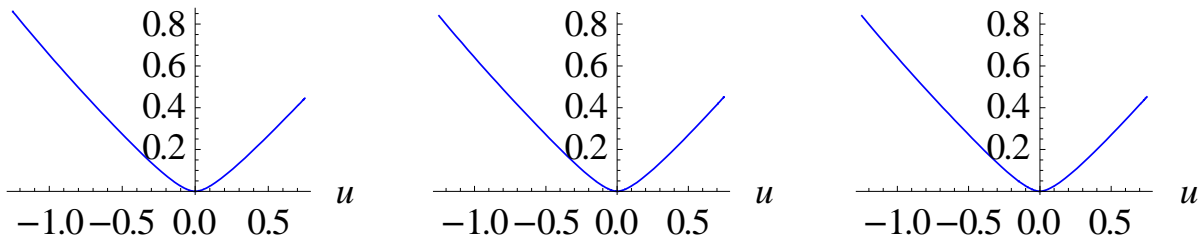

Figure 9.5. Integrand of (9.12) for the same BQ and OP as in Fig. 9.1. Left: $h = 0.1$. Center: $h = 10^{-5}$. Right: $h = 10^{-19}$.

Table 9.4. Integral in (9.12) evaluated for the second BQ using GKQ and DEQ ($p' = q' = 0.25$).

| | SECOND BQ, INTEGRAL IN (9.12) | | | | | |
|---|---|---|---|---|---|---|
| h | GKQ | GKQ SD | GKQ TIME* | DEQ | DEQ SD | DEQ TIME* |
| 1 | **6.13973701829767E-01** | 15 | 0.000 | 6.13973701829767E-01 | 15 | 0.031 |
| 1.E-01 | **6.20702546374996E-01** | 15 | 0.016 | 6.20702546374995E-01 | 15 | 0.031 |
| 1.E-03 | **6.14599145062923E-01** | 15 | 0.016 | 6.14599145062922E-01 | 15 | 0.031 |
| 1.E-05 | **6.14528139539690E-01** | 15 | 0.015 | 6.14528139539690E-01 | 15 | 0.031 |
| 1.E-07 | **6.14527428447519E-01** | 15 | 0.031 | 6.14527428447519E-01 | 15 | 0.015 |
| 1.E-09 | **6.14527421336494E-01** | 15 | 0.016 | 6.14527421336493E-01 | 15 | 0.016 |
| 1.E-11 | **6.14527421265384E-01** | 15 | 0.016 | 6.14527421265383E-01 | 15 | 0.016 |
| 1.E-13 | **6.14527421264673E-01** | 15 | 0.015 | 6.14527421264672E-01 | 15 | 0.015 |
| 1.E-15 | **6.14527421264666E-01** | 15 | 0.016 | 6.14527421264665E-01 | 15 | 0.016 |
| 1.E-17 | **6.14527421264666E-01** | 15 | 0.015 | 6.14527421264665E-01 | 15 | 0.016 |
| 1.E-19 | **6.14527421264666E-01** | 15 | 0.016 | 6.14527421264665E-01 | 15 | 0.016 |
| 0.E+00 | **6.14527421264666E-01** | 15 | 0.031 | 6.14527421264665E-01 | 15 | 0.016 |

*CPU time (in seconds) spent in Mathematica® kernel

The next integral in (2.45) is the second integral of either expression in (2.45). In order to take advantage of the results above, we choose the one in the second expression, namely

$$I_4 = \int_{-(1+p')}^{1-p'} u^2 du \int_{-(1+q')}^{1-q'} \frac{vdv}{R(u,v,h)}. \qquad (9.13)$$

This integral is similar to the integral in (9.5) withe exception that $u$ in the numerator appears to the second power rather than the first. We can take advantage of the result in (9.7) to write

$$I_4 = \int_{-(1+p')}^{1-p'} \frac{u^2 du}{F^2(u)} \Big[ R(u, 1-q', h) - R(u, -(1+q'), h) \Big] - \frac{1}{2} \int_{-(1+p')}^{1-p'} \frac{E(u,h)u^2 du}{F^2(u)} \int_{-(1+q')}^{1-q'} \frac{dv}{R(u,v,h)}. \qquad (9.14)$$

Following the development in (9.7), we write

$$I_{41} = \int_{-(1+p')}^{1-p'} \frac{u^2 du}{F^2(u)} \Big[ R(u, 1-q', h) - R(u, -(1+q'), h) \Big]. \qquad (9.15)$$

We present the results of the computation of the integrand and integral (under the same conditions as for the rest of this section) in Fig. 9.6 and Table 9.5, respectively. The two quadratures agree to at least 14 SD and, on average, they are equally fast. Despite appearances in Fig. 9.7, the origin is an inflection point, as we see in Fig. 9.8, where we have zoomed near the origin for $h = 10^{-19}$.

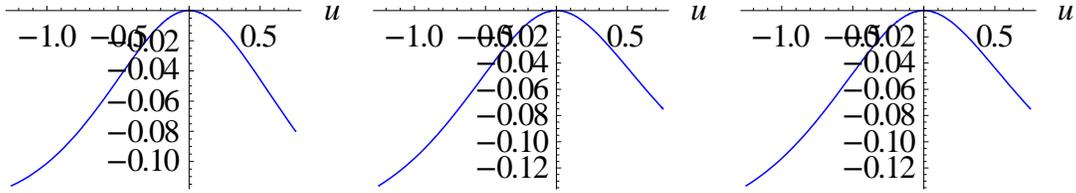

Figure 9.6. Integrand of (9.15) for the same BQ and OP as in Fig. 9.1. Left: $h = 0.1$. Center: $h = 10^{-5}$. Right: $h = 10^{-19}$.

Table 9.5. Integral in (9.15) evaluated for the second BQ using GKQ and DEQ ($p' = q' = 0.25$).

| | SECOND BQ, INTEGRAL IN (9.15) | | | | | |
|---|---|---|---|---|---|---|
| h | GKQ | GKQ SD | GKQ TIME* | DEQ | DEQ SD | DEQ TIME* |
| 1 | **-3.98493668638121E-02** | 15 | 0.015 | **-3.98493668638121E-02** | 15 | 0.000 |
| 1.E-01 | **-9.79517145971403E-02** | 15 | 0.000 | **-9.79517145971402E-02** | 15 | 0.015 |
| 1.E-03 | **-1.03987331842234E-01** | 15 | 0.016 | **-1.03987331842234E-01** | 15 | 0.016 |
| 1.E-05 | **-1.04047084513621E-01** | 15 | 0.000 | **-1.04047084513621E-01** | 15 | 0.016 |
| 1.E-07 | **-1.04047681977874E-01** | 15 | 0.000 | **-1.04047681977874E-01** | 15 | 0.016 |
| 1.E-09 | **-1.04047687952511E-01** | 15 | 0.000 | **-1.04047687952511E-01** | 15 | 0.016 |
| 1.E-11 | **-1.04047688012257E-01** | 15 | 0.015 | **-1.04047688012257E-01** | 15 | 0.000 |
| 1.E-13 | **-1.04047688012855E-01** | 15 | 0.000 | **-1.04047688012854E-01** | 15 | 0.000 |
| 1.E-15 | **-1.04047688012861E-01** | 15 | 0.000 | **-1.04047688012860E-01** | 15 | 0.000 |
| 1.E-17 | **-1.04047688012861E-01** | 15 | 0.000 | **-1.04047688012860E-01** | 15 | 0.000 |
| 1.E-19 | **-1.04047688012861E-01** | 15 | 0.016 | **-1.04047688012860E-01** | 15 | 0.000 |
| 0.E+00 | **-1.04047688012861E-01** | 15 | 0.016 | **-1.04047688012860E-01** | 15 | 0.000 |

*CPU time (in seconds) spent in Mathematica® kernel

Following (9.10), we write

$$I_{42} = \int_{-(1+p')}^{1-p'} \left\{ \frac{E(u,h)u^2 du}{F^3(u)} \right.$$
$$\left. \cdot \ln \left| \frac{\{2F(u)[R(u, 1-q', h) + F(u)(1-q')] + E(u,h)\}[R(u, -(1+q'), h) + F(u)(1+q')]}{2F(u)D(u,h) + E(u,h)[R(u, -(1+q'), h) - F(u)(1+q')]} \right| \right\}$$

(9.16)

We present the results of the computation of the integrand and integral (under the same conditions as for the rest of this section) in Fig. 9.7 and Table 9.6, respectively. The two quadratures agree to at least 14 SD and, on average, they are equally fast.

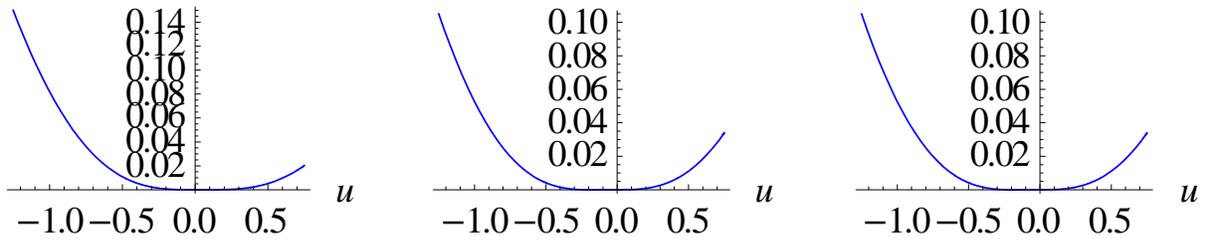

Figure 9.7. Integrand of (9.16) for the same BQ and OP as in Fig. 9.1. Left: $h = 0.1$. Center: $h = 10^{-5}$. Right: $h = 10^{-19}$.

Table 9.6. Integral in (9.16) evaluated for the second BQ using GKQ and DEQ ($p' = q' = 0.25$).

| | SECOND BQ, INTEGRAL IN (9.16) | | | | | |
|---|---|---|---|---|---|---|
| h | GKQ | GKQ SD | GKQ TIME* | DEQ | DEQ SD | DEQ TIME* |
| 1 | **2.10760323850730E-01** | 15 | 0.000 | **2.10760323850730E-01** | 15 | 0.015 |
| 1.E-01 | **5.31562115065914E-02** | 15 | 0.016 | **5.31562115065914E-02** | 15 | 0.016 |
| 1.E-03 | **3.78984658088151E-02** | 15 | 0.016 | **3.78984658088150E-02** | 15 | 0.016 |
| 1.E-05 | **3.77491975214039E-02** | 15 | 0.015 | **3.77491975214038E-02** | 15 | 0.015 |
| 1.E-07 | **3.77477051834983E-02** | 15 | 0.016 | **3.77477051834982E-02** | 15 | 0.016 |
| 1.E-09 | **3.77476902601537E-02** | 15 | 0.000 | **3.77476902601537E-02** | 15 | 0.015 |
| 1.E-11 | **3.77476901109203E-02** | 15 | 0.000 | **3.77476901109202E-02** | 15 | 0.016 |
| 1.E-13 | **3.77476901094279E-02** | 15 | 0.015 | **3.77476901094279E-02** | 15 | 0.016 |
| 1.E-15 | **3.77476901094130E-02** | 15 | 0.016 | **3.77476901094130E-02** | 15 | 0.000 |
| 1.E-17 | **3.77476901094129E-02** | 15 | 0.016 | **3.77476901094128E-02** | 15 | 0.015 |
| 1.E-19 | **3.77476901094129E-02** | 15 | 0.015 | **3.77476901094128E-02** | 15 | 0.000 |
| 0.E+00 | **3.77476901094129E-02** | 15 | 0.015 | **3.77476901094128E-02** | 15 | 0.015 |

*CPU time (in seconds) spent in Mathematica® kernel

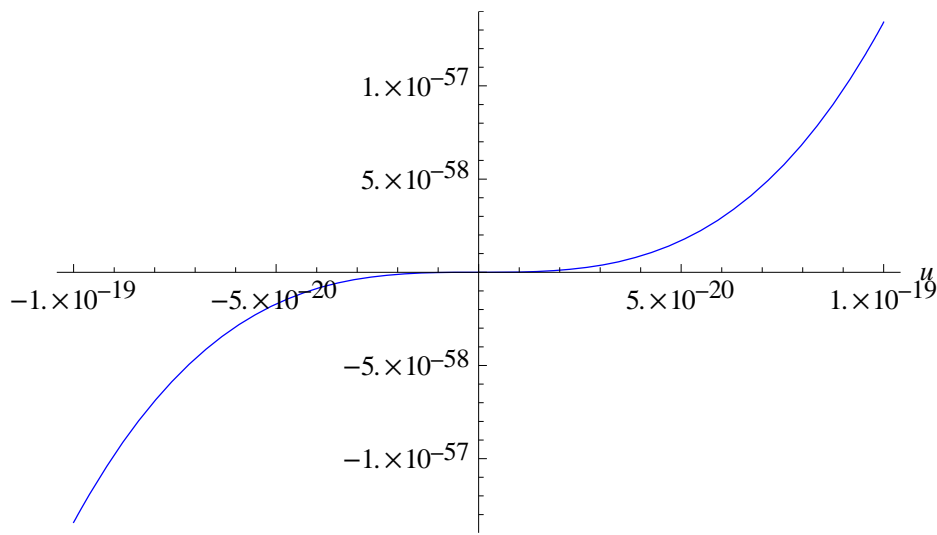

Figure 9.8. Graph on the right in Fig. 9.7 magnified near the origin.

The last integral in (2.45) is

$$I_5 = \int_{-(1+p')}^{1-p'} u^2 du \int_{-(1+q')}^{1-q'} \frac{v^2 dv}{R(u,v,h)}. \tag{9.17}$$

From [11], p. 83, Formula 2.264.3, we have

$$\int_{-(1+q')}^{1-q'} \frac{v^2 dv}{R(u,v,h)} = \frac{1}{2F^2(u)} \left[ v - \frac{3E(u,h)}{2F^2(u)} \right] R(u,v,h) \Big|_{-(1+q')}^{1-q'}$$

$$+ \frac{1}{2F^2(u)} \left[ \frac{3E^2(u,h)}{4F^2(u)} - D(u,h) \right] \int_{-(1+q')}^{1-q'} \frac{dv}{R(u,v,h)}$$

$$= \frac{1}{2F^2(u)} \left\{ \left[ 1 - q' - \frac{3E(u,h)}{2F^2(u)} \right] R(u, 1-q', h) + \left[ 1 + q' + \frac{3E(u,h)}{2F^2(u)} \right] R(u, -(1+q'), h) \right\}$$

$$+ \frac{1}{2F^2(u)} \left[ \frac{3E^2(u,h)}{4F^2(u)} - D(u,h) \right] \int_{-(1+q')}^{1-q'} \frac{dv}{R(u,v,h)}. \tag{9.18}$$

The remaining integral here can be handled as in (9.2). We thus have

$$I_5 = \frac{1}{2} \int_{-(1+p')}^{1-p'} \frac{u^2 du}{F^2(u)} \left\{ \left[ 1 - q' - \frac{3E(u,h)}{2F^2(u)} \right] R(u, 1-q', h) + \left[ 1 + q' + \frac{3E(u,h)}{2F^2(u)} \right] R(u, -(1+q'), h) \right\}$$

$$+ \frac{1}{2} \int_{-(1+p')}^{1-p'} \left\{ \frac{u^2 du}{F^3(u)} \left[ \frac{3E^2(u,h)}{4F^2(u)} - D(u,h) \right] \right.$$

$$\left. \cdot \ln \left| \frac{\{2F(u)[R(u, 1-q', h) + F(u)(1-q')] + E(u,h)\}[R(u, -(1+q'), h) + F(u)(1+q')]}{2F(u)D(u,h) + E(u,h)[R(u, -(1+q'), h) - F(u)(1+q')]} \right| \right\}$$

$$= \frac{1}{2} (I_{51} + I_{52}). \tag{9.19}$$

We have computed the integral

$$I_{51} = \int_{-(1+p')}^{1-p'} \frac{u^2 du}{F^2(u)} \left\{ \left[ 1 - q' - \frac{3E(u,h)}{2F^2(u)} \right] R(u, 1-q', h) + \left[ 1 + q' + \frac{3E(u,h)}{2F^2(u)} \right] R(u, -(1+q'), h) \right\}$$

$$\tag{9.20}$$

for the same conditions as above and we display the results in Fig. 9.9 and Table 9.7. We note that the two quadratures are in excellent agreement, that the influence of $h$ diminishes quickly with it, and that GKQ appears to run faster than DEQ.

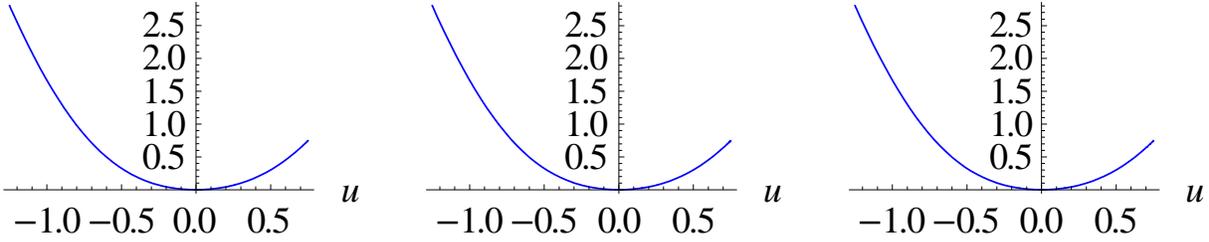

Figure 9.9. Integrand of (9.20) for the same BQ and OP as in Fig. 9.1. Left: $h = 0.1$. Center: $h = 10^{-5}$. Right: $h = 10^{-19}$.

Table 9.7. Integral in (9.20) evaluated for the second BQ using GKQ and DEQ ($p' = q' = 0.25$).

| | SECOND BQ, INTEGRAL IN (9.20) | | | | | |
|---|---|---|---|---|---|---|
| h | GKQ | GKQ SD | GKQ TIME* | DEQ | DEQ SD | DEQ TIME* |
| 1 | **1.22843441275541E+00** | 15 | 0.000 | **1.22843441275540E+00** | 15 | 0.000 |
| 1.E-01 | **1.22418844386084E+00** | 15 | 0.000 | **1.22418844386084E+00** | 15 | 0.015 |
| 1.E-03 | **1.22372160123087E+00** | 15 | 0.000 | **1.22372160123087E+00** | 15 | 0.016 |
| 1.E-05 | **1.22371695057202E+00** | 15 | 0.015 | **1.22371695057202E+00** | 15 | 0.015 |
| 1.E-07 | **1.22371690406827E+00** | 15 | 0.000 | **1.22371690406827E+00** | 15 | 0.000 |
| 1.E-09 | **1.22371690360323E+00** | 15 | 0.016 | **1.22371690360323E+00** | 15 | 0.000 |
| 1.E-11 | **1.22371690359858E+00** | 15 | 0.000 | **1.22371690359858E+00** | 15 | 0.015 |
| 1.E-13 | **1.22371690359854E+00** | 15 | 0.000 | **1.22371690359854E+00** | 15 | 0.016 |
| 1.E-15 | **1.22371690359854E+00** | 15 | 0.016 | **1.22371690359854E+00** | 15 | 0.015 |
| 1.E-17 | **1.22371690359854E+00** | 15 | 0.000 | **1.22371690359854E+00** | 15 | 0.016 |
| 1.E-19 | **1.22371690359854E+00** | 15 | 0.015 | **1.22371690359854E+00** | 15 | 0.000 |
| 0.E+00 | **1.22371690359854E+00** | 15 | 0.000 | **1.22371690359854E+00** | 15 | 0.000 |

*CPU time (in seconds) spent in Mathematica® kernel

We have also computed the second integral in (9.19), namely,

$$I_{52} = \int_{-(1+p')}^{1-p'} \left\{ \frac{u^2 du}{F^3(u)} \left[ \frac{3E^2(u,h)}{4F^2(u)} - D(u,h) \right] \right.$$
$$\left. \cdot \ln \left| \frac{\{2F(u)[R(u,1-q',h) + F(u)(1-q')] + E(u,h)\}[R(u,-(1+q'),h) + F(u)(1+q')]}{2F(u)D(u,h) + E(u,h)[R(u,-(1+q'),h) - F(u)(1+q')]} \right| \right\}$$
(9.21)

for the same conditions as previously and we exhibit the results in Table 9.8. We note that the two quadratures agree to 14 SD and that DEQ runs a little faster than GKQ. Since we have a removable singularity at the origin, we expect DEQ to perform better than DEQ. With this integral, we have concluded all the evaluations of the integral in (2.44) and (2.45) that have a $1 / R$ dependence.

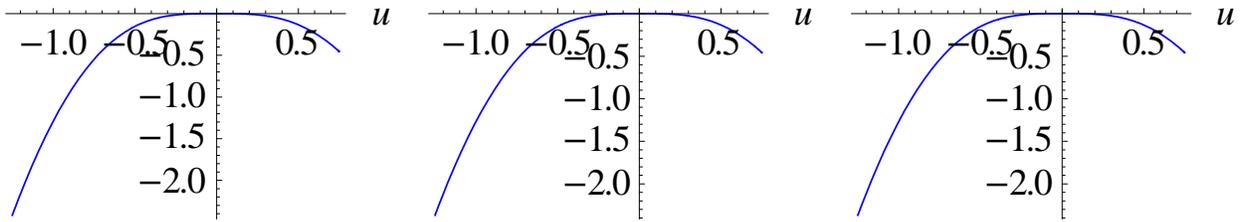

Figure 9.10. Integrand of (9.21) for the same BQ and OP as in Fig. 9.1. Left: $h = 0.1$. Center: $h = 10^{-5}$. Right: $h = 10^{-19}$.

Table 9.8. Integral in (9.21) evaluated for the second BQ using GKQ and DEQ ($p' = q' = 0.25$).

| h | GKQ | GKQ SD | GKQ TIME* | DEQ | DEQ SD | DEQ TIME* |
|---|---|---|---|---|---|---|
| 1 | **-9.70752066455716E-01** | 15 | 0.015 | **-9.70752066455715E-01** | 15 | 0.015 |
| 1.E-01 | **-8.60994565460606E-01** | 15 | 0.016 | **-8.60994565460605E-01** | 15 | 0.016 |
| 1.E-03 | **-8.47898333940654E-01** | 15 | 0.015 | **-8.47898333940653E-01** | 15 | 0.016 |
| 1.E-05 | **-8.47765615071529E-01** | 15 | 0.016 | **-8.47765615071528E-01** | 15 | 0.000 |
| 1.E-07 | **-8.47764287694453E-01** | 15 | 0.016 | **-8.47764287694452E-01** | 15 | 0.016 |
| 1.E-09 | **-8.47764274420663E-01** | 15 | 0.031 | **-8.47764274420662E-01** | 15 | 0.015 |
| 1.E-11 | **-8.47764274287926E-01** | 15 | 0.015 | **-8.47764274287924E-01** | 15 | 0.015 |
| 1.E-13 | **-8.47764274286598E-01** | 15 | 0.016 | **-8.47764274286597E-01** | 15 | 0.016 |
| 1.E-15 | **-8.47764274286585E-01** | 15 | 0.016 | **-8.47764274286584E-01** | 15 | 0.015 |
| 1.E-17 | **-8.47764274286585E-01** | 15 | 0.015 | **-8.47764274286584E-01** | 15 | 0.016 |
| 1.E-19 | **-8.47764274286585E-01** | 15 | 0.016 | **-8.47764274286584E-01** | 15 | 0.000 |
| 0.E+00 | **-8.47764274286585E-01** | 15 | 0.016 | **-8.47764274286584E-01** | 15 | 0.000 |

*CPU time (in seconds) spent in Mathematica® kernel

# 10. PROJECTION OF OBSERVATION POINT TO THE BOUNDARY

In this section we investigate the case in which the OP projects to a point on the boundary of the BQ. In this case, the OP is a distance $h$ removed from the integration BQ. Differently stated, for the self-cell of the impedance matrix, the exterior quadrature does not involve points that lie on the boundary of the BQ. Without loss of generality, we take the boundary point to be the point $p' = q' = -1$. Then the OP is along the normal to this point and at a distance $h > 0$. For our numerical investigation, we use the second BQ of Appendix B.

From (B.5) and (B.6), the position vector to a point of the BQ is

$$\mathbf{r}(p,q) = (3,0,0)p + (0,2,0)q + (0,0,1)pq, \quad |p| \leq 1, \quad |q| \leq 1 \tag{10.1}$$

from which we have that

$$\mathbf{r}_0 = \mathbf{r}(-1,-1) = (-3,-2,1). \tag{10.2}$$

From (2.12) and (2.14)

$$\mathbf{p}' = (3,0,-1), \quad \mathbf{q}' = (0,2,-1). \tag{10.3}$$

From (2.16), we can write the position vector in the form

$$\mathbf{r}(u,v) = \mathbf{r}_0 + \mathbf{p}'u + \mathbf{q}'v + \mathbf{r}_{pq}uv; \quad \mathbf{r}_{pq} = (0,0,1); \quad 0 \leq u \leq 2, \quad 0 \leq v \leq 2 \tag{10.4}$$

while, for the unit normal at $\mathbf{r}_0$, we have from (2.17)

$$\hat{n}' = \hat{n}(-1,-1) = \frac{\mathbf{p}' \times \mathbf{q}'}{|\mathbf{p}' \times \mathbf{q}'|} = \frac{2\hat{x} + 3\hat{y} + 6\hat{z}}{7}. \tag{10.5}$$

The first integral we examine is the one in (3.5)

$$I_1(-1,-1,h) = h\int_0^2 dv \int_0^2 \frac{du}{R^3(u,v,h)} = h\int_0^2 dv \left\{ \frac{2[2C^2(v)u + B(v,h)]}{\Delta(v,h)R(u,v,h)} \bigg|_0^2 \right\}$$

$$= 2h\int_0^2 \frac{dv}{\Delta(v,h)} \left\{ \frac{4C^2(v) + B(v,h)}{R(2,v,h)} - \frac{B(v,h)}{R(0,v,h)} \right\} \tag{10.6}$$

where $\Delta$ is given by (3.6). Instead of computing (10.6), we have used (3.5) with the parameters as defined in this section. We present the graph of the integrand (including the factor $2h$) for three values of $h$ in Fig. 10.1. We see that, in the present case, we have half of an impulse-like function. In table 10.1, we present the results of computing this expressions for a number of values of $h$ using GKQ and DEQ. Both quadratures agree to at least 14 SD but GKQ appears to have the time advantage. As we explained in Section 3, the value of (10.6) is zero when $h = 0$.

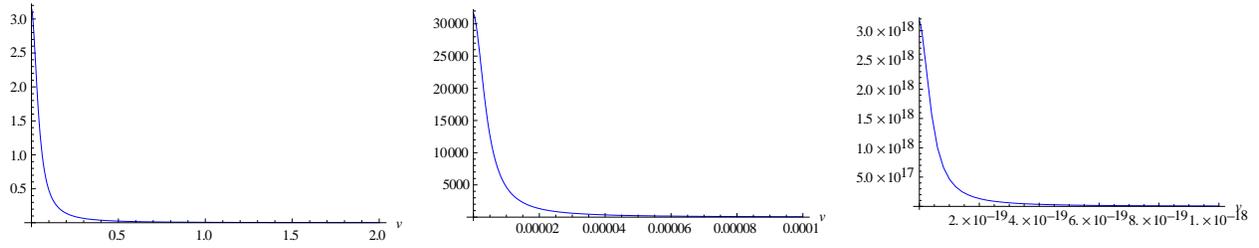

Figure 10.1. Integrand of (3.5) (including the factor $2h$) for the BQ and OP of the present section. Left: $h = 0.1$. Center: $h = 10^{-5}$. Right: $h = 10^{-19}$.

Table 10.1. Equation (3.5) (integral times $2h$) evaluated for the second BQ and for the OP (-1, -1, $h$).

| | SECOND BQ, EQ. (3.5) AT (-1, -1, $h$ ) | | | | | |
|---|---|---|---|---|---|---|
| h | GKQ | GKQ SD | GKQ TIME* | DEQ | DEQ SD | DEQ TIME* |
| 1 | **1.98931425647088E-01** | 15 | 0.016 | **1.98931425647088E-01** | 15 | 0.016 |
| 1.E-01 | 2.06228897168297E-01 | 15 | 0.015 | 2.06228897168297E-01 | 15 | 0.016 |
| 1.E-03 | 2.04202864837701E-01 | 15 | 0.016 | 2.04202864837701E-01 | 15 | 0.031 |
| 1.E-05 | 2.04129750969226E-01 | 15 | 0.031 | 2.04129750969225E-01 | 15 | 0.047 |
| 1.E-07 | 2.04128485688011E-01 | 15 | 0.031 | 2.04128485688011E-01 | 15 | 0.062 |
| 1.E-09 | 2.04128467692112E-01 | 15 | 0.063 | 2.04128467692112E-01 | 15 | 0.062 |
| 1.E-11 | 2.04128467458722E-01 | 15 | 0.078 | 2.04128467458722E-01 | 15 | 0.078 |
| 1.E-13 | 2.04128467455853E-01 | 15 | 0.015 | 2.04128467455853E-01 | 15 | 0.063 |
| 1.E-15 | 2.04128467455820E-01 | 15 | 0.016 | 2.04128467455819E-01 | 15 | 0.125 |
| 1.E-17 | 2.04128467455819E-01 | 15 | 0.015 | 2.04128467455819E-01 | 15 | 0.124 |
| 1.E-19 | 2.04128467455819E-01 | 15 | 0.031 | 2.04128467455819E-01 | 15 | 0.109 |

*CPU time (in seconds) spent in Mathematica® kernel

The second integral we examine is that in (4.4). For the present case, we have

$$I_1(-1,-1,h) = 4h \int_0^2 \frac{dv}{\mathsf{U}(v,h)} \left\{ \frac{A(v,h)}{R(0,v,h)} - \frac{A(v,h)+B(v,h)}{R(2,v,h)} \right\}. \tag{10.7}$$

Instead of working with this expression, we employ (4.4) with the settings of this section. We display the graph of the integrand (multiplied by $2h$) in Fig. 10.2 and the values of the integral (multiplied by $2h$) in Table 10.2. We see agreement between the two quadratures to at least 14 SD with GKQ having a time advantage for smaller values of $h$. Again, the value of (10.7) is zero when $h$ is zero. We will not examine (4.9) since we decided in Section 4 that (4.4) is more robust.

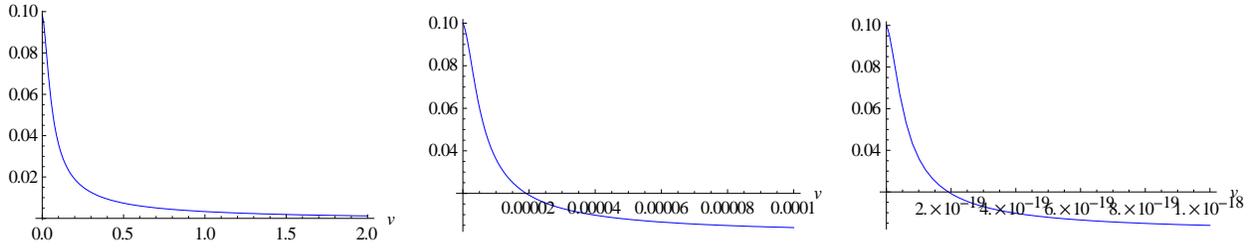

Figure 10.2. Integrand of (4.4) (including the factor 2*h*) for the BQ and OP of the present section. Left: $h = 0.1$. Center: $h = 10^{-5}$. Right: $h = 10^{-19}$.

Table 10.2. Equation (4.4) evaluated for the second BQ and for the OP (-1, -1, *h*).

| | SECOND BQ, EQ. (4.4) AT (-1, -1, *h* ) | | | | | |
|---|---|---|---|---|---|---|
| h | GKQ | GKQ SD | GKQ TIME* | DEQ | DEQ SD | DEQ TIME* |
| 1 | **8.01800114494501E-02** | 15 | 0.015 | **8.01800114494500E-02** | 15 | 0.016 |
| 1.E-01 | **1.67922365938930E-02** | 15 | 0.016 | **1.67922365938939E-02** | 15 | 0.015 |
| 1.E-03 | **3.47149941520428E-04** | 15 | 0.015 | **3.47149941520427E-04** | 15 | 0.016 |
| 1.E-05 | **5.27551546394241E-06** | 15 | 0.032 | **5.27551546394241E-06** | 15 | 0.047 |
| 1.E-07 | **7.07983515517956E-08** | 15 | 0.031 | **7.07983515517955E-08** | 15 | 0.031 |
| 1.E-09 | **8.88415976062065E-10** | 15 | 0.032 | **8.88415976062064E-10** | 15 | 0.062 |
| 1.E-11 | **1.06884844339842E-11** | 15 | 0.046 | **1.06884844339842E-11** | 15 | 0.063 |
| 1.E-13 | **1.24928091082149E-13** | 15 | 0.032 | **1.24928091082149E-13** | 15 | 0.062 |
| 1.E-15 | **1.42971337824561E-15** | 15 | 0.031 | **2.04128467455819E-01** | 15 | 0.063 |
| 1.E-17 | **1.61014584566974E-17** | 15 | 0.031 | **1.61014584566973E-17** | 15 | 0.062 |
| 1.E-19 | **1.79057831309386E-19** | 15 | 0.031 | **1.79057831309386E-19** | 15 | 0.124 |

*CPU time (in seconds) spent in Mathematica® kernel

The third integral is that of (5.2). Under the present conditions we write

$$I_1(-1,-1,h) = 4 \lim_{u \to 0^+} \int_u^2 \frac{v \, dv}{U(v,h)} \left\{ \frac{A(v,h)}{R(0,v,h)} - \frac{A(v,h)+B(v,h)}{R(2,v,h)} \right\}. \tag{10.8}$$

As above, we work with (5.2) since it is preferable to have one expression for all the points of the BQ rather than two. In Fig. 10.3, we present the graph of the integrand of (5.2) with the settings of this section while in Table 10.3, we display the values of the integral. The agreement between the two quadratures is not the best but it is according to the findings of Section 5. For this reason, and as in Section 5, we recommend use of DEQ with this integral.

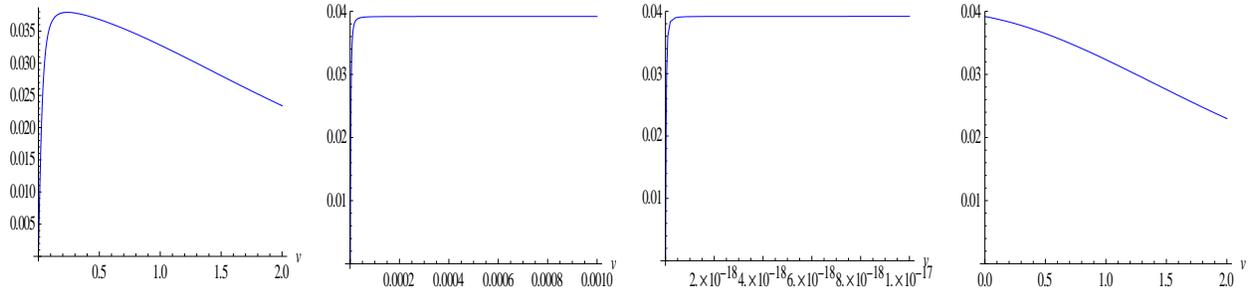

Figure 10.3. Integrand of (5.2) for the BQ and OP of the present section. From left to right: $h = 0.1$, $10^{-5}$, $10^{-19}$, $0.0$.

Table 10.3. Equation (5.2) evaluated for the second BQ and for the OP $(-1, -1, h)$.

| | SECOND BQ, EQ. (5.2) AT $(-1, -1, h)$ | | | | | |
|---|---|---|---|---|---|---|
| h | GKQ | GKQ SD | GKQ TIME* | DEQ | DEQ SD | DEQ TIME* |
| 1 | **5.48623249228077E-02** | 15 | 0.015 | **5.48623249228076E-02** | 15 | 0.016 |
| 1.E-01 | **6.31471881668210E-02** | 15 | 0.016 | **6.31471881668209E-02** | 15 | 0.000 |
| 1.E-03 | **6.38478555831382E-02** | 15 | 0.016 | **6.38478555831381E-02** | 15 | 0.016 |
| 1.E-05 | **6.38543411250010E-02** | 15 | 0.015 | **6.38543411250009E-02** | 15 | 0.015 |
| 1.E-07 | **6.38544058636068E-02** | 15 | 0.016 | **6.38544058636067E-02** | 15 | 0.016 |
| 1.E-09 | **6.38544065272168E-02** | 15 | 0.000 | **6.38544065109745E-02** | 15 | 0.015 |
| 1.E-11 | **6.38544065176107E-02** | 15 | 0.000 | **6.38544065174482E-02** | 15 | 0.032 |
| 1.E-13 | **6.38544065175146E-02** | 15 | 0.000 | **6.38544065175129E-02** | 15 | 0.015 |
| 1.E-15 | **6.38544065175136E-02** | 15 | 0.015 | **6.38544065175136E-02** | 15 | 0.000 |
| 1.E-17 | **6.38544065175136E-02** | 15 | 0.000 | **6.38544065175136E-02** | 15 | 0.000 |
| 1.E-19 | **6.38544065175136E-02** | 15 | 0.000 | **6.38544065175136E-02** | 15 | 0.016 |
| 0.E+00 | **6.38544065175136E-02** | 15 | 0.015 | **6.38544065175136E-02** | 15 | 0.000 |

*CPU time (in seconds) spent in Mathematica® kernel

The fourth integral is that of (6.2)

$$I_1(-1,-1,h) = 2\int_0^2 \frac{v^2 dv}{\Delta(v,h)} \left\{ \frac{[4C^2(v)+B(v,h)]}{R(2,v,h)} - \frac{B(v,h)}{R(0,v,h)} \right\}. \tag{10.9}$$

We present the graph of the integrand of (6.2) in Fig. 10.4, and the results of the integration in Table 10.4. We note that the two quadratures do not agree everywhere, something we also noticed and discussed in Section 6. As we argue there, we are inclined to trust the DEQ results more.

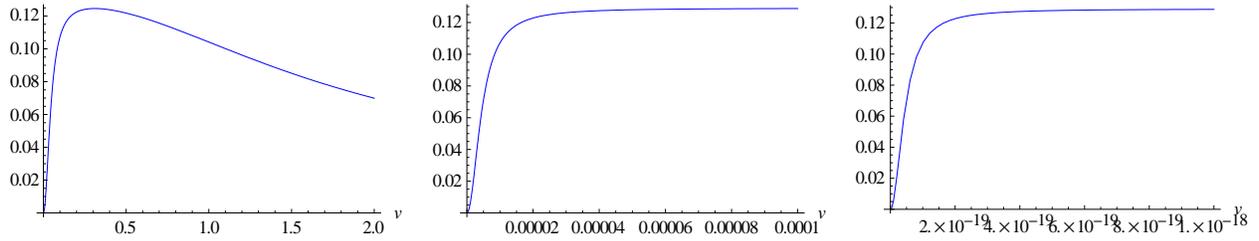

Figure 10.4. Integrand of (6.2) for the BQ and OP of the present section. Left: $h = 0.1$. Center: $h = 10^{-5}$. Right: $h = 10^{-19}$.

Table 10.4. Equation (6.2) evaluated for the second BQ and for the OP (-1, -1, $h$).

| | SECOND BQ, EQ. (6.2) AT (-1, -1, $h$) | | | | | |
|---|---|---|---|---|---|---|
| $h$ | GKQ | GKQ SD | GKQ TIME* | DEQ | DEQ SD | DEQ TIME* |
| 1 | **1.29643841378760E-01** | 15 | 0.015 | **1.29643841378760E-01** | 15 | 0.000 |
| 1.E-01 | **1.97637993630174E-01** | 15 | 0.000 | **1.97637993630173E-01** | 15 | 0.016 |
| 1.E-03 | **2.06902471722969E-01** | 15 | 0.016 | **2.06902471722968E-01** | 15 | 0.016 |
| 1.E-05 | **2.06997173180354E-01** | 15 | 0.016 | **2.06997173180354E-01** | 15 | 0.015 |
| 1.E-07 | **2.06998120377356E-01** | 15 | 0.015 | **2.06998120377356E-01** | 15 | 0.016 |
| 1.E-09 | **2.06998129849341E-01** | 15 | 0.016 | **2.06998129849341E-01** | 15 | 0.031 |
| 1.E-11 | **2.06998129944977E-01** | 15 | 0.000 | **2.06998129944060E-01** | 15 | 0.000 |
| 1.E-13 | **2.06998129945017E-01** | 15 | 0.000 | **2.06998129945008E-01** | 15 | 0.015 |
| 1.E-15 | **2.06998129945018E-01** | 15 | 0.000 | **2.06998129945017E-01** | 15 | 0.016 |
| 1.E-17 | **2.06998129945018E-01** | 15 | 0.015 | **2.06998129945017E-01** | 15 | 0.000 |
| 1.E-19 | **2.06998129945018E-01** | 15 | 0.000 | **2.06998129945017E-01** | 15 | 0.015 |

*CPU time (in seconds) spent in Mathematica® kernel

The fifth integral is that of (7.2)

$$I_1(-1,-1,h) = 4\int_0^2 \frac{v^2 dv}{\mathsf{U}(v,h)} \left\{ \frac{A(v,h)}{R(0,v,h)} - \frac{A(v,h)+B(v,h)}{R(1-p',v,h)} \right\}. \tag{10.10}$$

We did not encounter any issues in computing (7.2) for the values of the parameters as in (10.10). We present the results in Fig. 10.5 and Table 10.5.

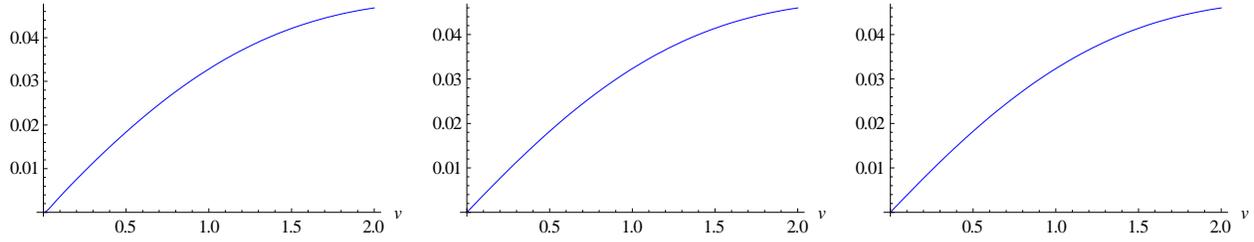

Figure 10.5. Integrand of (7.2) for the BQ and OP of the present section. Left: $h = 0.1$. Center: $h = 10^{-5}$. Right: $h = 10^{-19}$.

Table 10.5. Equation (7.2) evaluated for the second BQ and for the OP (-1, -1, h).

| | SECOND BQ, EQ. (7.2) AT (-1, -1, $h$) | | | | | |
|---|---|---|---|---|---|---|
| $h$ | GKQ | GKQ SD | GKQ TIME* | DEQ | DEQ SD | DEQ TIME* |
| 1 | 5.96506173743192E-02 | 15 | 0.015 | 5.96506173743191E-02 | 15 | 0.000 |
| 1.E-01 | 5.90033718075497E-02 | 15 | 0.016 | 5.90033718075496E-02 | 15 | 0.000 |
| 1.E-03 | 5.81997713359904E-02 | 15 | 0.016 | 5.81997713359903E-02 | 15 | 0.015 |
| 1.E-05 | 5.81902436219309E-02 | 15 | 0.000 | 5.81902436219308E-02 | 15 | 0.016 |
| 1.E-07 | 5.81901480417572E-02 | 15 | 0.000 | 5.81901480417571E-02 | 15 | 0.000 |
| 1.E-09 | 5.81901470859095E-02 | 15 | 0.000 | 5.81901470859094E-02 | 15 | 0.000 |
| 1.E-11 | 5.81901470763510E-02 | 15 | 0.000 | 5.81901470763510E-02 | 15 | 0.015 |
| 1.E-13 | 5.81901470762554E-02 | 15 | 0.000 | 5.81901470762554E-02 | 15 | 0.000 |
| 1.E-15 | 5.81901470762545E-02 | 15 | 0.000 | 5.81901470762544E-02 | 15 | 0.000 |
| 1.E-17 | 5.81901470762545E-02 | 15 | 0.000 | 5.81901470762544E-02 | 15 | 0.000 |
| 1.E-19 | 5.81901470762545E-02 | 15 | 0.000 | 5.81901470762544E-02 | 15 | 0.000 |

*CPU time (in seconds) spent in Mathematica® kernel

The sixth integral is that in (8.8)

$$I_{11}(-1,-1,h) = \lim_{u \to 0} \int_{u}^{2} \frac{-2u^2 du}{F^2(u,h)\Delta(u,h)} \left\{ \frac{D(u,h)E(u,h)}{R(u,0,h)} + \frac{[\Delta(u,h) - E^2(u,h)] - D(u,h)E(u,h)}{R(u,2,h)} \right\}$$

(10.11)

When in (8.8) we use the parameters of this section, we get the results shown in Fig. 10.6 and Table 10.6.

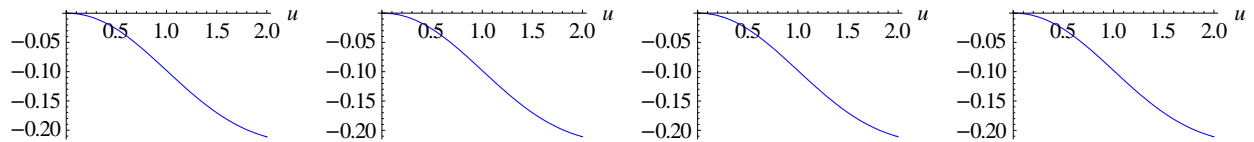

Figure 10.6. Integrand of (8.8) for the BQ and OP of the present section. From left to right: $h = 0.1$, $10^{-5}$, $10^{-19}$, 0.0.

Table 10.6. Equation (8.8) evaluated for the second BQ and for the OP (-1, -1, h).

| | SECOND BQ, EQ. (8.8) AT (-1, -1, h ) | | | | | |
|---|---|---|---|---|---|---|
| h | GKQ | GKQ SD | GKQ TIME* | DEQ | DEQ SD | DEQ TIME* |
| 1.E+00 | **-1.98001506170773E-01** | 15 | 0.016 | -1.98001506170773E-01 | 15 | 0.016 |
| 1.E-01 | **-1.99536743757196E-01** | 15 | 0.000 | -1.99536743757196E-01 | 15 | 0.000 |
| 1.E-03 | **-1.99406920428378E-01** | 15 | 0.000 | -1.99406920428377E-01 | 15 | 0.016 |
| 1.E-05 | **-1.99405269708712E-01** | 15 | 0.000 | -1.99405269708712E-01 | 15 | 0.015 |
| 1.E-07 | **-1.99405253165721E-01** | 15 | 0.000 | -1.99405253165720E-01 | 15 | 0.016 |
| 1.E-09 | **-1.99405253000287E-01** | 15 | 0.000 | -1.99405253000287E-01 | 15 | 0.016 |
| 1.E-11 | **-1.99405252998633E-01** | 15 | 0.000 | -1.99405252998633E-01 | 15 | 0.000 |
| 1.E-13 | **-1.99405252998616E-01** | 15 | 0.016 | -1.99405252998616E-01 | 15 | 0.000 |
| 1.E-15 | **-1.99405252998616E-01** | 15 | 0.015 | -1.99405252998616E-01 | 15 | 0.015 |
| 1.E-17 | **-1.99405252998616E-01** | 15 | 0.000 | -1.99405252998616E-01 | 15 | 0.016 |
| 1.E-19 | **-1.99405252998616E-01** | 15 | 0.000 | -1.99405252998616E-01 | 15 | 0.000 |
| 0.E+00 | **-1.99405252998616E-01** | 15 | 0.016 | -1.99405252998616E-01 | 15 | 0.015 |

*CPU time (in seconds) spent in Mathematica® kernel

The seventh integral is that in (8.19) in Section 8,

$$I_{12}(-1,-1,h) = \lim_{u \to 0} \int_u^2 \frac{u^2 du}{F^3(u)} \ln\left|\frac{\{2F(u)[R(u,2,h)+2F(u)]+E(u,h)\}R(u,0,h)}{2F(u)D(u,h)+E(u,h)R(u,0,h)}\right|. \quad (10.12)$$

We display the integrand of (8.19) for the present case in Fig. 10.7. It has a removable singularity at the origin when $h$ is zero and the limit there is zero. We present integration results in Table 10.7. For this calculation, we have specified the origin as a singular point of the integrand. If we do not do so, then the execution time rises by about an order of magnitude, as we can see from Table 10.8. This is a good example of the savings in time and accuracy we can obtain through careful algorithm design. As an aside, Mathematica® claims to look for singularities at the endpoints of the interval of integration. The last two tables seem to contradict this claim. In all cases below that involve the logarithmic function, we will present results with the origin specified as a singular point.

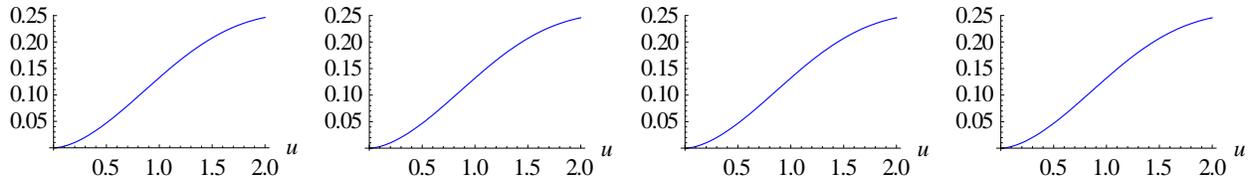

Figure 10.7. Integrand of (8.19) for the BQ and OP of the present section. From left to right: $h = 0.1, 10^{-5}, 10^{-19}, 0.0$.

Table 10.7. Equation (8.19) evaluated for the second BQ and for the OP (-1, -1, *h*). The left endpoint of the integration interval has been specified as having a singularity.

| | SECOND BQ, EQ. (8.19) AT (-1, -1, *h*) WITH SINGULARITY SPECIFIED | | | | | |
|---|---|---|---|---|---|---|
| *h* | GKQ | GKQ SD | GKQ TIME* | DEQ | DEQ SD | DEQ TIME* |
| 1.E+00 | 2.57604371771849E-01 | 15 | 0.016 | 2.57604371771849E-01 | 15 | 0.016 |
| 1.E-01 | 2.54768304196336E-01 | 15 | 0.015 | 2.54768304196336E-01 | 15 | 0.000 |
| 1.E-03 | 2.53736656700580E-01 | 15 | 0.016 | 2.53736656700579E-01 | 15 | 0.016 |
| 1.E-05 | 2.53725568236583E-01 | 15 | 0.000 | 2.53725568236583E-01 | 15 | 0.000 |
| 1.E-07 | 2.53725457274432E-01 | 15 | 0.016 | 2.53725457274432E-01 | 15 | 0.015 |
| 1.E-09 | 2.53725456164803E-01 | 15 | 0.000 | 2.53725456164803E-01 | 15 | 0.016 |
| 1.E-11 | 2.53725456153707E-01 | 15 | 0.015 | 2.53725456153706E-01 | 15 | 0.000 |
| 1.E-13 | 2.53725456153596E-01 | 15 | 0.016 | 2.53725456153596E-01 | 15 | 0.016 |
| 1.E-15 | 2.53725456153595E-01 | 15 | 0.000 | 2.53725456153594E-01 | 15 | 0.015 |
| 1.E-17 | 2.53725456153595E-01 | 15 | 0.015 | 2.53725456153594E-01 | 15 | 0.000 |
| 1.E-19 | 2.53725456153595E-01 | 15 | 0.016 | 2.53725456153594E-01 | 15 | 0.000 |
| 0.E+00 | 2.53725456153595E-01 | 15 | 0.015 | 2.53725456153594E-01 | 15 | 0.015 |

*CPU time (in seconds) spent in Mathematica® kernel

Table 10.8. Equation (8.19) evaluated for the second BQ and for the OP (-1, -1, *h*). The removable singularity at the left endpoint has not been specified.

| | SECOND BQ, EQ. (8.19) AT (-1, -1, *h*) WITH SINGULARITY NOT SPECIFIED | | | | | |
|---|---|---|---|---|---|---|
| *h* | GKQ | GKQ SD | GKQ TIME* | DEQ | DEQ SD | DEQ TIME* |
| 1.E+00 | 2.57604371771849E-01 | 15 | 0.281 | 2.57604371771849E-01 | 15 | 0.265 |
| 1.E-01 | 2.54768304196336E-01 | 15 | 0.125 | 2.54768304196336E-01 | 15 | 0.124 |
| 1.E-03 | 2.53736656700580E-01 | 15 | 0.140 | 2.53736656700579E-01 | 15 | 0.125 |
| 1.E-05 | 2.53725568236583E-01 | 15 | 0.140 | 2.53725568236583E-01 | 15 | 0.141 |
| 1.E-07 | 2.53725457274432E-01 | 15 | 0.141 | 2.53725457274432E-01 | 15 | 0.140 |
| 1.E-09 | 2.53725456164803E-01 | 15 | 0.156 | 2.53725456164803E-01 | 15 | 0.156 |
| 1.E-11 | 2.53725456153707E-01 | 15 | 0.156 | 2.53725456153706E-01 | 15 | 0.156 |
| 1.E-13 | 2.53725456153596E-01 | 15 | 0.171 | 2.53725456153596E-01 | 15 | 0.172 |
| 1.E-15 | 2.53725456153595E-01 | 15 | 0.203 | 2.53725456153594E-01 | 15 | 0.187 |
| 1.E-17 | 2.53725456153595E-01 | 15 | 0.156 | 2.53725456153594E-01 | 15 | 0.156 |
| 1.E-19 | 2.53725456153595E-01 | 15 | 0.172 | 2.53725456153594E-01 | 15 | 0.171 |
| 0.E+00 | 2.53725456153595E-01 | 15 | 0.125 | 2.53725456153594E-01 | 15 | 0.094 |

*CPU time (in seconds) spent in Mathematica® kernel

The eighth integral is that in (9.4), Section 9. With the present values of the parameters, it reads

$$I_1(-1,-1,h) = h\int_0^2 \frac{u\,du}{F(u)} \ln\left|\frac{\{2F(u)[R(u,2,h)+2F(u)]+E(u,h)\}R(u,0,h)}{2F(u)D(u,h)+E(u,h)R(u,0,h)}\right|. \quad (10.13)$$

We display the graph of the integrand (multiplied by *h*) in Fig. 10.8 and the numerical values of the integral in Table 10.9. We have specified the origin as a singular point.

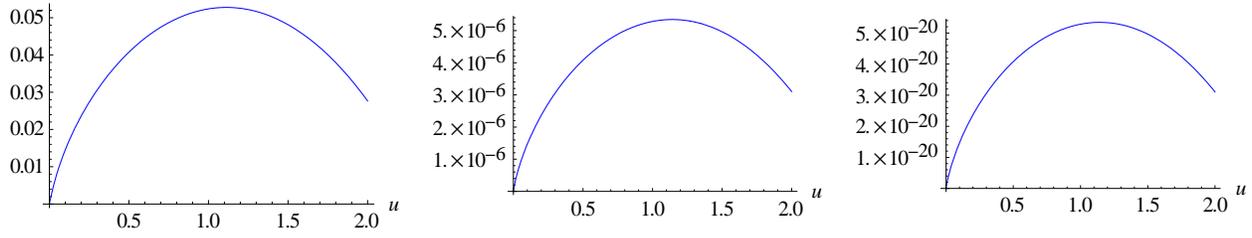

Figure 10.8. Integrand of (9.4) (multiplied by $h$) for the BQ and OP of the present section. From left to right: $h = 0.1$, $10^{-5}$, $10^{-19}$.

Table 10.9. Equation (9.4) evaluated for the second BQ and for the OP $(-1, -1, h)$.

| | SECOND BQ, EQ. (9.4) AT (-1, -1, h) | | | | | |
|---|---|---|---|---|---|---|
| $h$ | GKQ | GKQ SD | GKQ TIME* | DEQ | DEQ SD | DEQ TIME* |
| 1.E+00 | 5.59246708299679E-01 | 15 | 0.015 | 5.59246708299679E-01 | 15 | 0.015 |
| 1.E-01 | 8.20892237353126E-02 | 15 | 0.016 | 8.20892237353125E-02 | 15 | 0.000 |
| 1.E-03 | 8.38813455330103E-04 | 15 | 0.016 | 8.38813455330102E-04 | 15 | 0.016 |
| 1.E-05 | 8.38978912139656E-06 | 15 | 0.015 | 8.38978912139656E-06 | 15 | 0.015 |
| 1.E-07 | 8.38980564458723E-08 | 15 | 0.000 | 8.38980564458722E-08 | 15 | 0.000 |
| 1.E-09 | 8.38980580981599E-10 | 15 | 0.016 | 8.38980580981598E-10 | 15 | 0.000 |
| 1.E-11 | 8.38980581146828E-12 | 15 | 0.015 | 3.89805811468270E-13 | 15 | 0.000 |
| 1.E-13 | 8.38980581148480E-14 | 15 | 0.000 | 8.38980581148479E-14 | 15 | 0.016 |
| 1.E-15 | 8.38980581148497E-16 | 15 | 0.016 | 8.38980581148496E-16 | 15 | 0.000 |
| 1.E-17 | 8.38980581148497E-18 | 15 | 0.016 | 8.38980581148496E-18 | 15 | 0.000 |
| 1.E-19 | 8.38980581148497E-20 | 15 | 0.000 | 8.38980581148496E-20 | 15 | 0.000 |
| 0.E+00 | 0.00000000000000E+00 | 15 | 0.000 | 0.00000000000000E+00 | 15 | 0.000 |

*CPU time (in seconds) spent in Mathematica® kernel

The ninth integral is that in (9.8). It now reads

$$I_{21}(-1,-1,h) = \int_0^2 \frac{u\,du}{F^2(u)}\left[R(u,2,h) - R(u,0,h)\right]. \tag{10.14}$$

We present the graph of the integrand in Fig. 10.9 and the results for the integral in Table 10.10. Since we are dealing with a continuous integrand, we do not encounter any issues here.

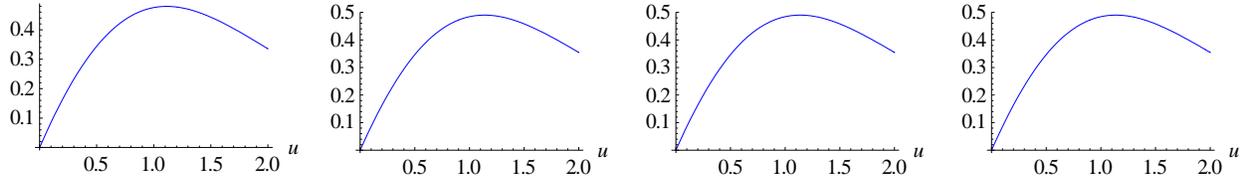

Figure 10.9. Integrand of (9.8) for the BQ and OP of the present section. From left to right: $h =$ 0.1, $10^{-5}$, $10^{-19}$, 0.0.

Table 10.10. Equation (9.8) evaluated for the second BQ and for the OP (-1, -1, h).

| | SECOND BQ, EQ. (9.8) AT (-1, -1, $h$ ) | | | | | |
|---|---|---|---|---|---|---|
| $h$ | GKQ | GKQ SD | GKQ TIME* | DEQ | DEQ SD | DEQ TIME* |
| 1.E+00 | 5.52495645404374E-01 | 15 | 0.000 | 5.52495645404374E-01 | 15 | 0.016 |
| 1.E-01 | 7.38834671557692E-01 | 15 | 0.016 | 7.38834671557691E-01 | 15 | 0.000 |
| 1.E-03 | 7.56512873020039E-01 | 15 | 0.000 | 7.56512873020039E-01 | 15 | 0.016 |
| 1.E-05 | 7.56685960277542E-01 | 15 | 0.000 | 7.56685960277541E-01 | 15 | 0.015 |
| 1.E-07 | 7.56687690771802E-01 | 15 | 0.000 | 7.56687690771801E-01 | 15 | 0.000 |
| 1.E-09 | 7.56687708076707E-01 | 15 | 0.015 | 7.56687708076706E-01 | 15 | 0.000 |
| 1.E-11 | 7.56687708249756E-01 | 15 | 0.000 | 7.56687708249755E-01 | 15 | 0.000 |
| 1.E-13 | 7.56687708251486E-01 | 15 | 0.000 | 7.56687708251486E-01 | 15 | 0.016 |
| 1.E-15 | 7.56687708251504E-01 | 15 | 0.016 | 7.56687708251503E-01 | 15 | 0.000 |
| 1.E-17 | 7.56687708251504E-01 | 15 | 0.015 | 7.56687708251503E-01 | 15 | 0.000 |
| 1.E-19 | 7.56687708251504E-01 | 15 | 0.016 | 7.56687708251503E-01 | 15 | 0.000 |
| 0.E+00 | 7.56687708251504E-01 | 15 | 0.015 | 7.56687708251503E-01 | 15 | 0.015 |

*CPU time (in seconds) spent in Mathematica® kernel

The tenth integral is that of (9.10)

$$I_{22}(-1,-1,h) = \int_0^2 \ln\left|\frac{\{2F(u)[R(u,2,h)+2F(u)]+E(u,h)\}R(u,0,h)}{2F(u)D(u,h)+E(u,h)R(u,0,h)}\right| \frac{E(u,h)u\,du}{F^3(u)}. \quad (10.15)$$

We present the graph of the integrand in Fig. 10.10 and evaluate the integral in Table 10.11. As in every case in which the logarithm appears, we have specified the lower limit as a singular point.

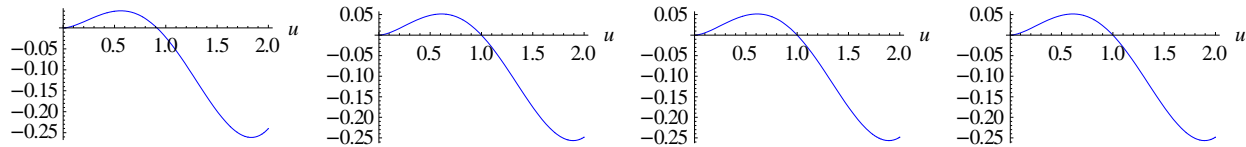

Figure 10.10. Integrand of (9.10) for the BQ and OP of the present section. From left to right: $h =$ 0.1, $10^{-5}$, $10^{-19}$, 0.0.

Table 10.11. Equation (9.10) evaluated for the second BQ and for the OP (-1, -1, h).

| | SECOND BQ, EQ. (9.10) AT (-1, -1, h) | | | | | |
|---|---|---|---|---|---|---|
| h | GKQ | GKQ SD | GKQ TIME* | DEQ | DEQ SD | DEQ TIME* |
| 1.E+00 | **-2.26678662933492E-01** | 15 | 0.016 | **-2.26678662933492E-01** | 15 | 0.015 |
| 1.E-01 | **-1.55488561430462E-01** | 15 | 0.000 | **-1.55488561430462E-01** | 15 | 0.000 |
| 1.E-03 | **-1.28342764132335E-01** | 15 | 0.000 | **-1.28342764132335E-01** | 15 | 0.015 |
| 1.E-05 | **-1.28058439603203E-01** | 15 | 0.015 | **-1.28058439603203E-01** | 15 | 0.016 |
| 1.E-07 | **-1.28055595123248E-01** | 15 | 0.016 | **-1.28055595123248E-01** | 15 | 0.000 |
| 1.E-09 | **-1.28055566678325E-01** | 15 | 0.000 | **-1.28055566678325E-01** | 15 | 0.016 |
| 1.E-11 | **-1.28055566393876E-01** | 15 | 0.015 | **-1.28055566393876E-01** | 15 | 0.000 |
| 1.E-13 | **-1.28055566391031E-01** | 15 | 0.016 | **-1.28055566391031E-01** | 15 | 0.016 |
| 1.E-15 | **-1.28055566391003E-01** | 15 | 0.000 | **-1.28055566391003E-01** | 15 | 0.000 |
| 1.E-17 | **-1.28055566391003E-01** | 15 | 0.016 | **-1.28055566391003E-01** | 15 | 0.015 |
| 1.E-19 | **-1.28055566391003E-01** | 15 | 0.000 | **-1.28055566391003E-01** | 15 | 0.000 |
| 0.E+00 | **-1.28055566391003E-01** | 15 | 0.016 | **-1.28055566391003E-01** | 15 | 0.016 |

*CPU time (in seconds) spent in Mathematica® kernel

The eleventh integral is that in (9.12)

$$I_3(-1,-1,h) = \int_0^2 \frac{u^2 du}{F(u)} \ln\left|\frac{\{2F(u)[R(u,2,h)+2F(u)]+E(u,h)\}R(u,0,h)}{2F(u)D(u,h)+E(u,h)R(u,0,h)}\right|. \quad (10.16)$$

We display the graph of the integrand and numerical results for the integral in Fig. 10.11 and Table 10.12, respectively. In Mathematica®, we have specified the origin as a singular point.

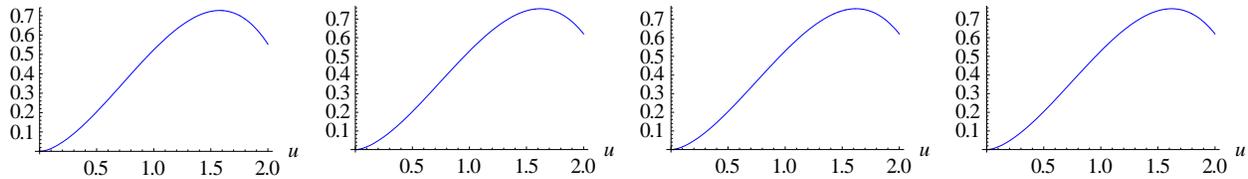

Figure 10.11. Integrand of (9.12) for the BQ and OP of the present section. From left to right: $h$ = 0.1, $10^{-5}$, $10^{-19}$, 0.0.

The twelfth integral is the one in (9.15), which now reads

$$I_{41}(-1,-1,h) = \int_0^2 \frac{u^2 du}{F^2(u)}[R(u,2,h) - R(u,0,h)]. \quad (10.17)$$

It is very similar to the one in (10.14) and does not present any problems. We display the graph of the integrand and numerical results for the integral in Fig. 10.12 and Table 10.13, respectively.

Table 10.12. Equation (9.12) evaluated for the second BQ and for the OP (-1, -1, $h$).

| | SECOND BQ, EQ. (9.12) AT (-1, -1, $h$) | | | | | |
|---|---|---|---|---|---|---|
| $h$ | GKQ | GKQ SD | GKQ TIME* | DEQ | DEQ SD | DEQ TIME* |
| 1.E+00 | 5.02286657894506E-01 | 15 | 0.015 | 5.02286657894505E-01 | 15 | 0.000 |
| 1.E-01 | 8.82844070725320E-01 | 15 | 0.000 | 8.82844070725319E-01 | 15 | 0.016 |
| 1.E-03 | 9.10805431835656E-01 | 15 | 0.000 | 9.10805431835655E-01 | 15 | 0.000 |
| 1.E-05 | 9.11073818922245E-01 | 15 | 0.016 | 9.11073818922245E-01 | 15 | 0.000 |
| 1.E-07 | 9.11076501688671E-01 | 15 | 0.015 | 9.11076501688670E-01 | 15 | 0.000 |
| 1.E-09 | 9.11076528516225E-01 | 15 | 0.000 | 9.11076528516224E-01 | 15 | 0.000 |
| 1.E-11 | 9.11076528784501E-01 | 15 | 0.000 | 9.11076528784500E-01 | 15 | 0.015 |
| 1.E-13 | 9.11076528787183E-01 | 15 | 0.016 | 9.11076528787183E-01 | 15 | 0.016 |
| 1.E-15 | 9.11076528787210E-01 | 15 | 0.000 | 9.11076528787209E-01 | 15 | 0.000 |
| 1.E-17 | 9.11076528787210E-01 | 15 | 0.000 | 9.11076528787210E-01 | 15 | 0.000 |
| 1.E-19 | 9.11076528787210E-01 | 15 | 0.015 | 9.11076528787210E-01 | 15 | 0.000 |
| 0.E+00 | 9.11076528787210E-01 | 15 | 0.000 | 9.11076528787210E-01 | 15 | 0.000 |

*CPU time (in seconds) spent in Mathematica® kernel

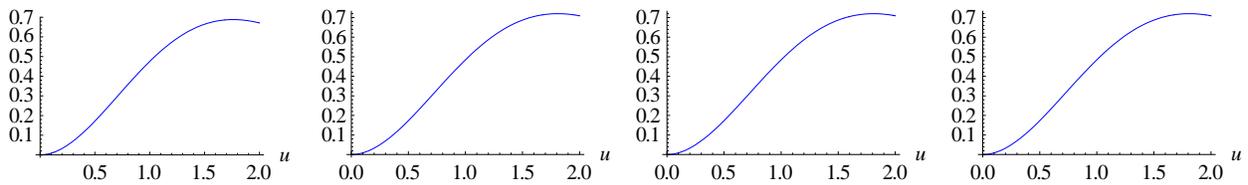

Figure 10.12. Integrand of (9.15) for the BQ and OP of the present section. From left to right: $h = 0.1$, $10^{-5}$, $10^{-19}$, $0.0$.

Table 10.13. Equation (9.15) evaluated for the second BQ and for the OP (-1, -1, $h$).

| | SECOND BQ, EQ. (9.15) AT (-1, -1, $h$) | | | | | |
|---|---|---|---|---|---|---|
| $h$ | GKQ | GKQ SD | GKQ TIME* | DEQ | DEQ SD | DEQ TIME* |
| 1.E+00 | 5.76861171125890E-01 | 15 | 0.000 | 5.76861171125889E-01 | 15 | 0.015 |
| 1.E-01 | 8.26941057116909E-01 | 15 | 0.000 | 8.26941057116908E-01 | 15 | 0.016 |
| 1.E-03 | 8.51928152176996E-01 | 15 | 0.016 | 8.51928152176995E-01 | 15 | 0.015 |
| 1.E-05 | 8.52175007767223E-01 | 15 | 0.000 | 8.52175007767222E-01 | 15 | 0.000 |
| 1.E-07 | 8.52177476018408E-01 | 15 | 0.015 | 8.52177476018406E-01 | 15 | 0.000 |
| 1.E-09 | 8.52177500700889E-01 | 15 | 0.000 | 8.52177500700888E-01 | 15 | 0.000 |
| 1.E-11 | 8.52177500947714E-01 | 15 | 0.000 | 8.52177500947713E-01 | 15 | 0.000 |
| 1.E-13 | 8.52177500950182E-01 | 15 | 0.000 | 8.52177500950181E-01 | 15 | 0.000 |
| 1.E-15 | 8.52177500950207E-01 | 15 | 0.000 | 8.52177500950206E-01 | 15 | 0.000 |
| 1.E-17 | 8.52177500950207E-01 | 15 | 0.016 | 8.52177500950206E-01 | 15 | 0.000 |
| 1.E-19 | 8.52177500950207E-01 | 15 | 0.016 | 8.52177500950206E-01 | 15 | 0.000 |
| 0.E+00 | 8.52177500950207E-01 | 15 | 0.015 | 8.52177500950206E-01 | 15 | 0.000 |

*CPU time (in seconds) spent in Mathematica® kernel

The thirteenth integral is that in (9.16) and it now reads

$$I_{42}(-1,-1,h) = \int_0^2 \ln\left|\frac{\{2F(u)[R(u,2,h)+2F(u)]+E(u,h)\}R(u,0,h)}{2F(u)D(u,h)+E(u,h)R(u,0,h)}\right|\frac{E(u,h)u^2 du}{F^3(u)}. \quad (10.18)$$

We display the behavior of the integrand in Fig. 10.13 and that of the integral in Table 10.14. We have specified the origin as a singular point.

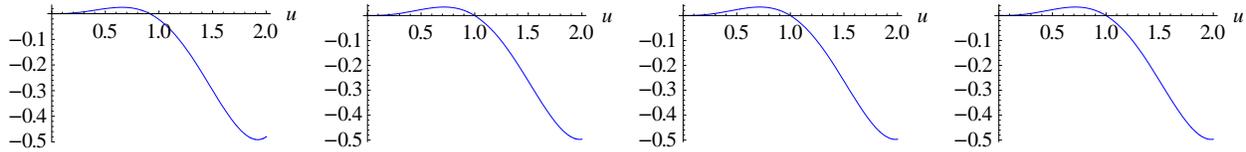

Figure 10.13. Integrand of (9.16) for the BQ and OP of the present section. From left to right: $h = 0.1$, $10^{-5}$, $10^{-19}$, $0.0$.

Table 10.14. Equation (9.16) evaluated for the second BQ and for the OP (-1, -1, h).

| | SECOND BQ, EQ. (9.16) AT (-1, -1, h) | | | | | |
|---|---|---|---|---|---|---|
| h | GKQ | GKQ SD | GKQ TIME* | DEQ | DEQ SD | DEQ TIME* |
| 1.E+00 | -2.68763264129815E-01 | 15 | 0.016 | -2.68763264129814E-01 | 15 | 0.016 |
| 1.E-01 | -2.75969746168586E-01 | 15 | 0.015 | -2.75969746168586E-01 | 15 | 0.016 |
| 1.E-03 | -2.44792059720471E-01 | 15 | 0.000 | -2.44792059720471E-01 | 15 | 0.000 |
| 1.E-05 | -2.44458303780912E-01 | 15 | 0.000 | -2.44458303780912E-01 | 15 | 0.015 |
| 1.E-07 | -2.44454964098009E-01 | 15 | 0.016 | -2.44454964098009E-01 | 15 | 0.016 |
| 1.E-09 | -2.44454930700967E-01 | 15 | 0.016 | -2.44454930700967E-01 | 15 | 0.000 |
| 1.E-11 | -2.44454930366997E-01 | 15 | 0.000 | -2.44454930366997E-01 | 15 | 0.015 |
| 1.E-13 | -2.44454930363657E-01 | 15 | 0.015 | -2.44454930363657E-01 | 15 | 0.000 |
| 1.E-15 | -2.44454930363624E-01 | 15 | 0.000 | -2.44454930363624E-01 | 15 | 0.016 |
| 1.E-17 | -2.44454930363624E-01 | 15 | 0.000 | -2.44454930363623E-01 | 15 | 0.000 |
| 1.E-19 | -2.44454930363624E-01 | 15 | 0.015 | -2.44454930363623E-01 | 15 | 0.015 |
| 0.E+00 | -2.44454930363624E-01 | 15 | 0.031 | -2.44454930363623E-01 | 15 | 0.000 |

*CPU time (in seconds) spent in Mathematica® kernel

The fourteenth integral is the one in (9.20)

$$I_{51}(-1,-1,h) = \int_0^2 \frac{u^2 du}{F^2(u)}\left\{\left[2-\frac{3E(u,h)}{2F^2(u)}\right]R(u,2,h)+\frac{3E(u,h)}{2F^2(u)}R(u,0,h)\right\}. \quad (10.19)$$

We display the behavior of the integrand in Fig. 10.14 and that of the integral in Table 10.15. This is a simple case since the integrand is continuous over the interval of integration.

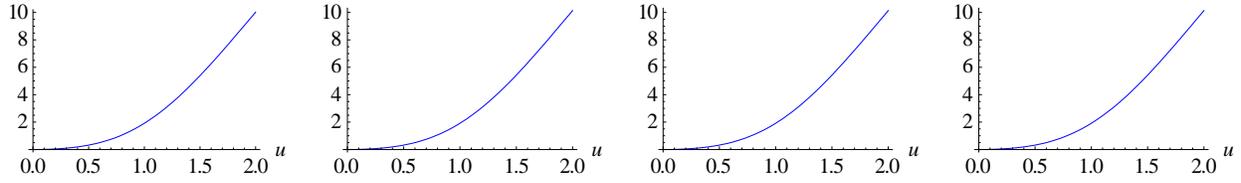

Figure 10.14. Integrand of (9.20) for the BQ and OP of the present section. From left to right: $h = 0.1$, $10^{-5}$, $10^{-19}$, $0.0$.

Table 10.15. Equation (9.20) evaluated for the second BQ and for the OP (-1, -1, h).

| | SECOND BQ, EQ. (9.20) AT (-1, -1, h ) | | | | | |
|---|---|---|---|---|---|---|
| h | GKQ | GKQ SD | GKQ TIME* | DEQ | DEQ SD | DEQ TIME* |
| 1.E+00 | 5.60854439808463E+00 | 15 | 0.000 | 5.60854439808462E+00 | 15 | 0.016 |
| 1.E-01 | 6.11755945508924E+00 | 15 | 0.000 | 6.11755945508923E+00 | 15 | 0.015 |
| 1.E-03 | 6.16183103052166E+00 | 15 | 0.000 | 6.16183103052165E+00 | 15 | 0.016 |
| 1.E-05 | 6.16226077809064E+00 | 15 | 0.000 | 6.16226077809063E+00 | 15 | 0.000 |
| 1.E-07 | 6.16226507425985E+00 | 15 | 0.000 | 6.16226507425985E+00 | 15 | 0.000 |
| 1.E-09 | 6.16226511722141E+00 | 15 | 0.000 | 6.16226511722141E+00 | 15 | 0.000 |
| 1.E-11 | 6.16226511765103E+00 | 15 | 0.000 | 6.16226511765102E+00 | 15 | 0.016 |
| 1.E-13 | 6.16226511765533E+00 | 15 | 0.015 | 6.16226511765532E+00 | 15 | 0.015 |
| 1.E-15 | 6.16226511765537E+00 | 15 | 0.016 | 6.16226511765536E+00 | 15 | 0.000 |
| 1.E-17 | 6.16226511765537E+00 | 15 | 0.016 | 6.16226511765536E+00 | 15 | 0.000 |
| 1.E-19 | 6.16226511765537E+00 | 15 | 0.015 | 6.16226511765536E+00 | 15 | 0.000 |
| 0.E+00 | 6.16226511765537E+00 | 15 | 0.016 | 6.16226511765536E+00 | 15 | 0.000 |

*CPU time (in seconds) spent in Mathematica® kernel

The fifteenth and last integral is the one in (9.21)

$$I_{52}(-1,-1,h) = \int_0^2 \frac{u^2 du}{F^3(u)} \left[ \frac{3E^2(u,h)}{4F^2(u)} - D(u,h) \right] \ln \left| \frac{\{2F(u)[R(u,2,h) + 2F(u)] + E(u,h)\} R(u,0,h)}{2F(u)D(u,h) + E(u,h)R(u,0,h)} \right|$$

(10.20)

We display the behavior of the integrand in Fig. 10.15 and that of the integral in Table 10.16. The origin has been specified in Mathematica® as a singular point.

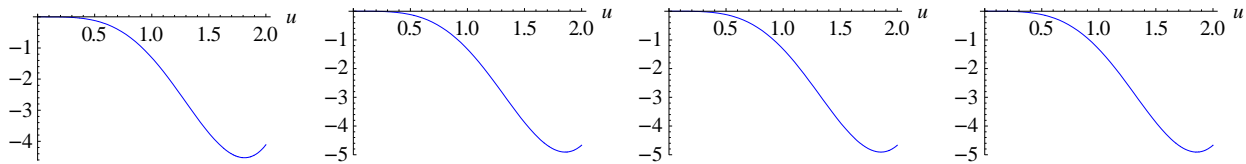

Figure 10.15. Integrand of (9.21) for the BQ and OP of the present section. From left to right: $h = 0.1$, $10^{-5}$, $10^{-19}$, $0.0$.

Table 10.16. Equation (9.21) evaluated for the second BQ and for the OP (-1, -1, *h*).

| | SECOND BQ, EQ. (9.21) AT (-1, -1, *h*) | | | | | |
|---|---|---|---|---|---|---|
| *h* | GKQ | GKQ SD | GKQ TIME* | DEQ | DEQ SD | DEQ TIME* |
| 1.E+00 | **-1.51467528654500E+00** | 15 | 0.016 | **-1.51467528654499E+00** | 15 | 0.000 |
| 1.E-01 | **-3.70097698584184E+00** | 15 | 0.000 | **-3.70097698584184E+00** | 15 | 0.000 |
| 1.E-03 | **-3.88962192550836E+00** | 15 | 0.015 | **-3.88962192550836E+00** | 15 | 0.016 |
| 1.E-05 | **-3.89146716923316E+00** | 15 | 0.016 | **-3.89146716923315E+00** | 15 | 0.015 |
| 1.E-07 | **-3.89148561762424E+00** | 15 | 0.000 | **-3.89148561762424E+00** | 15 | 0.000 |
| 1.E-09 | **-3.89148580210775E+00** | 15 | 0.016 | **-3.89148580210774E+00** | 15 | 0.016 |
| 1.E-11 | **-3.89148580395258E+00** | 15 | 0.015 | **-3.89148580395258E+00** | 15 | 0.015 |
| 1.E-13 | **-3.89148580397103E+00** | 15 | 0.000 | **-3.89148580397103E+00** | 15 | 0.016 |
| 1.E-15 | **-3.89148580397122E+00** | 15 | 0.016 | **-3.89148580397121E+00** | 15 | 0.016 |
| 1.E-17 | **-3.89148580397122E+00** | 15 | 0.000 | **-3.89148580397121E+00** | 15 | 0.015 |
| 1.E-19 | **-3.89148580397122E+00** | 15 | 0.015 | **-3.89148580397121E+00** | 15 | 0.016 |
| 0.E+00 | **-3.89148580397122E+00** | 15 | 0.000 | **-3.89148580397121E+00** | 15 | 0.016 |

*CPU time (in seconds) spent in Mathematica® kernel

# 11. SUMMARY AND CONCLUSIONS

In this report we have explored ways to compute the singular part of the integral of MFIE over a BQ. In doing so we have followed the approach for the corresponding integral of EFIE [5]. In Section 2 we isolated the singular part of the integral and we broke it down into a number of sub-integrals the integrands of which appear in (2.44) and (2.45). In Sections 3 through 10, we proceeded to show how to evaluate these integrals. These are two-dimensional integrals and, in all cases, they can be evaluated analytically along one of the two dimensions This results in one-dimensional integrals with a more complicated integrand. Some of these integrals may be evaluated analytically. We tried that and found that we had to locate poles of the integrand numerically and that, depending on the values of the parameters involved, we would have to use a number of different integration formulas. For these reasons we abandoned this approach in favor of a numerical evaluation of the integrals.

In performing the numerical evaluation of the one-dimensional integrals, we used two numerical algorithms in Mathematica® 7: GKQ and DEQ. We performed tests on the second BQ of Appendix B and we used OPs both interior to the BQ and on its boundary. On most occasions, we had the two quadratures agreeing to at least 14 SD. If we felt that one quadrature was preferable to the other for a specific integral, we stated so. Thus, if the interested reader has the development of a software package based on MFIE, s/he can design an algorithm based on one of the two approaches. Our experience in working with these two algorithms is that they can go a long way in providing desired accuracy if they are thoughtfully designed.

The trade-off in reducing the original two-dimensional integral to a one-dimensional one through exact analytical integration is that, in the process, we made the integrand more time consuming to compute numerically. Although we have not run tests, we do not believe that the one dimensional integral comes close to requiring the time that the two-dimensional integral requires for a numerical evaluation; moreover, a two-dimensional numerical subroutine would have to be designed very carefully to deal with the integrals encountered in this report.

In concluding, we point out that more tests can be performed as, for example, the case of two flat BQs with the OP on one and near the common boundary of the two BQ. One may also wish to consider higher-power terms in (2.44) and (2.45), generated by higher-order basis functions. They can be easily dealt with using integration by parts. Formulas 2.263.1, p. 82, in [11] does just that.

**APPENDIX A:** GEOMETRY OF A BILINEAR QUADRILATERAL

In this appendix we provide a basic description of a BQ. With respect to a rectangular coordinate system *xyz*, we define a BQ by

$$\mathbf{r}(p,q) = \frac{1}{4}\left[\mathbf{r}_{11}(1-p)(1-q) + \mathbf{r}_{12}(1-p)(q+1) + \mathbf{r}_{21}(p+1)(1-q) + \mathbf{r}_{22}(p+1)(q+1)\right],$$

$$|p| \leq 1, \quad |q| \leq 1. \tag{A.1}$$

The four constant vectors $\mathbf{r}_{ij}$ denote the four corners of the BQ. We note that (-1, -1) maps to $\mathbf{r}_{11}$, (-1, 1) maps to $\mathbf{r}_{12}$, (1, 1) maps to $\mathbf{r}_{22}$ and (1, -1) maps to $\mathbf{r}_{21}$. We show this in Fig. (A.1).

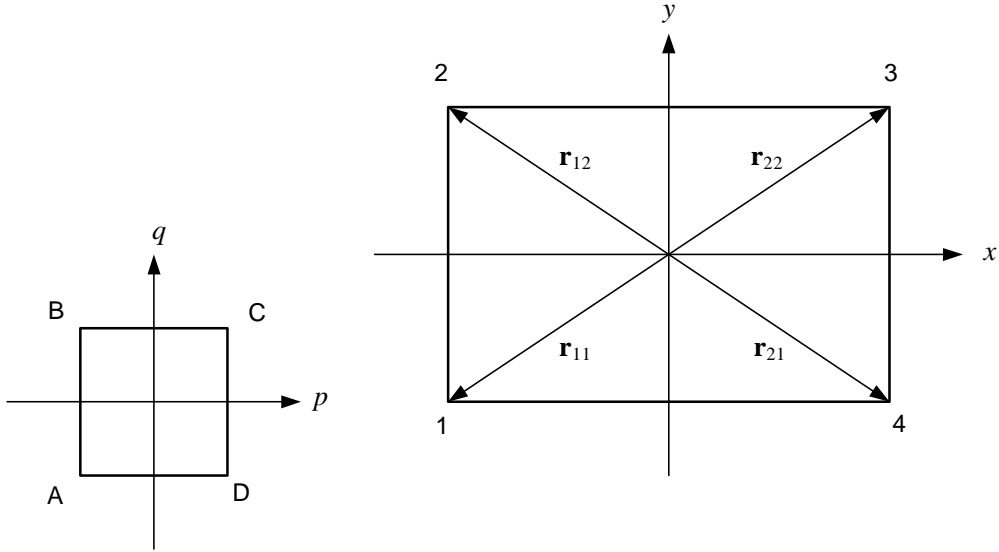

Figure A.1. Example of how the basic square in the *pq*-plane maps to a BQ in the *xy*-plane that occupies the region $|x| \leq 3$, $|y| \leq 2$. The point A maps to 1, the point B to 2, point C to 3 and point D to 4.

We collect terms in *p*, *q* and their product in (A.1) to write

$$\mathbf{r}(p,q) = \mathbf{r}_{00} + \mathbf{r}_p p + \mathbf{r}_q q + \mathbf{r}_{pq} pq, \quad |p| \leq 1, \quad |q| \leq 1 \tag{A.2}$$

where

$$\mathbf{r}_{00} = \frac{1}{4}\left[\mathbf{r}_{11} + \mathbf{r}_{12} + \mathbf{r}_{21} + \mathbf{r}_{22}\right] \tag{A.3}$$

$$\mathbf{r}_p = \frac{1}{4}\left[-\mathbf{r}_{11} - \mathbf{r}_{12} + \mathbf{r}_{21} + \mathbf{r}_{22}\right], \quad \mathbf{r}_q = \frac{1}{4}\left[-\mathbf{r}_{11} + \mathbf{r}_{12} - \mathbf{r}_{21} + \mathbf{r}_{22}\right] \tag{A.4}$$

$$\mathbf{r}_{pq} = \frac{1}{4}\left[\mathbf{r}_{11} - \mathbf{r}_{12} - \mathbf{r}_{21} + \mathbf{r}_{22}\right]. \tag{A.5}$$

For the example of Fig. A.1, $\mathbf{r}_{00} = \mathbf{0}$, $\mathbf{r}_p = 3\hat{x}$, $\mathbf{r}_q = 2\hat{y}$, $\mathbf{r}_{pq} = \mathbf{0}$.

The function in (A.2) maps points from the square $|p| \leq 1$, $|q| \leq 1$ to points in the *xyz*-space. The matrix of this transformation is

$$M = \begin{Vmatrix} \frac{\partial}{\partial p}(\hat{x} \cdot \mathbf{r}) & \frac{\partial}{\partial p}(\hat{y} \cdot \mathbf{r}) & \frac{\partial}{\partial p}(\hat{z} \cdot \mathbf{r}) \\ \frac{\partial}{\partial q}(\hat{x} \cdot \mathbf{r}) & \frac{\partial}{\partial q}(\hat{y} \cdot \mathbf{r}) & \frac{\partial}{\partial q}(\hat{z} \cdot \mathbf{r}) \end{Vmatrix} = \begin{Vmatrix} \hat{x} \cdot (\mathbf{r}_p + \mathbf{r}_{pq}q) & \hat{y} \cdot (\mathbf{r}_p + \mathbf{r}_{pq}q) & \hat{z} \cdot (\mathbf{r}_p + \mathbf{r}_{pq}q) \\ \hat{x} \cdot (\mathbf{r}_q + \mathbf{r}_{pq}p) & \hat{y} \cdot (\mathbf{r}_q + \mathbf{r}_{pq}p) & \hat{z} \cdot (\mathbf{r}_q + \mathbf{r}_{pq}p) \end{Vmatrix} \tag{A.6}$$

and it must be of rank 2 if we are not to have any singular points. This means that

$$\frac{\partial \mathbf{r}(p,q)}{\partial p} \times \frac{\partial \mathbf{r}(p,q)}{\partial q} \neq \mathbf{0} \tag{A.7}$$

at every non-singular point $(p, q)$. This follows from the fact that the three 2x2 sub-matrices must have a non-zero determinant. Since the first of these vectors is tangent to a constant *q*-curve while the second to a constant *p*-curve, condition (A.7) says that the two vectors should not be collinear at a point ([13], pp. 56 – 57). What does this mean geometrically? From (A.2) we have that

$$\frac{\partial \mathbf{r}(p,q)}{\partial p} \times \frac{\partial \mathbf{r}(p,q)}{\partial q} = \left(\mathbf{r}_p + \mathbf{r}_{pq}q\right) \times \left(\mathbf{r}_q + \mathbf{r}_{pq}p\right) = \mathbf{r}_p \times \mathbf{r}_q + \mathbf{r}_{pq} \times \left(\mathbf{r}_q q - \mathbf{r}_p p\right). \tag{A.8}$$

If we assume that $\mathbf{r}_p$ and $\mathbf{r}_q$ are not collinear (and, hence, their cross product is not zero), then they define a plane in space. The vector $\left(\mathbf{r}_q q - \mathbf{r}_p p\right)$ lies on that plane. Then, a necessary condition for (A.8) to be equal to zero is that the vector $\mathbf{r}_{pq}$ does not have a component normal to that plane. Differently stated, for (A.8) to become zero, we must have the last three vectors in (A.2) lying on the same plane. In such a case, (A.2) describes a planar BQ. From this discussion, it is clear that the test for determining whether a BQ is planar is that $\mathbf{r}_{pq} \cdot \left(\mathbf{r}_p \times \mathbf{r}_q\right) = 0$. It is obvious from this condition that if $\mathbf{r}_p$ and $\mathbf{r}_q$ are collinear, we then have an irregular BQ.

Although (A.8) can be equal to zero only for a planar BQ, the converse is not true, *i.e.*, (A.8) is not zero for *all* planar BQ. We investigate this further. For a planar BQ, we can resolve the last vector in (A.2) along the directions of the preceding two vectors

$$\mathbf{r}_{pq} = \mathsf{r}\mathbf{r}_p + \mathsf{s}\mathbf{r}_q \tag{A.9}$$

where $\mathsf{r}$ and $\mathsf{s}$ are scalars. In place of (A.2), we now have

$$\mathbf{r}(p,q) = \mathbf{r}_{00} + \mathbf{r}_p p(1 + \mathsf{r}q) + \mathbf{r}_q q(1 + \mathsf{s}p), \quad |p| \leq 1, \quad |q| \leq 1. \tag{A.10}$$

Using this, we have

$$\frac{\partial \mathbf{r}(p,q)}{\partial p} \times \frac{\partial \mathbf{r}(p,q)}{\partial q} = \left[\mathbf{r}_p(1+\mathsf{r}q)+\mathbf{r}_q\mathsf{s}q\right] \times \left[\mathbf{r}_p\mathsf{r}p+\mathbf{r}_q(1+\mathsf{s}p)\right]$$
$$= (\mathbf{r}_q \times \mathbf{r}_p)\mathsf{rs}\,pq + (\mathbf{r}_p \times \mathbf{r}_q)(1+\mathsf{r}q)(1+\mathsf{s}p) = (\mathbf{r}_p \times \mathbf{r}_q)(1+\mathsf{r}q+\mathsf{s}p). \qquad (A.11)$$

If the cross product is zero, then we have a straight line in space. We exclude this case. When is the scalar part zero? We let

$$P = \mathsf{s}p, \quad Q = \mathsf{r}q. \qquad (A.12)$$

From the conditions on $p$ and $q$ in (A.2), we have that

$$|P| = |\mathsf{s}||p| \le |\mathsf{s}|, \quad |Q| = |\mathsf{r}||q| \le |\mathsf{r}|. \qquad (A.13)$$

We also set the scalar factor in (A.11) equal to zero

$$1 + Q + P = 0. \qquad (A.14)$$

We want to know whether this equation is satisfied with values of $P$ and $Q$ in the range of (A.13). In Fig. A.2, the lower left corner of the rectangle is the point closest to the straight line. The value of $P$ there is $-|\mathsf{s}|$ and the value of $Q$ for the straight line is $-1+|\mathsf{s}|$. For no intersection, we want $-1+|\mathsf{s}|$ to be smaller than $-|\mathsf{r}|$; for one intersection point, we want $-1+|\mathsf{s}|$ to be equal to $-|\mathsf{r}|$; and, for an uncountable number of intersection points, we want $-1+|\mathsf{s}|$ to be greater than $-|\mathsf{r}|$. We summarize these results in Table A.1.

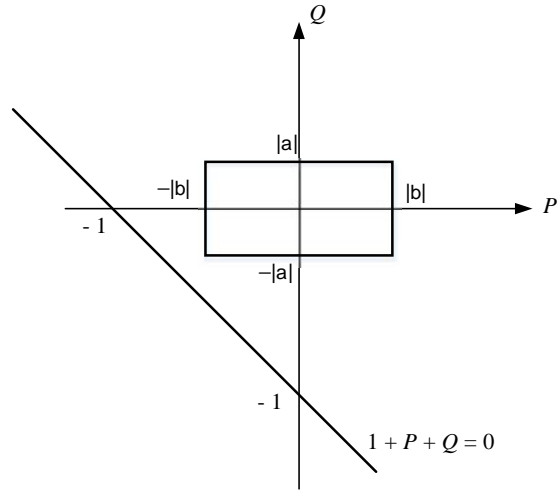

Figure A.2. Geometry for the intersection of the straight line with the rectangle.

Table A.1. Solutions of (A.14) subject to (A.13).

| | | | |
|---|---|---|---|
| $|r|+|s|<1$ | No intersection points | No solution of (A.14) | Regular BQ |
| $|r|+|s|=1$ | One intersection point | One solution of (A.14) | Irregular BQ |
| $|r|+|s|>1$ | Many intersection points | Many solutions of (A.14) | Irregular BQ |

We demonstrate these results using three BQs. The position vector for the first is

$$\mathbf{r}(p,q) = \mathbf{r}_p p(1+0.2q) + \mathbf{r}_q q(1+0.3p), \quad (r=0.2, s=0.3) \tag{A.15}$$

and its graph is shown in Fig. A.3. This is a regular BQ. The position vector of the second BQ is

$$\mathbf{r}(p,q) = \mathbf{r}_p p(1+0.5q) + \mathbf{r}_q q(1+0.5p), \quad (r=0.5, s=0.5) \tag{A.16}$$

and its graph is shown in Fig. A.4. This is an irregular BQ in the form of a triangle. The position vector of the third BQ is

$$\mathbf{r}(p,q) = \mathbf{r}_p p(1+2q) + \mathbf{r}_q q(1+3p), \quad (r=2, s=3). \tag{A.17}$$

Its graph is given in Fig. A.5. This is also an irregular BQ. In all three cases, $\mathbf{r}_p = (3, 0, -0.5)$ while $\mathbf{r}_q = (0, 2, -0.5)$.

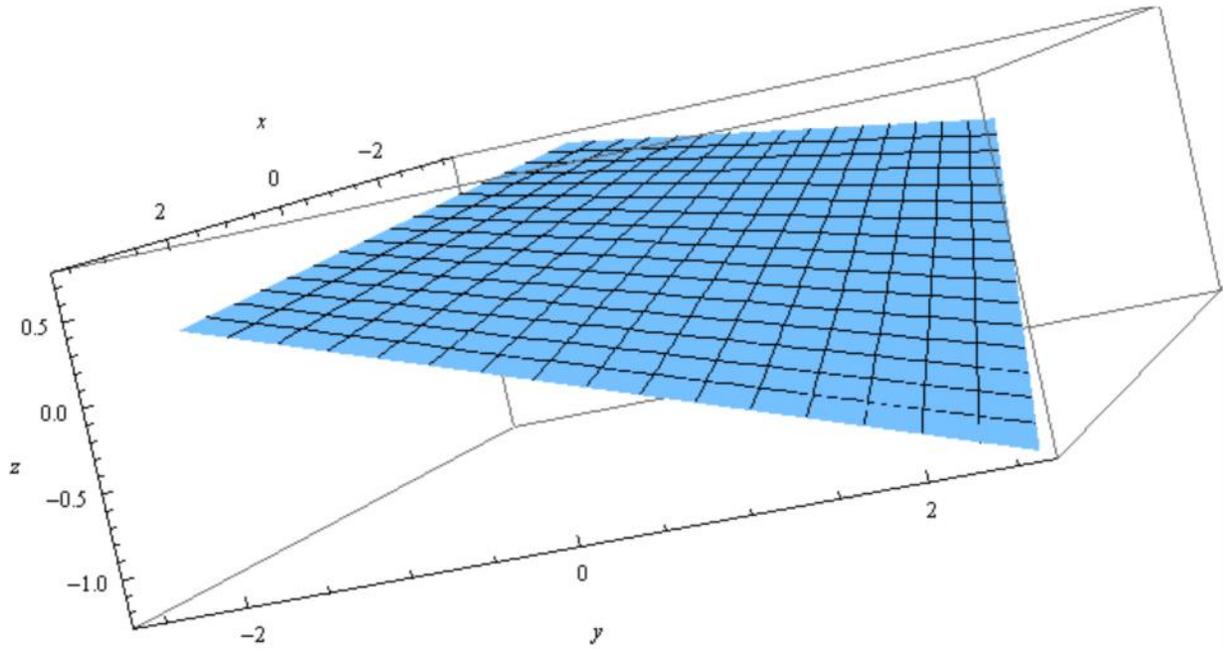

Figure A.3. Graph of (A.15), a planar, regular BQ.

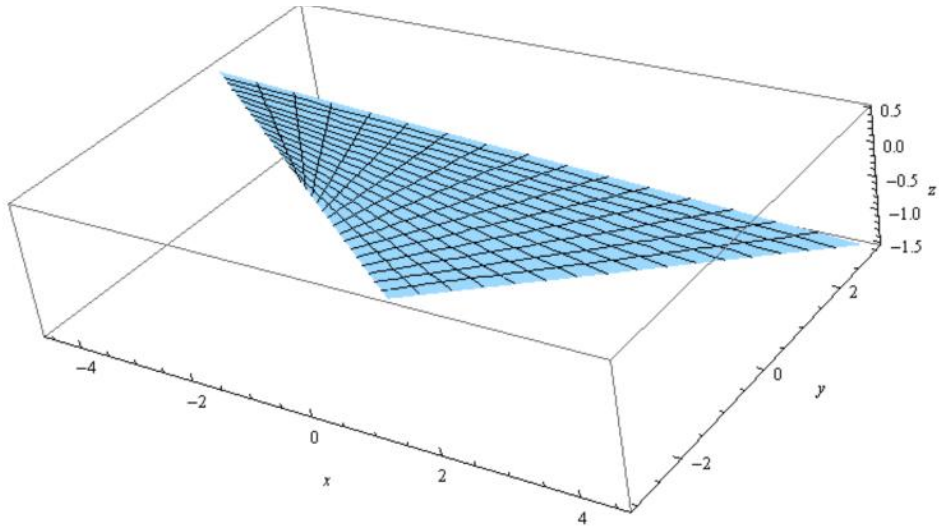

Figure A.4. Graph of (A.16), a planar, barely irregular BQ.

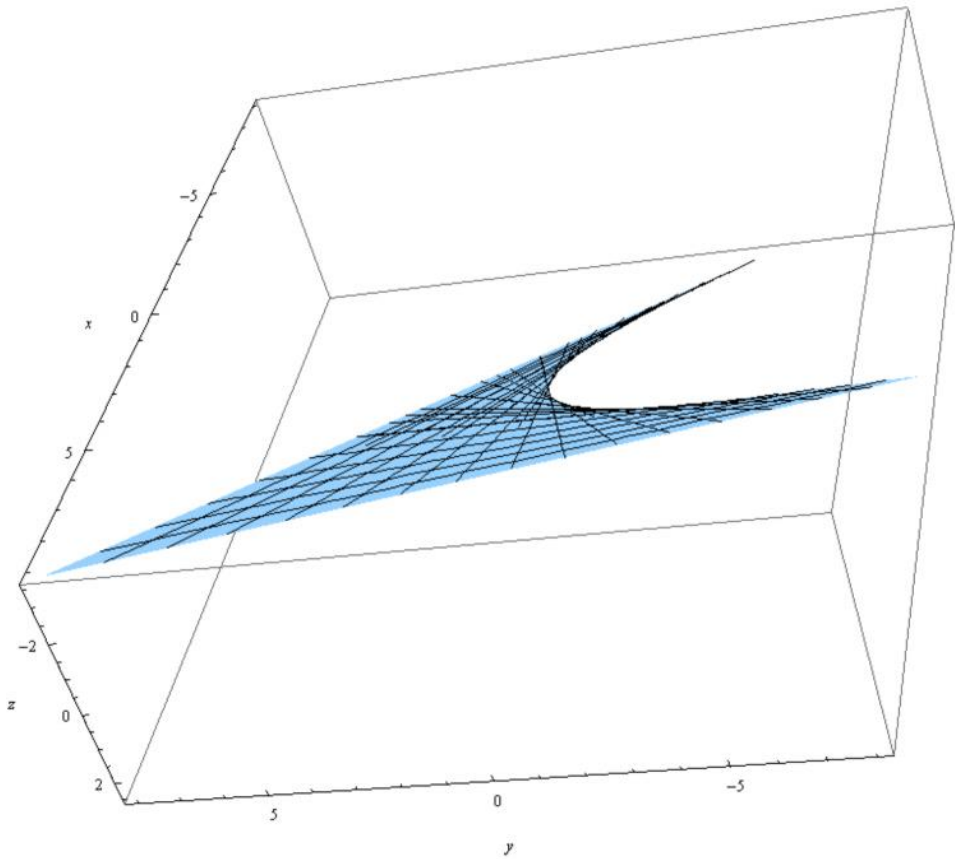

Figure A.5. Graph of (A.17), a planar, highly irregular BQ.

The unit normal at a point on the surface is given by ([13], p. 62)

$$\hat{n} = \frac{\dfrac{\partial \mathbf{r}(p,q)}{\partial p} \times \dfrac{\partial \mathbf{r}(p,q)}{\partial q}}{\left| \dfrac{\partial \mathbf{r}(p,q)}{\partial p} \times \dfrac{\partial \mathbf{r}(p,q)}{\partial q} \right|} \tag{A.18}$$

and we note that we could also have defined it as the negative of this. The surface element at a point is defined by

$$dS = \left| \frac{\partial \mathbf{r}(p,q)}{\partial p} \times \frac{\partial \mathbf{r}(p,q)}{\partial q} \right| dpdq. \tag{A.19}$$

If we hold one of the parameters constant, we can measure arc length along the other according to the formula ([13], p. 58)

$$s = \int_{p_1}^{p_2} \left| \frac{\partial \mathbf{r}(p,q)}{\partial p} \right| dp = \int_{p_1}^{p_2} \left| \mathbf{r}_p + \mathbf{r}_{pq} q \right| dp = \left| \mathbf{r}_p + \mathbf{r}_{pq} q \right| (p_2 - p_1), \quad -1 \le p_1 < p_2 \le 1, \quad q = \text{const.}$$
$$\tag{A.20}$$

Similarly,

$$s = \int_{q_1}^{q_2} \left| \frac{\partial \mathbf{r}(p,q)}{\partial q} \right| dq = \left| \mathbf{r}_q + \mathbf{r}_{pq} p \right| (q_2 - q_1), \quad -1 \le q_1 < q_2 \le 1, \quad p = \text{const.} \tag{A.21}$$

We point out that if we have a curve on the BQ described parametrically by $p(t)$ and $q(t)$, then we have a more general formula for its arc length ([13], p. 58). In any case, the above formulas make sense since we are moving along straight lines.

We consider next a point $(p_0 \; q_0)$ in the square of definition of these variables. This maps to the point $\mathbf{r}_0$ of the BQ

$$\mathbf{r}_0 = \mathbf{r}(p_0, q_0) = \mathbf{r}_{00} + \mathbf{r}_p p_0 + \mathbf{r}_q q_0 + \mathbf{r}_{pq} p_0 q_0, \quad |p_0| \le 1, \quad |q_0| \le 1. \tag{A.22}$$

We can reference (A.2) to this point by expanding in a Taylor series about it

$$\mathbf{r}(p,q) = \mathbf{r}_0 + \left( \mathbf{r}_p + \mathbf{r}_{pq} q_0 \right)(p - p_0) + \left( \mathbf{r}_q + \mathbf{r}_{pq} p_0 \right)(q - q_0) + \mathbf{r}_{pq}(p - p_0)(q - q_0). \tag{A.23}$$

This expression is exact. We can easily show this by collecting terms in powers of $p$, $q$ and $pq$. At the point $\mathbf{r}_0$, we define coordinates $p_0 q_0 n$. The unit vectors are

$$\hat{p}_0 = \frac{\frac{\partial \mathbf{r}(p_0, q_0)}{\partial p_0}}{\left|\frac{\partial \mathbf{r}(p_0, q_0)}{\partial p_0}\right|}, \quad \hat{q}_0 = \frac{\frac{\partial \mathbf{r}(p_0, q_0)}{\partial q_0}}{\left|\frac{\partial \mathbf{r}(p_0, q_0)}{\partial q_0}\right|}, \quad \hat{n} = \frac{\frac{\partial \mathbf{r}(p_0, q_0)}{\partial p_0} \times \frac{\partial \mathbf{r}(p_0, q_0)}{\partial q_0}}{\left|\frac{\partial \mathbf{r}(p_0, q_0)}{\partial p_0} \times \frac{\partial \mathbf{r}(p_0, q_0)}{\partial q_0}\right|}. \tag{A.24}$$

Although $\hat{p}_0$ and $\hat{q}_0$ are perpendicular to $\hat{n}$, they are not necessarily perpendicular to each other. Letting

$$\mathbf{p}_0 = \frac{\partial \mathbf{r}(p_0, q_0)}{\partial p_0}, \quad \mathbf{q}_0 = \frac{\partial \mathbf{r}(p_0, q_0)}{\partial q_0} \tag{A.25}$$

we can write for (A.23)

$$\mathbf{r}(p, q) = \mathbf{r}_0 + \mathbf{p}_0 u + \mathbf{q}_0 v + \mathbf{r}_{pq} uv \tag{A.26}$$

where

$$u = p - p_0, \quad v = q - q_0; \quad -(1 + p_0) \leq u \leq 1 - p_0, \quad -(1 + q_0) \leq v \leq 1 - q_0. \tag{A.27}$$

## APPENDIX B: TEST BILINEAR QUADRILATERALS

We present here three BQs that may be used for testing purposes. We begin with a flat BQ whose corners are given by

$$\mathbf{r}_{11} = (-3,-2,1), \quad \mathbf{r}_{22} = (3,2,-1), \quad \mathbf{r}_{12} = (-3,2,0), \quad \mathbf{r}_{21} = (3,-2,0) \tag{B.1}$$

From (A.3) – (A.5)

$$\mathbf{r}_0 = (0,0,0), \quad \mathbf{r}_p = \frac{1}{2}(6,0,-1), \quad \mathbf{r}_q = \frac{1}{2}(0,4,-1), \quad \mathbf{r}_{pq} = (0,0,0). \tag{B.2}$$

From (A.2), we then have

$$\mathbf{r}(p,q) = \frac{1}{2}(6,0,-1)p + \frac{1}{2}(0,4,-1)q, \quad |p| \leq 1, \quad |q| \leq 1. \tag{B.3}$$

The graph of the BQ is shown in Fig. B.1. This is a flat BQ. The unit normal is given in (A.18).

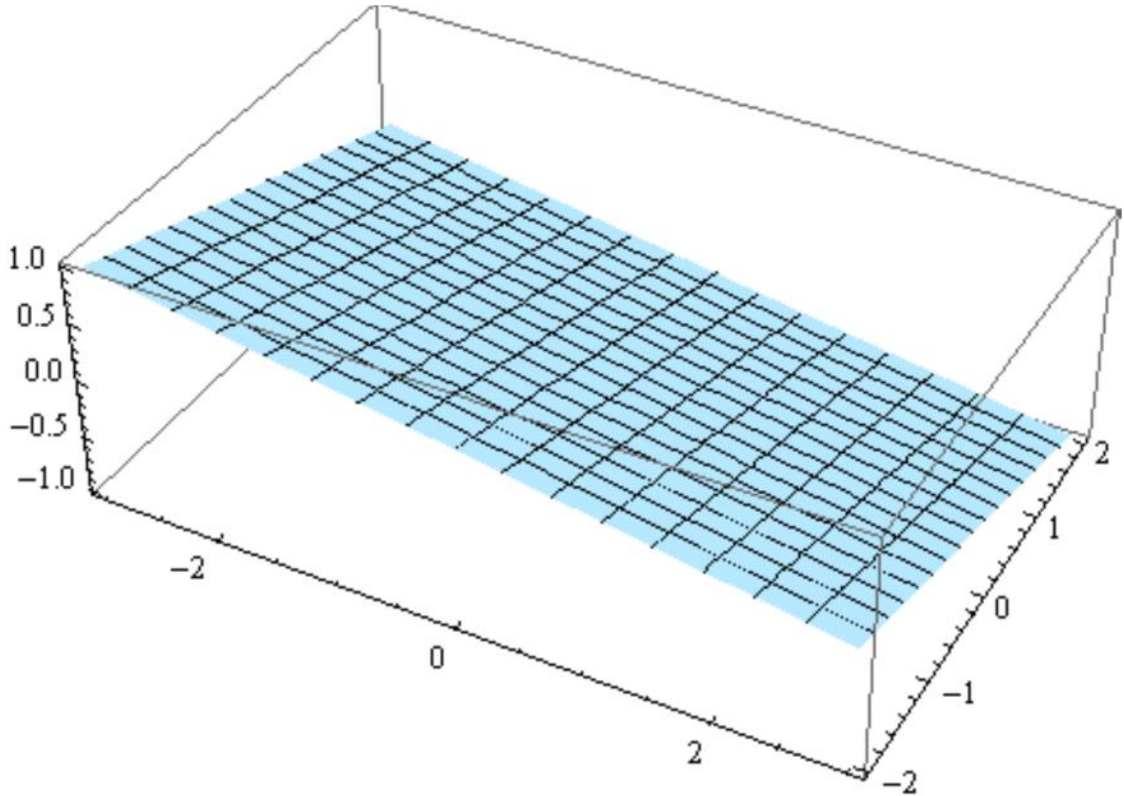

Figure B.1. BQ of Equation (B.3)

The second BQ has corner vectors

$$\mathbf{r}_{11} = (-3,-2,1), \quad \mathbf{r}_{22} = (3,2,1), \quad \mathbf{r}_{12} = (-3,2,-1), \quad \mathbf{r}_{21} = (3,-2,-1) \tag{B.4}$$

from which

$$\mathbf{r}_0 = (0,0,0), \quad \mathbf{r}_p = (3,0,0), \quad \mathbf{r}_q = (0,2,0), \quad \mathbf{r}_{pq} = (0,0,1) \tag{B.5}$$

so that

$$\mathbf{r}(p,q) = (3,0,0)p + (0,2,0)q + (0,0,1)pq, \quad |p| \leq 1, \quad |q| \leq 1. \tag{B.6}$$

The graph of this BQ is shown in Fig. B.2.

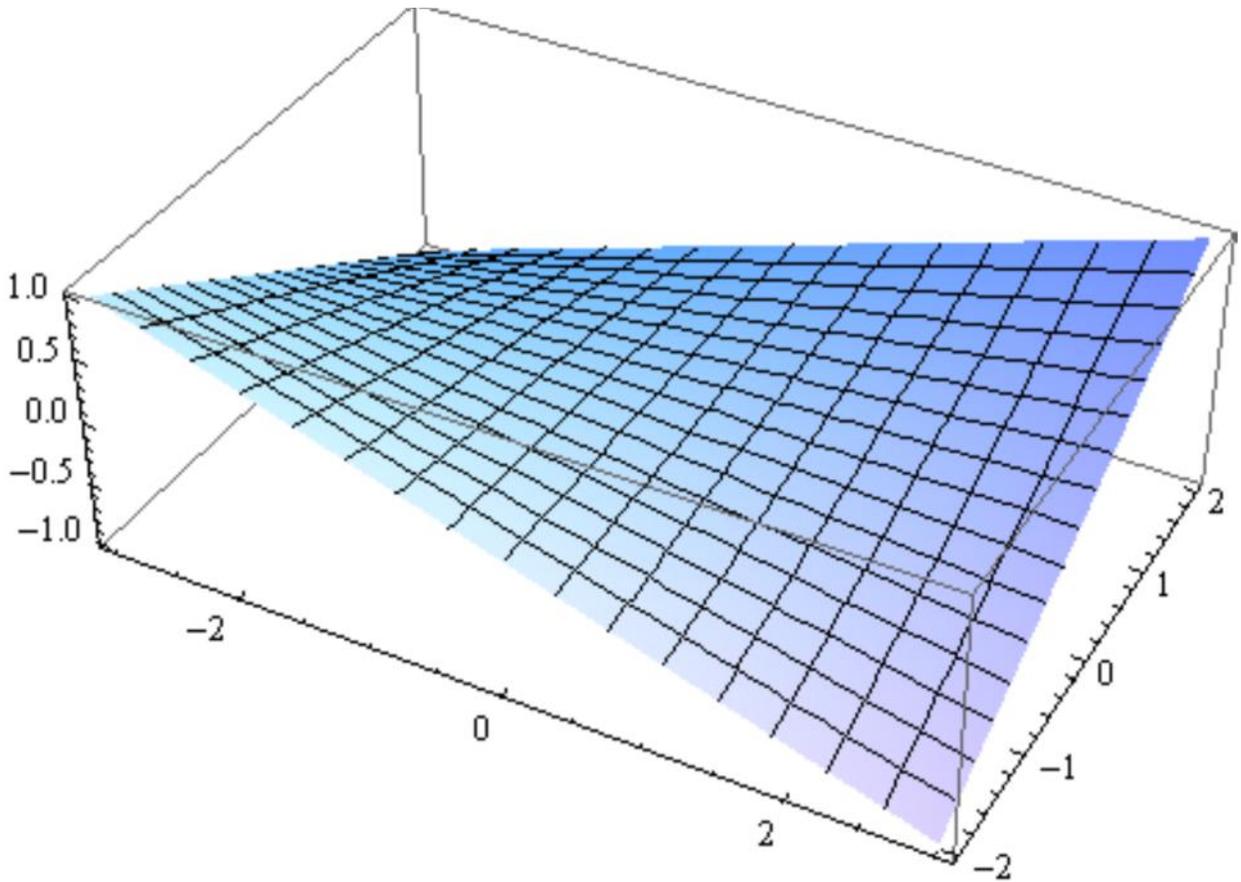

Figure B.2. BQ of Equation (B.6).

We go next to an extreme case

$$\mathbf{r}_{11} = (-3,-2,10), \quad \mathbf{r}_{22} = (3,2,10), \quad \mathbf{r}_{12} = (-3,2,-10), \quad \mathbf{r}_{21} = (3,-2,-10) \tag{B.7}$$

so that

$$\mathbf{r}_0 = (0,0,0), \quad \mathbf{r}_p = (3,0,0), \quad \mathbf{r}_q = (0,2,0), \quad \mathbf{r}_{pq} = (0,0,10) \tag{B.8}$$

and

$$\mathbf{r}(p,q) = (3,0,0)p + (0,2,0)q + (0,0,10)pq, \quad |p| \le 1, \quad |q| \le 1. \tag{B.9}$$

The graph of this BQ is shown in Figs. B.3 and B.4.

We note that condition (A.7) is satisfied at all points of these quadrilaterals. For example, for the last BQ we have

$$\frac{\partial \mathbf{r}(p,q)}{\partial p} \times \frac{\partial \mathbf{r}(p,q)}{\partial q} = \left[(3,0,0) + (0,0,10)q\right] \times \left[(0,2,0) + (0,0,10)p\right]$$
$$= (3,0,0) \times (0,2,0) + (3,0,0) \times (0,0,10)p + (0,0,10) \times (0,2,0)q$$
$$= 6\hat{z} - 30p\hat{y} - 20q\hat{x} \tag{B.10}$$

and the magnitude of this vector is always greater than zero. This means that we have a tangent plane at every point of the BQ with a normal given by (A.18). For the example

$$\hat{n} = \frac{-20q\hat{x} - 30p\hat{y} + 6\hat{z}}{\sqrt{(20q)^2 + (30p)^2 + 36}}. \tag{B.11}$$

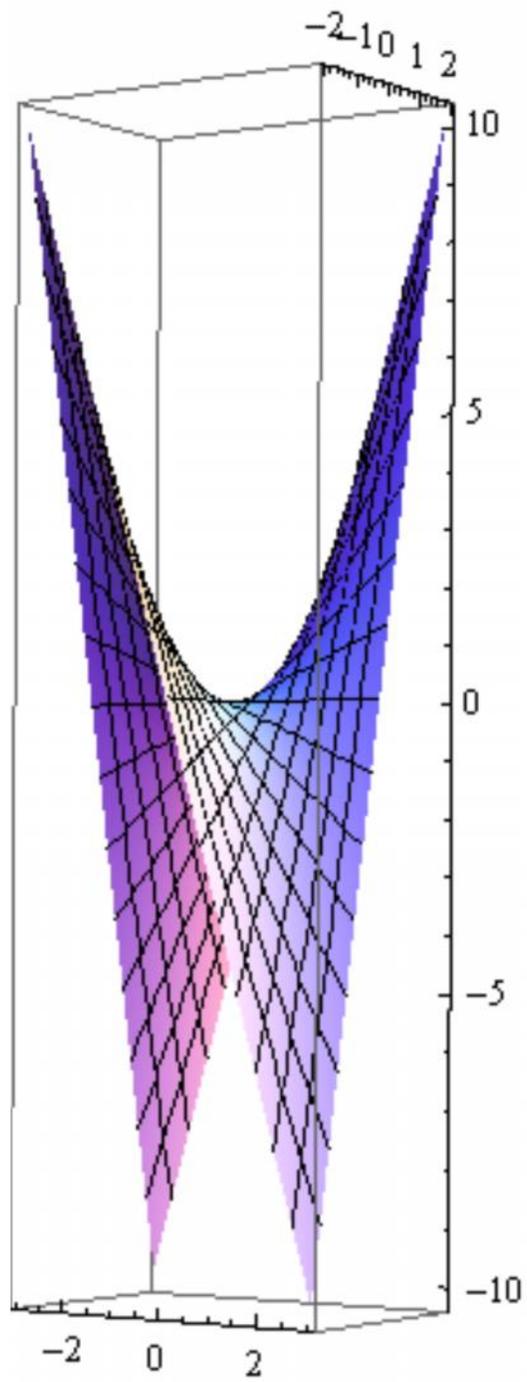

Figure B.3. BQ of Equation (B.9).

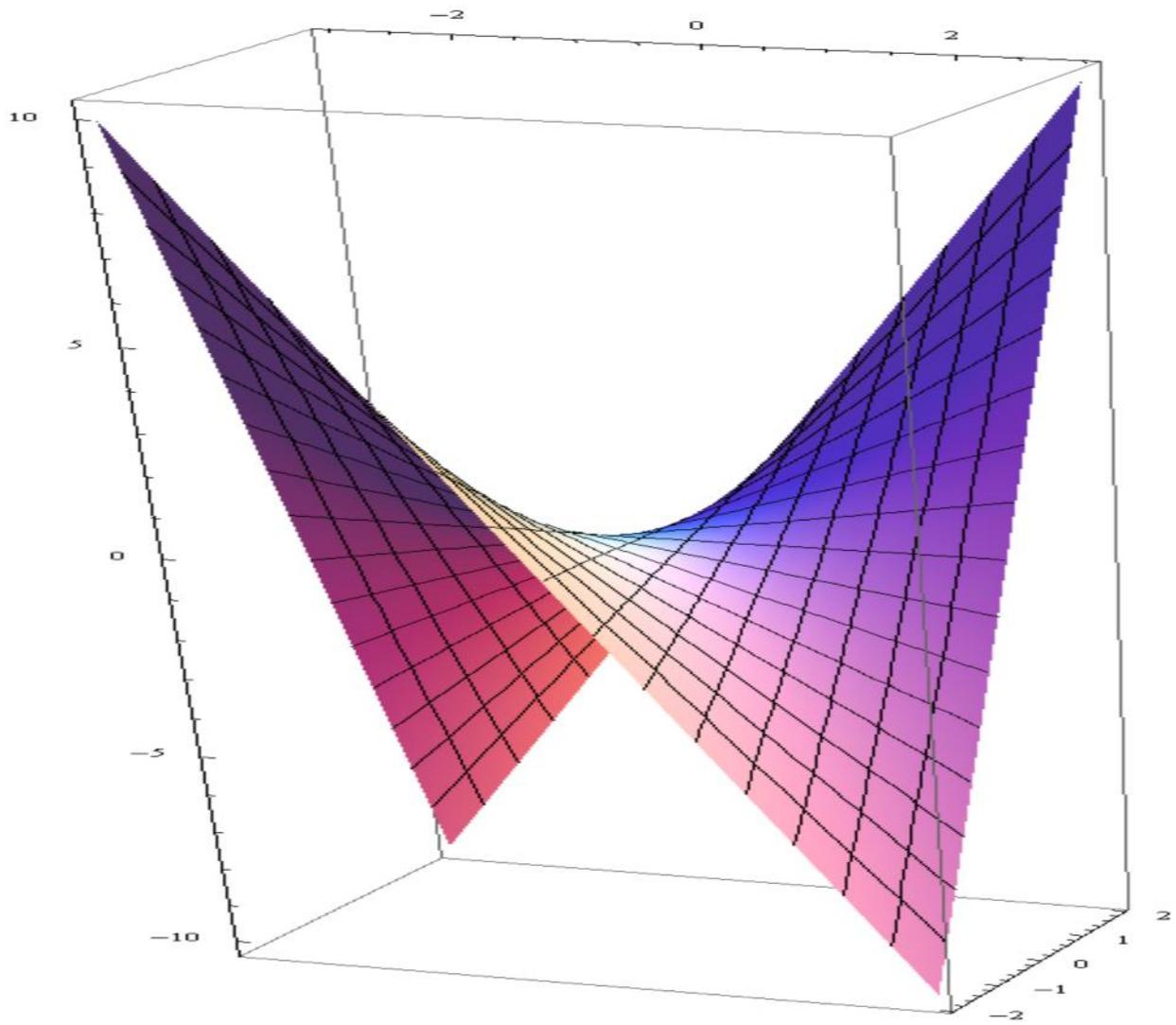

Figure B.4. Another view of the graph of the BQ of Equation (B.9).

# REFERENCES

1. J. Van Bladel, *Electromagnetic Fields*. New York: McGraw – Hill, 1964.
2. D. S. Jones, Methods in Electromagnetic Wave Propagation. Oxford: Clarendon Press, 1979.
3. B. M. Notaroš, "Higher Order Frequency-Domain Computational Electromagnetics", invited review paper, *IEEE Trans. Antennas Propagat.*, Vol. 56, No. 8, pp. 2251 - 2276, 2008.
4. B. M. Kolundzija and A. R. Djordjevi , *Electromagnetic Modeling of Composite Metallic and Dielectric Structures*. Boston: Artech House, 2002. Web site: http://www.wipl-d.com/
5. J. S. Asvestas, "Calculation of Moment Matrix Elements for Bilinear Quadrilaterals and Higher Order-Basis Functions", NAVAIR Technical Report NAWCADPAX/TR-2015-241, Patuxent River, MD, 6 Jan. 2016.
6. Wolfram Research, Inc., *Mathematica*, Version 7.0. Champaign, IL, 2008.
7. P. J. Davis and P. Rabinowitz, *Methods of Numerical Integration*, Second ed. Orlando, FL: Academic Press, 1984.
8. H. Takahasi and M. Mori, "Double Exponential Formulas for Numerical Integration", *Research Institute for Mathematical Sciences, Kyoto Univ.,* Vol. 9, pp. 721 – 741, 1974.
9. C. Müller, *Foundations of the Mathematical Theory of Electromagnetic Waves*. New York: Springer – Verlag, 1969.
10. J. A. Stratton, *Electromagnetic Theory*. New York: McGraw – Hill, 1941.
11. I. S. Gradshteyn and I. M. Ryzhik, *Table of Integrals, Series, and Products*. Corrected and Enlarged Edition. New York: Academic Press, 1980.
12. G. Dahlquist and Å. Björck, *Numerical Methods*. Englewood Cliffs, NJ: Prentice – Hall, 1974.
13. D. J. Struik, Classical Differential Geometry. Reading, MA: Addison – Wesley, 1961.